\documentclass[fleqn,usenatbib, useAMS]{mnras}

\usepackage{newtxtext,newtxmath}

\usepackage[T1]{fontenc}
\usepackage{ae,aecompl}
\usepackage[utf8]{inputenc}
\usepackage{threeparttable}
\usepackage{booktabs, caption, makecell}
\usepackage{float}

\usepackage{graphicx}	
\usepackage{amsmath}	

\usepackage{rotating}
\usepackage{float}
\usepackage{tabularx}
\usepackage{natbib}
\usepackage{hyperref}

\title[Globular Clusters UVIT Legacy Survey (GlobULeS)]{Globular Clusters UVIT Legacy Survey (GlobULeS) I. FUV-optical Color-Magnitude Diagrams for Eight Globular Clusters}

\author[Sahu et al.]{Snehalata Sahu,$^{1,2}$\thanks{E-mail: snehalatash30@gmail.com}
Annapurni Subramaniam$^{2}$\thanks{Email: purni@iiap.res.in}, Gaurav Singh$^{3}$, Ramakant Yadav$^{3}$,  Aldo R. Valcarce$^{4}$,  \newauthor Samyaday Choudhury$^{5, 6}$, Sharmila Rani$^{2}$, Deepthi S. Prabhu$^{2}$, Chul Chung$^{7}$, Patrick C\^ot\'e $^{8}$, Nathan Leigh$^{9,10}$, \newauthor Aaron M. Geller$^{11,12}$, Sourav Chatterjee$^{1}$, N. Kameswara Rao$^{2}$, Avrajit Bandyopadhyay$^{3}$, Michael Shara$^{10,13}$, \newauthor Emanuele Dalessandro$^{14}$, Gajendra Pandey$^{2}$, Joesph E. Postma$^{15}$, John Hutchings$^{16}$, Mirko Simunovic$^{17}$, \newauthor Peter B. Stetson$^{8}$, Sivarani Thirupathi$^{2}$, Thomas Puzia$^{18}$, Young-Jong Sohn$^{7}$\\
$^{1}$Tata Institute of Fundamental Research, Mumbai 400005\\
$^{2}$Indian Institute of Astrophysics, Koramangala II Block, Bangalore-560034, India\\
$^{3}$Aryabhatta Research Institute of Observational Sciences, Nainital, India\\
$^{4}$Pontificia Univ Catolica Chile, Inst Astrofis, Av Vicuna Mackenna 4860, Santiago 7820436, Chile\\
$^{5}$ Department of Physics and Astronomy, Macquarie University, Balaclava Road, Sydney, NSW 2109, Australia\\
$^{6}$ Research Centre for Astronomy, Astrophysics and Astrophotonics, Macquarie University, Balaclava Road, Sydney, NSW 2109, Australia\\
$^{7}$Department of Astronomy \& Center for Galaxy Evolution Research, Yonsei University, Seoul 03722, Republic of Korea\\
$^{8}$NRC, Herzberg Astronomy \& Astrophysics, 5071 West Saanich Road, Victoria, BC V9E 2E7, Canada\\
$^{9}$Departamento de Astronom\'ia, Facultad Ciencias F\'isicas y Matematicas, Universidad de Concepci\'on, Av. Esteban Iturra s/n Barrio Universitario, \\Casilla 160-C, Concepci\'on, Chile\\
$^{10}$Department of Astrophysics, American Museum of Natural History, Central Park West and 79th Street, New York, NY 10024-5192, USA 2\\
$^{11}$ Center for Interdisciplinary Exploration and Research in Astrophysics (CIERA) and Department of Physics and Astronomy, Northwestern University, \\1800 Sherman Ave., Evanston, IL 60201, USA\\
$^{12}$Adler Planetarium, Department of Astronomy, 1300 S. Lake Shore Drive, Chicago, IL 60605, USA.\\
$^{13}$ Institute of Astronomy, University of Cambridge, Madingley Road, Cambridge CB3 0HA, UK.\\
$^{14}$INAF-Astrophysics and Space Science Observatory Bologna, Via Gobetti 93/3 I-40129 Bologna, Italy.\\
$^{15}$Department of Physics and Astronomy, University of Calgary, Calgary, AB T2N 1N4, Canada.\\
$^{16}$Johns Hopkins University, Baltimore, MD, USA.\\
$^{17}$Subaru Telescope,National Astronomical Observatory of Japan, 650 N Aohoku Pl, Hilo, HI 96720, USA.\\
$^{18}$Institute of Astrophysics, Pontificia Universidad Cat\'olica de Chile, Av. Vicuña MacKenna 4860, 7820436, Santiago, Chile
} 

\date{Accepted 2022 April 21. Received 2022 April 18; in original form 2021 December 22}

\pubyear{2021}

\begin{document}
\label{firstpage}
\pagerange{\pageref{firstpage}--\pageref{lastpage}}
\maketitle

\begin{abstract}
We present the first results of eight Globular Clusters (GCs) from the AstroSat/UVIT Legacy Survey program GlobULeS based on the observations carried out in two FUV filters (F148W and F169M). The FUV-optical and FUV-FUV color-magnitude diagrams (CMDs) of GCs with the proper motion membership were constructed by combining the UVIT data with \textit{HST} UV Globular Cluster Survey (HUGS) data for inner regions and \textit{Gaia}~Early Data Release (EDR3) for regions outside the \textit{HST}'s field. We detect sources as faint as F148W $\sim$ 23.5~mag which are classified based on their locations in CMDs by overlaying stellar evolutionary models. The CMDs of 8 GCs are combined with the previous UVIT studies of 3 GCs to create stacked FUV-optical CMDs to highlight the features/peculiarities found in the different evolutionary sequences. The FUV (F148W) detected stellar  populations of 11 GCs comprises 2,816 Horizontal Branch (HB) stars (190 Extreme HB candidates), 46 post-HB (pHB), 221 Blue Straggler Stars (BSS), and 107 White Dwarf (WD) candidates. We note that the blue HB color extension obtained from F148W$-$G color and the number of FUV detected EHB candidates are strongly correlated with the maximum internal Helium (He) variation within each GC, suggesting that FUV-optical plane is the most sensitive to He abundance variations in the HB. We discuss the potential science cases that will be addressed using these catalogues including HB morphologies, BSSs, pHB, and, WD stars.  
\end{abstract}


\section{Introduction}
Globular clusters (GCs) are old roughly spherical agglomerations of stars that harbour a variety of hot and exotic stellar populations such as blue straggler stars (BSSs) \citep{Sandage1953}, Helium (He) white dwarfs (WDs), Cataclysmic Variables (CVs) \citep{Cool1995, Cool2002} all of which emit substantially in ultraviolet (UV) regime. Dynamical encounters involving binary stars in the crowded environments of GCs lead to the formation of such exotic systems \citep{Shara2006, Hurley2007}. Thus, studying them is crucial for understanding the connection between stellar evolution and dynamics \citep{Hut1992}. However, owing to their rarity compared to the total number of stars in the cluster, detecting and characterising them is not an easy task. These populations have bluer spectral energy distributions (SEDs) than the majority of the main sequence (MS) and red giant branch (RGB) stars. Also, since they dominantly emit in UV wavelengths, identifying them from a crowd of MS and RGB stars becomes easier at such wavelengths than in optical \citep{Ferraro1997, Ferraro1999}. 

Studies have established that Far-UV (FUV) observations are crucial for identifying and probing the nature of such UV bright exotic populations in dense environments of GCs \citep{Knigge2008}. Various space missions such as the Ultraviolet Imaging Telescope (UIT), Hubble Space Telescope (\textit{HST}), and, Galaxy Evolution Explorer (\textit{GALEX}) have highlighted the significance of FUV observations of GCs. Using deep FUV photometry and spectroscopic survey of the core of GC 47~Tuc with \textit{HST}, \cite{Knigge2002, Knigge2008} studied 48 blue sources leading to the discovery of exotic objects that include binary companions to WDs and a BSS. \cite{Dieball2005, Dieball2007, Dieball2010, Dieball2017} detected a large number of dynamically formed stellar populations (BSSs, CVs and WDs) in GCs NGC~2808, M15, M80, NGC~6397 from FUV$-$Near-UV (NUV) Color-Magnitude Diagrams (CMDs) using \textit{HST} observations. Thus, FUV-CMDs have proved to be a powerful tool for probing the nature of exotic objects. However, the above FUV studies with the \textit{HST} have mostly focused on the dense cores of GCs and not their outer radii.

Several literatures \citep{Milone2014, Brown2016, Dalessandro2013} of HB morphologies have shown that the FUV-optical and NUV-optical CMDs are extremely useful in understanding the HB discontinuities and characterising the He-enhanced sub-populations in GCs. FUV-NUV CMDs were also successfully used in identifying the otherwise optically faint Extreme HB (EHB) and Blue Hook (BHk) candidates with effective temperatures $(T_{\rm eff}) >$ 21,000~K and 32,000~K, respectively \citep{Momany2004}. Massive GCs such as NGC~2808, M15, and M80 \citep{Dalessandro2010, Dieball2005, Dieball2007, Dieball2010} host a sizeable population of EHB and BHk stars, which form well-separated clumps in FUV-NUV CMDs. However, the number of such stars identified in low-density GCs is small. Identifying and characterising such EHB stars are important as they are one of the major contenders to explain the 'UV-upturn' found in elliptical galaxies \citep{Laura1990, Connell1999}. Recent FUV studies extending to extra-galactic GCs \citep{Mark2018}, based on HST and GALEX observations, indicate that He enhanced HB and EHB stars might be responsible for the observed excess FUV emission. 

Wide-field UV observations are extremely useful for deriving a complete census of luminous and hot post-HB (pHB) stars that evolve away from the HB. They are comprised of AGB ma\'nque (AGBm), post-early AGB (PeAGB), and post-AGB (PAGB) stars. Using \textit{GALEX} observations, \cite{Schiavon2012} provided the catalogue of post He-core burning candidates (AGBm, PeAGB, PAGB) in 44 GCs classified from FUV-NUV CMDs. However, the cluster membership and evolutionary scenarios were not explored in their study. Recently, \cite{Moehler2019} studied the evolutionary phases of 19 reported PAGB stars in 17 GCs and pointed out that a complete sample of UV bright stars in a large number of GCs is required to test stellar evolution theoretical models. In all the above cases, FUV observations are necessary to efficiently detect and characterise the hot and luminous stellar populations, which otherwise suffer from crowding and have large bolometric corrections in the optical. 

In the present era, a detailed analysis of UV stellar populations in a large sample of GCs covering the full cluster region in FUV is still lacking. Previous missions dedicated to FUV studies of GCs have some limitations such as 1) \textit{HST}/WFPC2 field of view (FOV) is not large enough to cover the entire GC and its FUV filters suffer from red leak problem \citep{Holtz1995}, 2) on the other hand, the \textit{UIT} and \textit{GALEX} had a large FOV (40$'$ and 1.2$^\circ$). However, due to their poor resolution (3\arcsec and 5\arcsec respectively), they are incapable of resolving the FUV sources lying just outside the \textit{HST's} field. The advantage of the Ultra-Violet Imaging Telescope (UVIT) \citep{Subramaniam2016, Tandon2017} onboard the Indian space observatory \textit{AstroSat} is its larger FOV (28\arcmin) compared with the \textit{HST}, and higher spatial resolution ($\sim$1\farcs5) compared with the \textit{GALEX} and \textit{UIT}. In addition, multiple filters (five) in FUV can help in fine sampling the SEDs of hot stellar populations in GCs. 
The instrument and photometric calibration details of the UVIT are provided by \cite{Tandon2017}.

Several works on GCs with UVIT/FUV filters in the past five years have contributed to our understanding of HB morphologies \citep{Subramaniam2017,Sahu_ngc288, Rani2021, Ranjan2021}, detection of unusual and rare populations such as pHB stars \citep{Deepthi2021, 2021ApJ...923..162R}, and, the identification of BSSs and EHBs with hot companions \citep{Sahu_ngc5466, Singh2020}. To further expand our knowledge on UV bright stellar populations in GCs, we have conducted the UVIT legacy survey of GCs using two UVIT/FUV filters- F148W and F169M. This survey is designated as `GlobULeS', an acronym for Globular Cluster UVIT Legacy Survey. Here, we present some initial results from this survey, with the FUV-optical CMDs of eight GCs by combining UVIT/FUV observations with the \textit{HST} UV Globular Cluster Survey (HUGS) \citep{Nardiello2018, Piotto2015} and \textit{Gaia} Early Data Release 3 (EDR3) \citep{Gaia2020, Gaia2020_p} for the first time. In addition, we have created catalogues of hot stellar populations in these clusters with their membership probabilities. These catalogues will be useful for deriving the physical parameters (luminosities, effective temperatures, etc.) needed to understand the evolution of hot and exotic stellar populations.

The paper is arranged as follows. We describe the GC sample selection, along with the UVIT data reduction in Section~\ref{sec:sample_data}. The FUV catalogue and CMDs are presented in Section~\ref{sec:cmd_uvit}. We highlight the important features of the stacked CMDs in Section~\ref{sec:stack_cmd}. The discussion of our results and possible science cases are presented in Sections~\ref{sec:results} and \ref{sec:sc_case}, respectively. We summarise and conclude our study in Section~\ref{sec:conc}.

\section{GC sample and data reduction} \label{sec:sample_data}

\begin{table*}
\centering
\begin{threeparttable}[1pt]
\caption{Observation details of GCs with the UVIT under GlobULeS survey and other previous programs (marked in *).}
\begin{tabular}{lcccccc}
\hline
Cluster & RA (h m s)\tnote{*} & Dec (d m s)\tnote{*} & Cycle &  Observation Date &\multicolumn{2}{c}{t$_{\rm exp}$ (sec)\tnote{1}}\\
&  & & & &	UVIT/F148W & UVIT/F169M  \\\hline
NGC~362*	&	01 03 14.26	&	-70 50 55.6 	&	G06	&	2016-11-10	&	4614	&	3941\\
NGC~1904 (M79)*	&	05 24 11.09 	&	-24 31 29.0 	&	A02	&	2016-11-18	& 4901	&	3915\\
NGC~5272 (M3)*	&	13 42 11.62	&	+28 22 38.2    	&	A05	&	2019-03-08	&	3000	&	2883\\
NGC~5897	&	15 17 24.50 	&	-21 00 37.0  	&	A04	&	2018-06-16		&	13715	&	-	\\	
NGC~6205 (M13)	&	16 41 41.24  	&	+36 27 35.5	&	A05	&	2019-03-13	&	6657  &	 6657\\
NGC~6341 (M92)	&	17 17 07.39 	&	+43 08 09.4 	&	A04	&	2018-06-30	& 13726	&	-	\\
NGC~6809 (M55)	&	19 39 59.71 	&	-30 57 53.1 	&	A05	&	2018-10-01	&	6572	&	6630\\
NGC~7099 (M30)	&	21 40 22.12  	&	-23 10 47.5 	&	A06	&	2019-09-17	& 6623  &  7087 \\
\hline
\label{tab:gc_uvit_survey}
\end{tabular}
\begin{tablenotes}\footnotesize
    \item[*] Right Ascension and Declination \citep{Harris1996}(2010 edition)
    \item[1] Exposure time in F148W and F169M filters
\end{tablenotes}
\end{threeparttable}
\end{table*}

\begin{figure}
\centering
\includegraphics[width=\columnwidth]{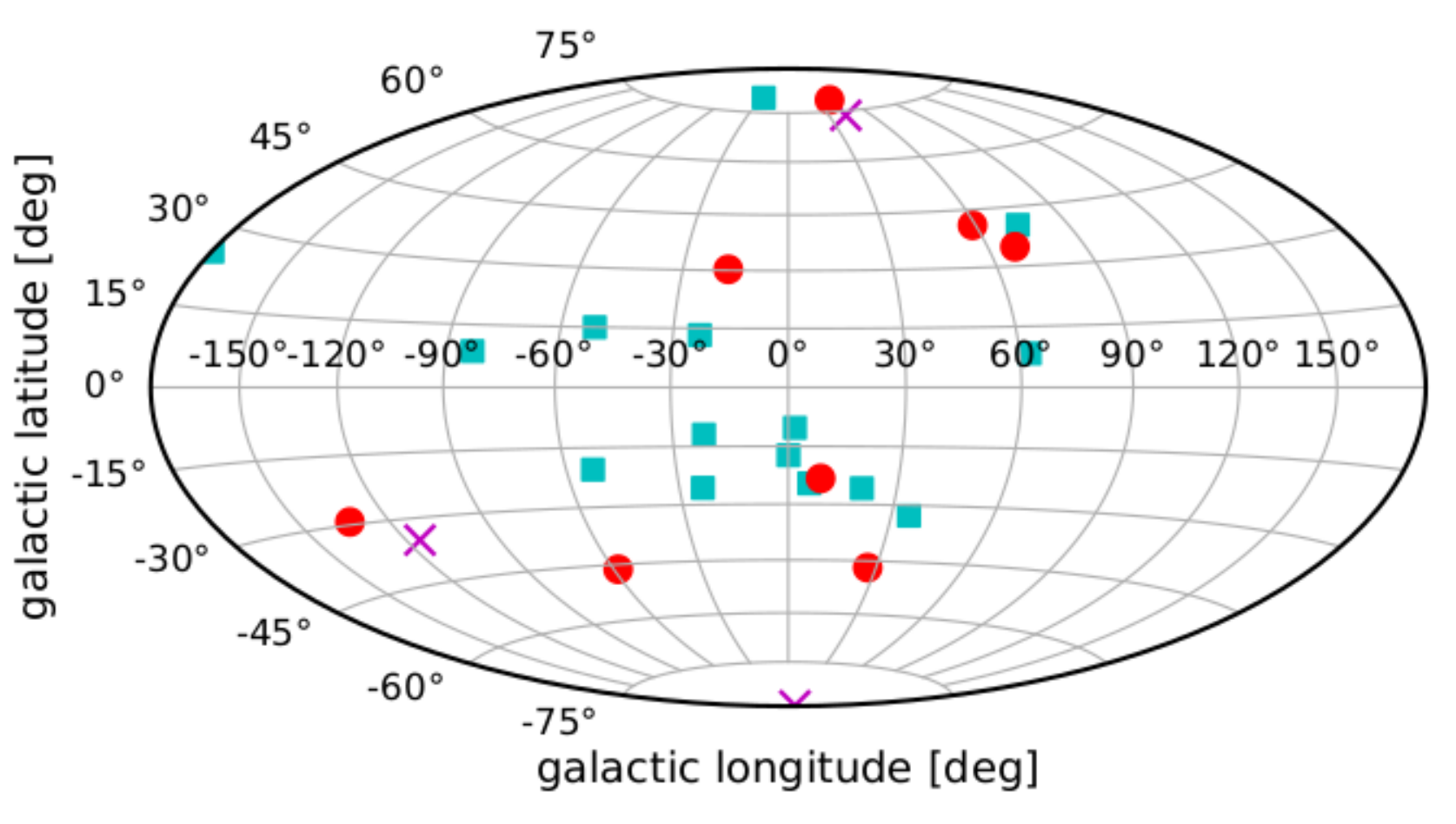}
\caption{Distribution of our GC sample in the galactic plane. The red dots are eight clusters under GlobULeS survey and other individual programs. The magenta crosses are GCs from our previous studies. The cyan square symbols are upcoming targets of the GlobULeS survey.}
\label{fig:gc_dist}
\end{figure}

We present the FUV study of eight clusters in this work for the first time. Out of these, five are observed under the GlobULeS survey and three are observed by other individual programs using the UVIT (see Table~\ref{tab:gc_uvit_survey} for details). The observations were carried out using the F148W and F169M filters with mean wavelengths ($\rm \lambda_{mean}$) 1481~\AA~ and 1608~\AA, respectively. The F148W is a broadband FUV filter with $\Delta \lambda \sim$ 500~\AA~ and F169M is a medium-band filter with $\Delta \lambda \sim$ 290~\AA. Out of eight GCs, six have observations in two FUV filters, whereas, two clusters have only F148W observations. We have combined this sample with previous UVIT studies of three GCs, NGC~288, NGC~1851, and NGC~5466 \citep{Subramaniam2017, Sahu_ngc288, Sahu_ngc5466} to analyse a total of 11 GCs in this work. The distribution of our GC sample in the galactic plane observed under the GlobULeS survey, other programs, and, previous UVIT studies are shown in Figure~\ref{fig:gc_dist}. 
There are 15 additional GCs scheduled for observations in the coming \textit{AstroSat} cycles. Their analysis will be presented in the forthcoming papers. Overall, the sample consists of 26 GCs (20 GCs from GlobULeS) with metallicities ranging from  $-2.3 \leq \rm [Fe/H] \leq -0.64$ dex and covering both the hemispheres. The sample selection was limited by various mission constraints, such as (i) avoiding observations in the Galactic plane, as the bright sources may harm the UV detectors; (ii) avoiding the GCs lying in the Dec range $-$10 to +10 due to RAM angle constraints, and; (iii) avoiding the GCs withing Galactic latitude $-30\deg < b < +30\deg$, since they do not have \textit{GALEX} observations. All the clusters in our sample have a central pointing with the UVIT, excluding NGC 1904 (M79). This cluster has an off-centred pointing due to a safe count limit in the FUV ($\leq$ 300 counts per second), set by the UVIT observations during the early observing period.

\begin{figure*}
\centering
\includegraphics[width=\textwidth]{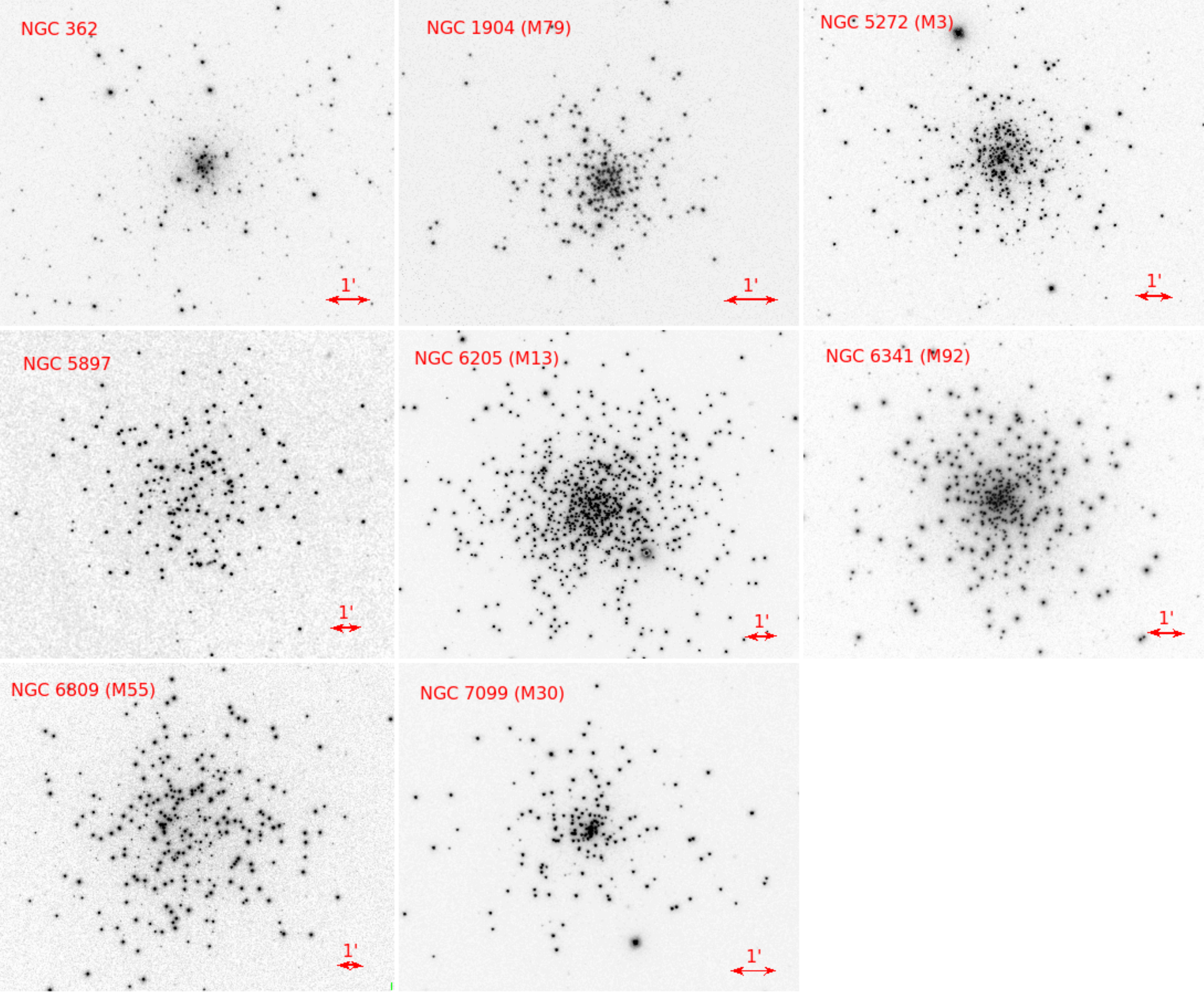}
\caption{The UVIT images of GCs NGC~362, NGC~1904, NGC~5272, NGC~5897, NGC~6205, NGC~6341, NGC~6809, NGC~7099, where black dots corresponds to detections in the F148W filters. North is up and East is left in the images.}
\label{fig:uv_image}
\end{figure*}

The Level~1 (L1) data of the sample were downloaded from ISRO science Data archive for \textit{AstroSat}. CCDLAB \citep{Postma2017} was used to generate the science images from L1 data of the UVIT. CCDLAB corrects for the satellite drift, flat field, distortion, fixed pattern noise, and cosmic rays. The images obtained in different orbits are aligned and combined to get the final deep exposure image of each cluster in both FUV filters. The images have sub-pixel sampling (1/8) with image dimensions 4096 $\times$ 4096. The science-ready images of the eight clusters are shown in Figure~\ref{fig:uv_image}. The UVIT was able to resolve most of the clusters except NGC~6341 (M92) and three core-collapsed GCs, NGC~362, NGC~1904 (M79), and, NGC~7099 (M30). From the total sample of 11 GCs, the UVIT's FOV covers only 6 clusters out to their tidal radii.

\subsection{Photometry}\label{phot}
Point spread function (PSF) photometry was performed on the science-ready images using the DAOPHOT package in IRAF \citep{Stetson1987}. PHOT task was used to perform aperture photometry. To plot the curve of growth and apply aperture correction, the magnitudes were obtained at different apertures. The photometry file generated at one aperture ($\sim$ FWHM) was fed to the PSTSELECT task and isolated stars in the field were chosen for generating a model PSF using the PSF task. The average PSF of the sources in the sample varies from $1\farcs6$ to $1\farcs8$. The model PSF was fitted to all the stars in the aperture photometry file to obtain the fluxes using the ALLSTAR task.

\begin{table}
\centering
\begin{minipage}{10cm}
\begin{threeparttable}[t]
\caption{Photometry details of GCs observed with the UVIT under GlobULeS survey and other programs.}
\begin{tabular}{lcccccc}
\hline
Cluster & $r_{c}$\tnote{1} & $r_{h}$\tnote{1} & N$_{\rm F148W}$\tnote{2}  & N$_{\rm F169M}$\tnote{2} & \multicolumn{2}{c}{Magnitude limit \tnote{3}}\\
& & & \multicolumn{2}{c}{$\sigma <$ 0.2} & F148W & F169M\\
\hline
NGC~362\tnote{*}	&	0.18	&	0.82	&	606	&	400	&	23.5	&	23	\\
NGC~1904\tnote{*}	&	0.16	&	0.65	&	280	&	244	&	22	&	21.5	\\
NGC~5272	&	0.37	&	2.31	&	456	&	366	&	23.5	&	23	\\
NGC~5897	&	1.4	&	2.06	&	213	&	-	&	22	&	-	\\
NGC~6205	&	0.62	&	1.69	&	882	&	856	&	22	&	21.5	\\
NGC~6341\tnote{*}	&	0.26	&	1.02	&	493	&	-	&	23.5	&	-	\\
NGC~6809	&	1.8	&	2.83	&	424	&	364	&	23	&	22	\\
NGC~7099\tnote{*}	&	0.06	&	1.03	&	248	&	240	&	23	&	22.5	\\
\hline
\label{tab:gc_uvit_survey_phot}
\end{tabular}
\begin{tablenotes}\footnotesize
    \item[1] Core radius ($r_{c}$), and half-light radius ($r_{h}$) in arcmin from \cite{Harris1996} (2010 edition)
    \item[2] Number of stars detected in the F148W and F169M with fit errors $<$ 0.2.
    \item[3] The UVIT detection limit (AB mag).
    \item[*] Cores of these GCs are not resolved by the UVIT.
\end{tablenotes}
\end{threeparttable}
\end{minipage}  
\end{table}

The PSF magnitudes were converted to aperture photometry scale, and aperture correction was applied by choosing isolated bright stars in the field. Finally, the instrumental magnitudes are calibrated to the AB mag system ($m_{AB}$) \citep{Tandon2017} using the relation:

\begin{equation}
    m_{AB} = -2.5\rm log(CPS) + ZP  
\end{equation}
where CPS is the counts per second in the FUV filters and ZP is the zero point defined as the AB magnitude corresponding to Unit conversion (UC) given by:
\begin{equation}
    ZP = (-2.5\rm log(UC) \times (\lambda_{mean})^{2})-2.407
\end{equation}

\begin{figure}
\centering
\includegraphics[width=0.9\columnwidth]{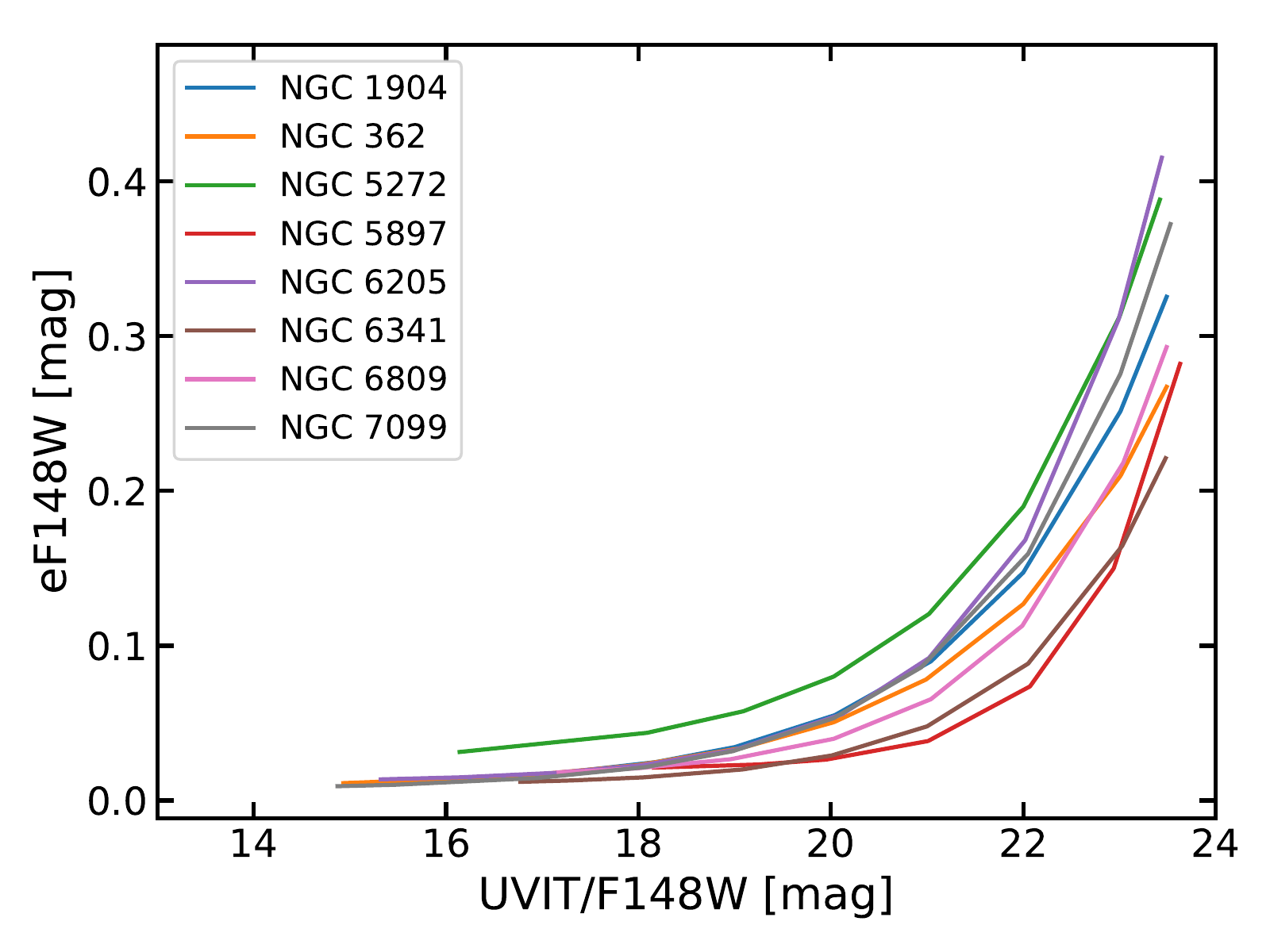}
\includegraphics[width=0.9\columnwidth]{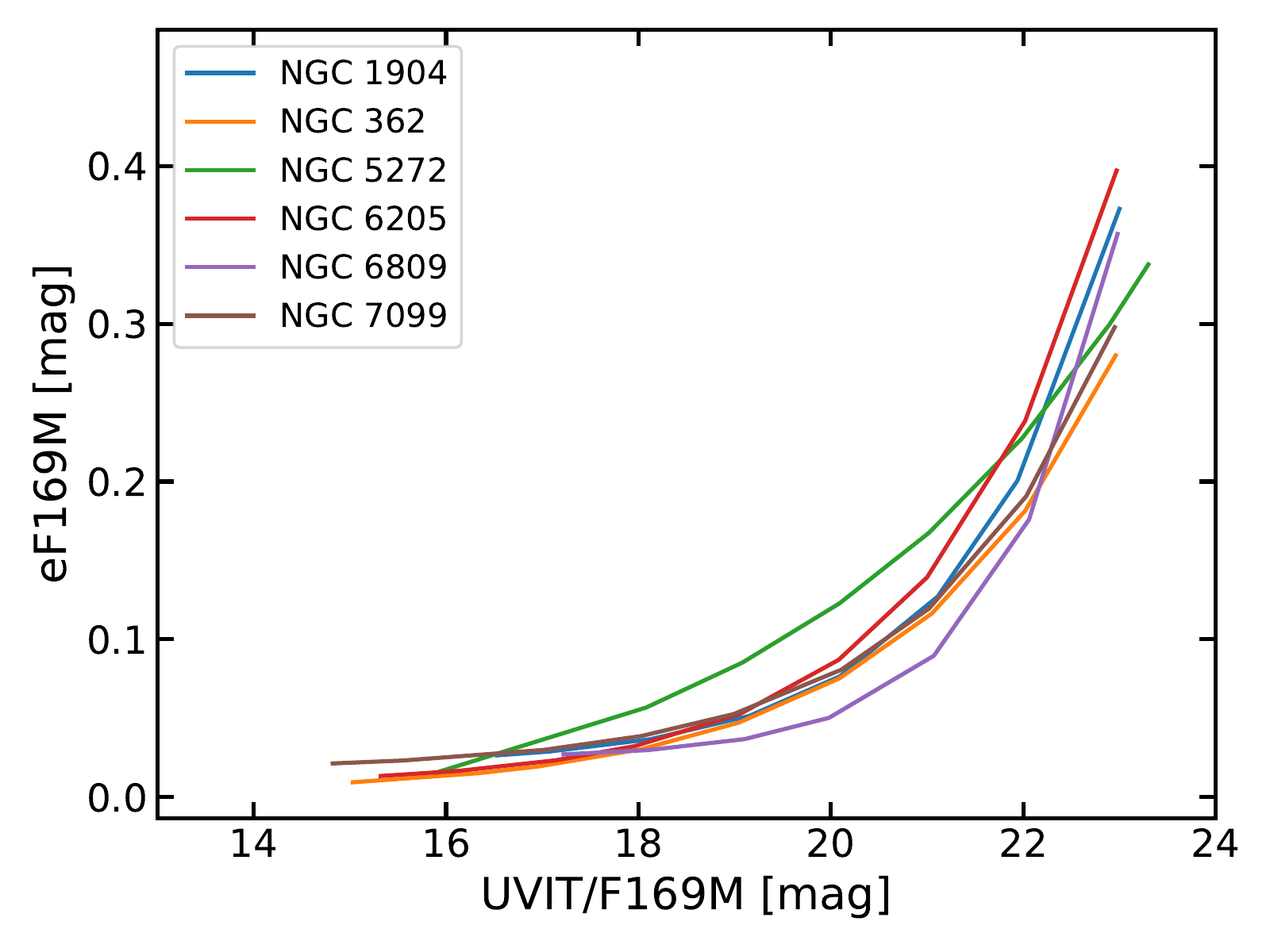}
\caption{The PSF fitting errors (median) of the magnitudes obtained from PSF photometry for eight clusters in F148W filter (top panel) and six clusters in F169M filter (bottom panel).}
\label{fig:error}
\end{figure}

The ZPs of F148W and F169M filters are 18.003 and 17.453 mag, respectively \citep{Tandon2017}. The magnitudes obtained in the UVIT filters are corrected for saturation following the steps provided in \cite{Tandon2017}. A plot of PSF fit errors (median) as a function of magnitude for eight clusters in the F148W filter and six in the F169M is shown in Figure~\ref{fig:error}. Mostly, we detect stars up to 22~mag with fit errors less than 0.1 and 0.2~mag in F148W and F169M filters, respectively. The number of detected sources with PSF fit errors less than 0.2 along with the magnitude detection limit in the observed filters are provided in Table~\ref{tab:gc_uvit_survey_phot}.

WCS registration of the UVIT images was carried out using the CCMAP task of IRAF. Typical root mean square (RMS) errors for the astrometric calibration of images of seven clusters in F148W and F169M filters is $\sim 0\farcs1-0\farcs2$ both in RA and Dec except for NGC 6341 ($\sim0\farcs35$). The X and Y positions of sources in the images were converted to RA and Dec coordinates using the WCSCTRAN task of IRAF. 

The extinction coefficients $A_{F148W}$ and $A_{F169M}$ in F148W and F169M filters were calculated to be 2.64 $A_{V}$ and 2.56 $A_{V}$, respectively, using Fitzpatrick extinction law \citep{Fitz1999}. Here, $A_{V}=R \times E(B-V)$ where $E(B-V)$ is the reddening of the cluster considered from \citep{Harris1996} (Table~\ref{tab:uv_models}). These values agree with the UVIT extinction coefficients by \cite{Chen2019}. 

\subsection{Artificial Star Test}
Since the completeness of detected sources (in the fainter magnitude end) are affected by crowding in the cores of GCs, we performed artificial star tests (AST) on the UVIT images to check the completeness of the detected sources at different magnitudes and locations. Sources ($\leq$ 30\% of the total detections) with a spatial density distribution similar to that of the cluster were simulated and added to the science image by keeping their magnitudes fixed using the ADDSTAR task of DAOPHOT \citep{Stetson1987}. The model PSF constructed for performing the photometry of real stars is used in the ADDSTAR task. Once the simulated image is generated, photometry was performed adopting the same parameters and methods as used for the real stars in the science image (described in Section~\ref{phot}). We considered a star to be recovered in the test when its spatial difference is less than 1\farcs5 from the added location and its magnitude difference is $<$ 0.8 mag. These steps were repeated by varying the XY positions of the added stars for a fixed magnitude. For checking the completeness, the AST was performed at different magnitudes with the faintest ones reaching the observational detection limit in each filter. Note that the artificial stars are added and recovered at once in the UVIT/FUV images, as crowding is less severe in FUV unlike the NUV and optical images.

\begin{figure}
\centering
\includegraphics[scale=0.5]{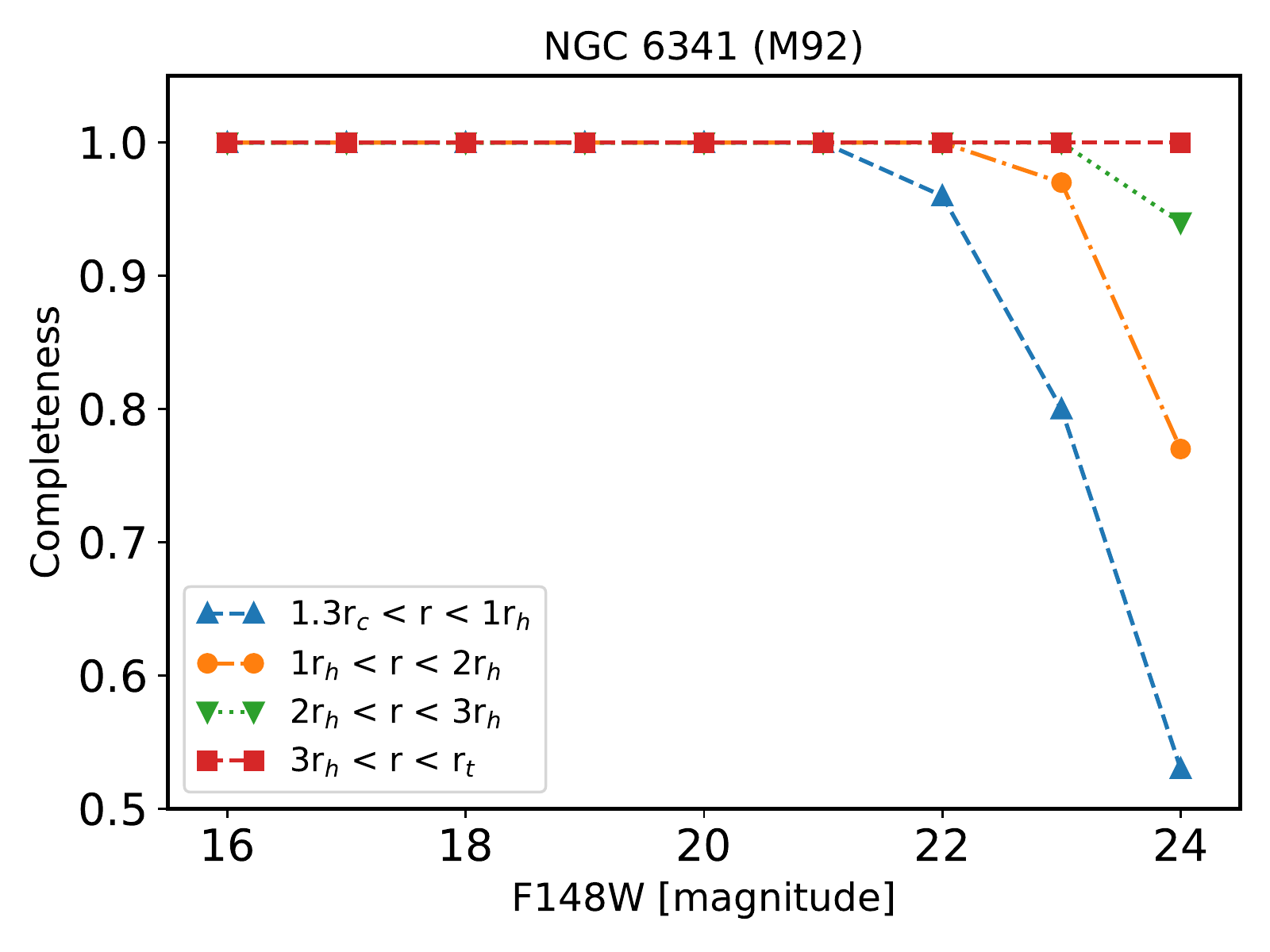}
\caption{Completeness of NGC~6341 as a function of F148W  magnitude for different radial bins as marked in the figure where $r_{c}, r_{h}$, and $r_{t}$ are the core radius, half-light radius and tidal radius of the cluster respectively \citep{Harris1996}.}
\label{fig:ast_ngc6341}
\end{figure}

To check the variation of completeness with the magnitude at increasing radii from the cluster centre, we divided the clusters into different concentric annuli and calculated the number of recovered artificial stars from photometry with the number of added stars in the image at each annulus. For example, the completeness plot of cluster NGC~6341 for F148W magnitude at different annuli is shown in Figure~\ref{fig:ast_ngc6341}. We note that the data is 100\% complete at 21~mag at all the radial bins. By contrast, the completeness drops to 50\% at 24~mag within half-light radius of the cluster ($r_{h} \sim 60''$). The innermost radial bin chosen for this cluster is 20$''$, as this cluster is not resolved by UVIT inside that radius. 

Using the above method, AST was performed for all the clusters in the sample and completeness was calculated mostly in the three annular regions, $0 < r < r_{c}$, $r_{c} < r < r_{h}$, and, $r > r_{h}$, where $r_{c}$ and $r_{h}$ are the core radius and half-light radius of the clusters respectively \citep{Harris1996}. The completeness plot of these clusters for sources lying inside the core radius and within the core to half-light radius are shown in Figure \ref{fig:ast_all} (appendix). We note that the completeness is above 90\% for all the sources lying in the regions outside $r_{h}$, and, that are brighter than the UVIT magnitude detection limit (Table \ref{tab:gc_uvit_survey_phot}). For sources lying in regions $r_{c} < r < r_{h}$, the completeness is $\sim$80-90\% in clusters  NGC 5897, NGC 6205, NGC 6809, and, NGC 7099. This reduces to 60-70\% in case of NGC 362 and NGC 5272 with comparatively less exposure times than other GCs in the sample. Considering the core regions ($0 < r < r_{c}$) that are majorly affected by crowding, the completeness drops down to $\sim$70-80\% in clusters NGC 5897 and NGC 6809 which further reduces to $\sim$50\% in NGC 5272 and NGC 6205. 
The completeness of sources lying within the core radius is not calculated for four clusters (NGC~362, NGC~1904, NGC~6341, and NGC~7099) which are not resolved by UVIT. 

\begin{figure*}
\centering
\includegraphics[scale=0.24]{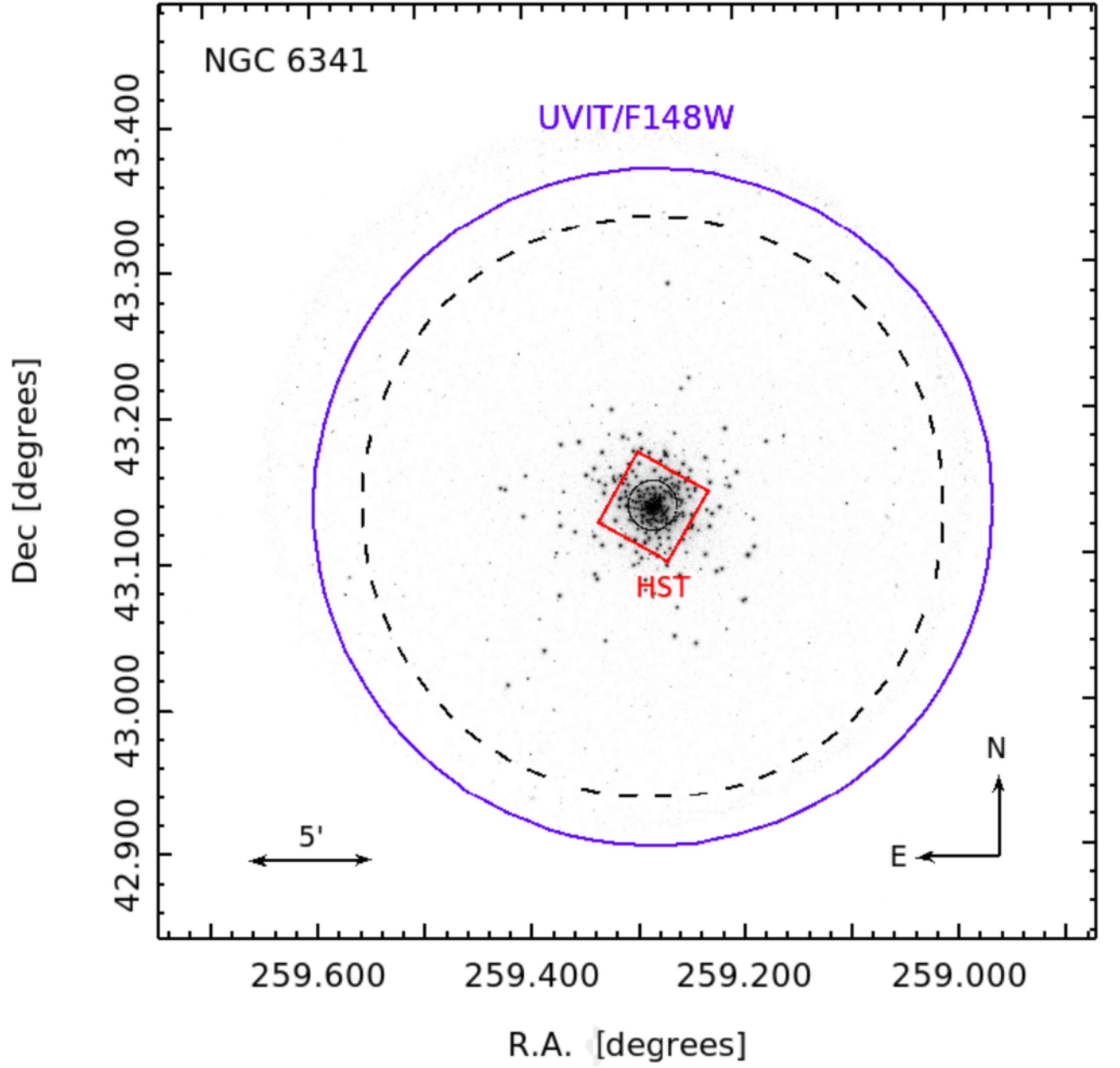}
\includegraphics[scale=0.44]{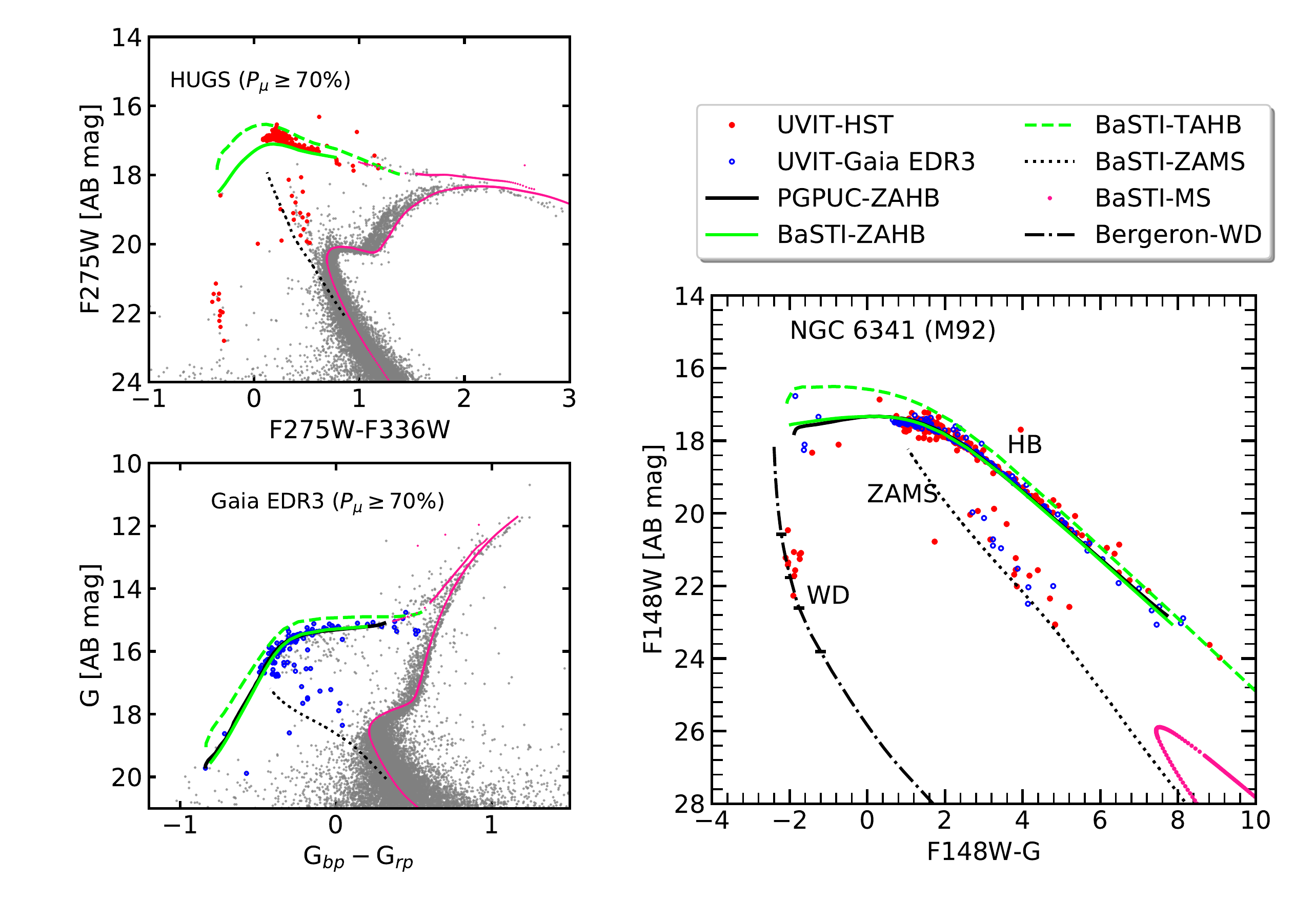}
\caption{Left Panel: Map of the UVIT (blue) and \textit{HST} (red) fields of view plotted over the F148W filter image of NGC~6341. The cluster half-light (1$\farcm$02) and tidal radii (12$\farcm$4) are shown in black solid and dashed lines, respectively. Middle upper Panel: F275W, F275W$-$F336W CMD of NGC~6341 from HUGS survey where the UVIT-\textit{HST} common detections are shown in red dots. Middle lower Panel: (G, $G_{bp}-G_{rp}$) CMD from \textit{Gaia} EDR3 where blue open circles are UVIT-\textit{Gaia} common detections. Right Panel: (F148W, F148W$-$G/F606W) CMD over-plotted with stellar evolutionary models.}
\label{fig:fov_meth}
\end{figure*}

\section{FUV Catalogue and CMDs} \label{sec:cmd_uvit}
To construct the FUV-optical CMDs, we cross-matched the UVIT data with the HUGS data of GCs \citep{Nardiello2018} for the central regions covering 202$'' \times 202''$. Figure~\ref{fig:fov_meth} (left panel) shows the area covered by the \textit{HST} and the UVIT on top of the UVIT/F148W image of NGC~6341. We note that many UV bright stars lie outside the FOV of the \textit{HST}. To complete the coverage of outer regions, we cross-matched the UVIT data with \textit{Gaia}~EDR3 and estimated their proper motion (PM) membership probabilities as described below. The \textit{HST} data used in the catalogue has PM accuracy of about 0.6~mas yr$^{-1}$, with a time baseline of 7-9 years \citep{Soto2017}. For comparison, the PM uncertainty for \textit{Gaia} EDR3 is around 1.4~mas yr$^{-1}$ at G=21~mag.

We used Topcat \citep{Taylor2005} to cross-match the positions (R.A. and Dec) of sources detected in the UVIT filters with the \textit{HST} and \textit{Gaia}~EDR3 with a maximum separation of $2''$, which is the typical FWHM of the PSF for the UVIT filters. However, a majority of the cross-matched sources ($>90\%$) have match radii of $<1''$ (Figure~\ref{fig:sep}). 

\subsection{UVIT-\textit{HST} data}
We used the HUGS catalogues \citep{Nardiello2018} from method-1 photometry which gives good measurements for sources with F275W $<$ 23~mag to cross-match all the FUV stars detected with the UVIT. The catalogues also provide PM membership probabilities ($P_{\mu}$). We have considered sources with $P_{\mu}\geq70\%$ for analysis. However, most of the UVIT-\textit{HST} cross-matched sources ($\sim96\%$) have $P_{\mu}\geq 90\%$. Since NGC~1904 was not included in the HUGS program, we used the data provided by \cite{Lanzoni2007} for cross-matching with the UVIT. This cluster has observations in F160B, F218W, F336W, F439W and F555W filters of \textit{HST}-WFPC2. However, the $P_{\mu}$ of this cluster is not available.

As the \textit{HST} has a higher resolution ($< 0\farcs1$) compared with the UVIT, their direct cross-matching for the cluster members may lead to incorrect matches. Since most stars bluer than MS and RGB stars are hot enough to emit in the FUV wavelengths, we have excluded MS, SGB, RGB, and AGB stars from the cross-match to reduce the crowding effects, and selected the rest of the sources from the F275W vs F275W$-$F336W CMD plane. In addition, we also did a visual inspection with eye to make sure that the cross-matched sources are unique. The \textit{HST} CMD of NGC~6341 is shown in the middle panel of Figure~\ref{fig:fov_meth} where the UVIT-\textit{HST} common detections are marked in red. The membership probabilities of WDs are unavailable in the \textit{HST} catalogues, so we relied on F275W vs F275W$-$F336W plane for their selection as they form a well-defined sequence and show FUV emission (Figure~\ref{fig:fov_meth}). 

\subsection{UVIT-\textit{Gaia} EDR3 data}\label{sec:gaiaedr3}

\begin{figure*}
\centering
\includegraphics[width=0.47\textwidth]{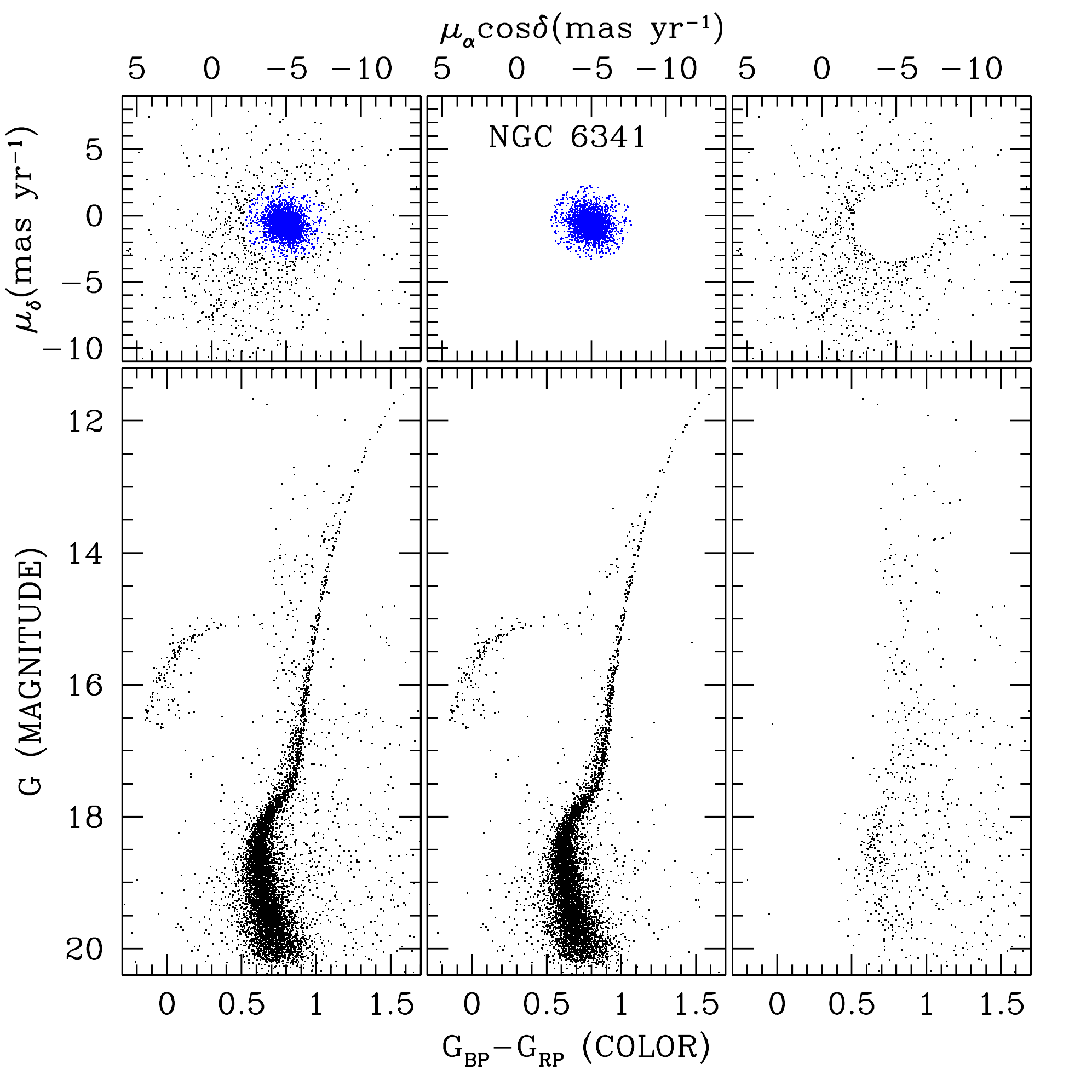}
\includegraphics[width=0.452\textwidth]{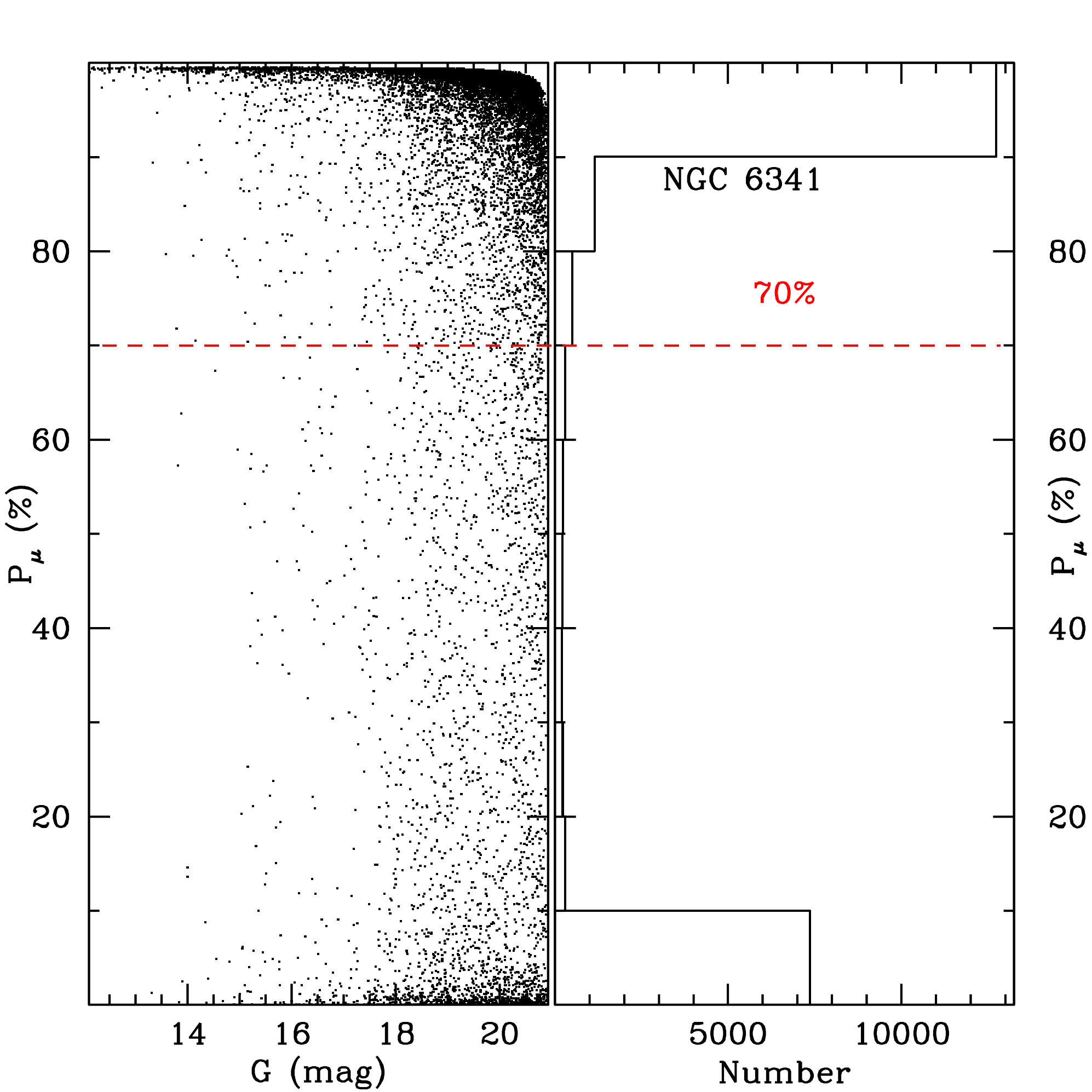}
\caption{Left Panel: VPD of NGC~6341 where blue dots are the sources selected for PM membership analysis, and black dots are the field stars. The stars with good astrometric measurements, i.e., with \textit{nper} $>$ 8 and \textit{nal} $>$ 120 are used to obtain the mean and standard deviation of the cluster distribution. Right Panel: P$_{\mu}$ with respect to the histogram of the number of stars and G magnitude in the y-axis. The red dashed line marks the stars where most of the cluster members lie with $P_{\mu} > 70\%$.} 
\label{fig:vpd_ngc6341}
\end{figure*}

To select probable cluster members using \textit{Gaia} proper motion data, we initially used the mean PMs of clusters from \cite{Helmi2018}. We considered the sources out to the tidal radius of the cluster for $P_{\mu}$ determination. The vector point diagram (VPD) and the respective \textit{Gaia} CMD of NGC~6341 are shown, as an example, in the left panel of Figure~\ref{fig:vpd_ngc6341}. The VPDs of the rest of the clusters are shown in Figure~\ref{fig:vpds_gaiadr3}. The VPD selection criteria adopted for each GC show a clear separation of the probable cluster members from the field members. To find the centre of the elliptical distribution, we rotated the semi-major axis of the elliptical distribution to the normal elliptical distribution along their respective centre. Then, we obtained the mean and standard deviation of PMs for each GC from their histograms and Gaussian fits. 
The probable cluster members selected from the VPD show an elliptical distribution for most of the clusters. This distribution is possibly due to systematics in the PM, arising from the non-uniform sampling of the sky produced by the \textit{Gaia} scanning law \citep{Fab2021}. Although \textit{Gaia}~EDR3 has significantly improved the systematics by a factor of 2.5 in comparison to \textit{Gaia}~DR2, they are still present. In order to reduce the effect of systematics, we considered stars with good astrometric measurements i.e., with \textit{nal} $>$ 120 and \textit{nper} $>$ 8, where, \textit{nal} is the total number of observations and \textit{nper} is the number of visibility periods used in the astrometric solution \citep{Kim2019}. 

To estimate $P_{\mu}$, we used the method of \cite{Bala1998}. We first estimate the frequency distribution of cluster stars ($\phi^{\nu}_{c}$) and field stars ($\phi^{\nu}_{f}$). The frequency distribution function for the $i^{th}$ star of a cluster and the field can be given as follows:

\begin{multline}
\phi^{\nu}_{c} = \frac{1}{{2\pi}\surd{(\sigma^{2}_{xc}+ \epsilon^{2}_{xi})(\sigma^{2}_{yc}+ \epsilon^{2}_{yi})}} \\
\times exp \left \{ -\frac{1}{2} \Bigg[ \frac{(\mu_{xi}-\mu_{xc})^{2}}{\sigma^{2}_{xc}+ \epsilon^{2}_{xi}} + \frac{(\mu_{yi}-\mu_{yc})^{2}}{\sigma^{2}_{yc}+ \epsilon^{2}_{yi}}
\Bigg] \right \} 
\end{multline}

\begin{multline}
\phi^{\nu}_{f} = \frac{1}{{2\pi}\surd{(1-\gamma^{2})}\surd{(\sigma^{2}_{xf}+ \epsilon^{2}_{xi})(\sigma^{2}_{yf}+ \epsilon^{2}_{yi})}} exp \Bigg \{ -\frac{1}{2(1-\gamma^{2})}\\ 
\Bigg[ \frac{(\mu_{xi}-\mu_{xf})^{2}}{\sigma^{2}_{xf}+ \epsilon^{2}_{xi}} - \frac{2\gamma(\mu_{xi}-\mu_{xf})(\mu_{yi}-\mu_{yf})}{\surd{(\sigma^{2}_{xf}+ \epsilon^{2}_{xi})(\sigma^{2}_{yf}+ \epsilon^{2}_{yi})}} + \frac{(\mu_{yi}-\mu_{yf})^{2}}{\sigma^{2}_{yf}+ \epsilon^{2}_{yi}}
\Bigg] \Bigg \} 
\end{multline}

where $\mu_{xi}$ and $\mu_{yi}$ denote the proper motions of the $i^{th}$ star. The $\mu_{xc}$ and $\mu_{yc}$ represents the cluster's proper motion centre, $\mu_{xf}$ and $\mu_{yf}$ represents the field proper motion centre, $\epsilon_{xi}$ and $\epsilon_{yi}$ are the observed errors in proper motion components, $\sigma_{xc}$ and $\sigma_{yc}$ denotes the cluster intrinsic proper motion dispersion, while $\sigma_{xf}$ and $\sigma_{yf}$ denotes the field intrinsic proper motion dispersion and $\gamma$ is the correlation coefficient.

The value of $\gamma$ can be estimated as

\begin{gather}
\gamma = \frac{(\mu_{xi}-\mu_{xf})(\mu_{yi}-\mu_{yf})}{\sigma_{xf}\sigma_{yf}}
\end{gather}

For the distribution function $\phi^{\nu}_{c}$ and $\phi^{\nu}_{f}$, we considered the stars with proper motion error better than 1 mas yr$^{-1}$.\\

To obtain the distribution of stars, we use 

\begin{gather}
\phi = n_{c} \phi^{\nu}_{c} + n_{f} \phi^{\nu}_{f}
\end{gather}

where $n_{c}$ and $n_{f}$ are the normalized number of stars found for cluster and field ($n_{c}$ + $n_{f}$ = 1). Hence, the membership probability
for $i_{th}$ star is given by

\begin{gather}
P^{\mu} (i) = \frac{\phi_{c} (i)}{\phi (i)}
\end{gather}

In the Gaia EDR3 catalog, all sources have been treated like single stars i.e., in the case of binary and multiple stellar systems, the astrometry is obtained for either component or from the photo-centre of the binary/multiple systems. For a binary located at a distance, d $\approx$ 500 pc, the predicted deviation for an angular separation of 1\arcsec is on the order of $\mu$as, well below the sensitivity of Gaia EDR3 \citep{Badry2021}. Since GCs are located at a much farther distance, therefore, the uncertainty due to consideration of single stars would have a negligible effect on the membership determination. 

To include the fainter members of the clusters having a larger error in PM, we selected sources with a probability above $70\%$. The right panel of Figure~\ref{fig:vpd_ngc6341} shows the distribution of $P_{\mu}$ with respect to G magnitude. From this figure, we notice that most of the stars in NGC~6341 have $P_{\mu}>70\%$. 

We further note that almost all the UVIT-\textit{Gaia} cross-matched sources have $P_{\mu}>70\%$, except in a few clusters. In six clusters, 13 stars lie between $30\%<P_{\mu}<70\%$. In NGC~6205 (M13), 28 stars in the UVIT-\textit{Gaia} EDR3 catalogue have $P_{\mu}<50\%$, and 7 do not have PM measurements. A comparison of this cluster data with the recently published PM catalogue of GCs from \textit{Gaia} EDR3 \citep{Vasilev2021} shows that 11 of these stars have $P_{\mu}>50\%$ but are of the lowest quality flag. Since, these stars with bluer color $(G_{bp}-G_{rp}) < 0.6$ are located in the high density regions ($<2 r_{h}$), the PM measurements might not be reliable. However, these stars mostly lying at the bluer end of the HB region (near EHB and pHB) are bright in FUV; hence, we have included them in the NGC~6205 catalog. 

\begin{table*}
\centering
\begin{threeparttable}
\addtolength{\tabcolsep}{3pt}
\caption{List of model parameters of the studied clusters.}
\begin{tabular}{lcccccccc}
\hline
\hline
Cluster & E(B$-$V)\tnote{1} & $(m-M)_{V}$\tnote{2} & \multicolumn{3}{c}{PGPUC ($[\alpha/Fe]$ = 0.3)\tnote{3}}& \multicolumn{3}{c}{BaSTI ($[\alpha/Fe]$ = 0.4) \tnote{4}} \\\hline
& & & [Fe/H] & Z & Y & [Fe/H] & Z & Y  \\\hline	
  NGC~288 & 0.03 & 14.84 & -1.32 & 1.39E-3 & 0.245 & -1.3 & 1.57E-3 & 0.249\\
  NGC~362 & 0.05 & 14.83 & -1.26 & 1.6E-3 & 0.245 & -1.3 & 1.57E-3 & 0.249\\
  NGC~1851 & 0.02 & 15.47 & -1.18 & 1.92E-3 & 0.245 & -1.2 & 1.97E-3 & 0.249\\
  NGC~1904 & 0.01 & 15.59 & -1.6 & 7.3E-4 & 0.245 & -1.55 & 8.86E-4 & 0.248\\
  NGC~5272 & 0.01 & 15.07 & -1.5 & 9.2E-4 & 0.245 & -1.55 & 8.86E-4 & 0.248\\
  NGC~5466 & 0.00 & 16.02 & -1.98 & 3.1E-4 & 0.245 & -2.2 & 1.98E-4 & 0.247\\
  NGC~5897 & 0.09 & 15.76 & -1.9 & 3.7E-4 & 0.245 & -1.9 & 3.97E-4 & 0.247\\
  NGC~6205 & 0.02 & 14.33 & -1.53 & 8.6E-4 & 0.245 & -1.55 & 8.86E-4 & 0.248\\
  NGC~6341 & 0.02 & 14.65 & -2.31 & 1.4E-4 & 0.245 & -2.2 & 1.98E-4 & 0.247\\
  NGC~6809 & 0.08 & 13.89 & -1.94 & 3.4E-4 & 0.245 & -1.9 & 3.97E-4 & 0.247\\
  NGC~7099 & 0.03 & 14.64 & -2.27 & 1.6E-4 & 0.245 & -2.2 & 1.98E-4 & 0.247\\
\hline
\label{tab:uv_models}
\end{tabular}
\begin{tablenotes}\footnotesize
    \item[1] Reddening \citep{Harris1996} (2010 edition)
    \item[2] Distance modulus \citep{Harris1996} (2010 edition)
    \item[3] PGPUC model parameters [Fe/H], Z and Y \citep{Aldo2012}.
    \item[4] BaSTI model parameters [Fe/H], Z and Y \citep{Piet2020}.
\end{tablenotes}
\end{threeparttable}
\end{table*}

We were unable to select the WD sequence as it is fainter than the magnitude limit of \textit{Gaia} EDR3 (G$\sim$21 mag). We checked the expected G magnitudes for the brightest WDs in our sample from the cooling models and found that they lie at around G$\sim$ 21-22~mag which is near or just below the \textit{Gaia} detection limit. In addition, since the WDs located in the outer regions of the clusters are affected more by background sources as compared to the core regions, their membership determination is extremely important to identify them. However, due to their fainter G magnitudes, their PM information is either unavailable or not reliable. The UVIT and \textit{Gaia} common detections are shown in blue in the middle panel of Figure~\ref{fig:fov_meth}.

In order to plot the FUV-optical CMD of the entire cluster region in a similar plane, we have transformed the \textit{HST} filter F606W to G band of \textit{Gaia}~EDR3 by using the relation below (Figure~\ref{fig:hst_g_relt}):

\begin{equation}
\centering
y=0.01+0.05x-0.11x^2
\label{eq:hst_mag_gaia}
\end{equation}

where y=(G$-$F606W) and x=(F606W$-$F814W). The stars used for deriving the relationship cover the entire HB colour range, from RHB to EHB stars, and with good astrometric measurements. 

The FUV-optical CMD of NGC~6341 after transforming the \textit{HST} to \textit{Gaia} plane is shown in the rightmost panel of Figure~\ref{fig:fov_meth} where the UVIT-\textit{HST} common detections are marked in red and the UVIT-\textit{Gaia} in blue. Thus, the complete CMD includes HB, BSS and WD detected from inner regions (the UVIT-\textit{HST} cross-matched stars) and, HB and BSS from the outer regions (the UVIT-\textit{Gaia}~EDR3 cross-matched stars).

\begin{table*}
\centering
\begin{threeparttable}
\addtolength{\tabcolsep}{-2.6pt}
\caption{UVIT catalogue of three clusters NGC 288, NGC 1851 and NGC 5466 where first three rows of each cluster are shown for reference. The full table will be made available online through Vizier catalogue access tool.}
\begin{tabular}{ccccccccccccc}
\hline
\hline
Cluster &  UVIT\_ID & RA [deg] & Dec [deg] & F148W & e\_F148W & F169M & e\_F169M & G \tnote{1} & mem\_prob\tnote{2} & unique\_ID\tnote{3} & Classification & Cross-match data \\\hline
NGC1851 & UV18510001 & 78.509010 & -40.065076 & 18.472 & 0.025 & 18.359 & 0.035 & 17.095 & 97.9 & R0000775 & HB & HST\\
NGC1851 & UV18510002 & 78.501673 & -40.061816 & 19.809 & 0.047 & 19.541 & 0.062 & 16.307 & 97.7 & R0001235 & HB & HST\\
NGC1851 & UV18510003 & 78.520213 & -40.064029 & 21.039 & 0.116 & - & - & 16.199 & 96.4 & R0000896 & HB & HST\\\hline
NGC288 & UV2880001 & 13.197465 & -26.549876 & 18.1 & 0.019 & 17.915 & 0.028 & 16.063 & 95.3 & R0001732 & HB & HST\\
NGC288 & UV2880002 & 13.202455 & -26.554893 & 17.601 & 0.023 & 17.533 & 0.023 & 17.312 & 96.8 & R0001666 & HB & HST\\
NGC288 & UV2880003 & 13.197637 & -26.554042 & 20.191 & 0.061 & 19.719 & 0.057 & 15.42 & 97.3 & R0001681 & HB & HST\\\hline
NGC5466 & UV54660001 & 211.3581 & 28.55536 & 20.652 & 0.102 & 20.267 & 0.066 & 16.74 & 96.9 & R0001393 & HB & HST\\
NGC5466 & UV54660002 & 211.3822 & 28.54057 & 21.376 & 0.154 & 21.43 & 0.114 & 23.534 & -1.0 & R0021345 & WD & HST\\
NGC5466 & UV54660003 & 211.3753 & 28.544 & 22.37 & 0.207 & 22.432 & 0.192 & 18.618 & 96.8 & R0001100 & BSS & HST\\\hline
\label{tab:catalog}
\end{tabular}
\begin{tablenotes}\footnotesize
    \item[1] G mag in AB system from Gaia EDR3 data \citep{Gaia2020} for UVIT-Gaia sources, whereas, transformed from HST filters for UVIT-HST sources (refer Equation \ref{eq:hst_mag_gaia}).
    \item[2] Membership probability from HUGS survey \citep{Nardiello2018} for UVIT-HST sources, whereas, derived in this work for UVIT-Gaia EDR3 sources (refer Section \ref{sec:gaiaedr3}).
    \item[3] Unique ID corresponding to the column name \textit{ID\_number} from HUGS catalogue \citep{Nardiello2018} for UVIT-HST sources, and, \textit{source\_id} from Gaia EDR3 data for UVIT-Gaia sources.
\end{tablenotes}
\end{threeparttable}
\end{table*}

\subsection{Stellar Models}\label{sec:models}
The HB, BSS, WDs, etc. are most prominently observed in the FUV-optical and FUV-FUV CMDs. We used various stellar evolutionary models to identify and classify their sequences. Reddening and distance modulus of individual clusters from \cite{Harris1996} (2010 edition) were adopted in the models to fit the observed CMDs. As an example, the rightmost panel of Figure~\ref{fig:fov_meth} shows an over plot of all the models described below for FUV-optical CMD of NGC~6341. 

\begin{itemize}
    \item \textit{PGPUC HB models}: We generated zero-age HB (ZAHB) loci using Princeton-Goddard-PUC (PGPUC) stellar evolution code for the UVIT filters \citep{Aldo2012} \footnote{\url{http://www2.astro.puc.cl/pgpuc/}}.  
    This code is an updated version of the original Princeton code by \cite{schwarz1965}. The physics incorporated in the code are described in detail in \cite{Aldo2012}. The ZAHB models are created for an $\alpha$-element enhancement value [$\alpha$/Fe] of 0.3 and initial He abundance (Y$_{ini}$) of 0.245 \citep{Cassisi2003} where atomic diffusion is also taken into account. This is typical [$\alpha/Fe$] value found in GCs \citep{Dias2016}. \\
    
    \item \textit{BaSTI-IAC models}: We generated the ZAHB and Terminal-age HB (TAHB) evolutionary tracks for the UVIT and \textit{Gaia}~EDR3 filters from an updated BaSTI-IAC online database \citep{Hidalgo2018, Piet2020} \footnote{\url{http://basti-iac.oa-abruzzo.inaf.it/hbmodels.html}}. We considered $\alpha$-enhanced models with [$\alpha/Fe$]=0.3, Y$_{ini} \sim$ 0.247, mass-loss parameter $\eta=0.3$. Atomic diffusion is not included. The BaSTI ZAHB and TAHB models which corresponds to core-He exhaustion is shown in the rightmost panel of Figure~\ref{fig:fov_meth}.\\ 
    In the case of BSSs, we used the zero-age MS (ZAMS) isochrones of 0.5~Gyr generated from BaSTI-IAC models for the UVIT and \textit{Gaia}~EDR3 filters for reference to show the extended ZAMS, by keeping the model's parameters fixed as that chosen for the HB tracks. The model parameters chosen cover the mass range from MSTO of GCs ($\sim 0.8 M_{\odot}$) to twice its mass at MSTO. This range approximately corresponds to the expected locations of BSSs according to their distribution in the CMDs of several GCs \citep{Ferraro2003, Raso2017}.
    The 0.5 Gyr isochrone corresponding to $\rm [Fe/H]=-2.2$ dex and model mass range of 0.8-1.6~$M_{\odot}$ is shown in the rightmost panel of Figure~\ref{fig:fov_meth}.  The BSS sequence extends more than 4 magnitude in FUV-optical CMDs. \\
    
    \item \textit{WD cooling Models}: DA spectral type WD models \citep{Bergeron1995, Tremblay2011, Bedard2020} of mass 0.5 $M_{\odot}$ \citep{Renzini1988, Richer1997, Moehler2004} with pure hydrogen (H) grid and thick H layers are kindly provided and transformed into UVIT filter systems by Pierre Bergeron \textit{Gaia}~DR2\footnote{\url{http://www.astro.umontreal.ca/~bergeron/CoolingModels}}. The models are shown in black dash-dotted lines in the CMDs.

\end{itemize}
We note that the ZAHB, BSSs, and WD stars appear well-matched with the locations predicted by stellar models.

The model parameters adopted for the studied clusters are given in Table~\ref{tab:uv_models}. The metallicities in PGPUC models are interpolated to the exact values provided by \cite{Harris1996} (2010 edition), whereas, for BaSTI-IAC models, the values are chosen close to the cluster metallicities as reported by \cite{Harris1996} (2010 edition). 

\subsection{Classification of FUV sources}\label{sec:classify}
For the UVIT-\textit{HST} cross-matched data, we plotted the matched stars in F275W vs F275W$-$F336W plane, whereas, for the UVIT-\textit{Gaia} cases were plotted in G vs G$_{bp}-$G$_{rp}$ plane. Their locations were simultaneously checked in the FUV-optical CMDs and classifications were assigned. The classified HB, BSS, and WD stars from the FUV-optical CMDs lie at locations as predicted by their respective models. The stars that lie near or above the TAHB sequence, and are bluer than F148W$-$G=2~mag, were classified as pHB stars. .

To select EHB stars, we estimated the temperatures of the HB stars using the F148W$-$G vs T$_{\rm eff}$ relation from the PGPUC models. Those HB stars with T$_{\rm eff}>$ 23,000~K were classified as candidate EHB stars.

The number of BSSs, HBs, pHB, and WDs detected are listed in Table~\ref{tab:number_stars}. We compared the number of detected HB stars in the UVIT/F148W with those selected for cross-match from the \textit{HST} and \textit{Gaia} CMDs. We found the HB detections to be $>90\%$ after cross-match for seven clusters except for NGC~5272 (M3), and three dense clusters NGC~362, NGC~1851, NGC~1904. As M3 has a shorter exposure time compared with other clusters in our survey, we recover 73\% of the HB stars in the UVIT/F148W after cross-matching. To check for the variables detected in FUV, we cross-matched the UVIT data with the catalogue of variable stars in GCs by \cite{Clement2001}. The number of FUV detected variables of 11 clusters are given in Table~\ref{tab:number_stars}, where the classification according to \cite{Clement2001}. The probable FUV counterparts to the known X-ray sources in GCs (which includes CVs, and other X-ray binaries) and their SEDs will be studied in our forthcoming work.

The final catalogue consists of cluster name, UVIT ID, RA, Dec, magnitudes, and errors in their respective F148W and F169M filters, G magnitudes from \textit{Gaia}~EDR3 data and \textit{HST} transformed filters, membership probability, and their classification, as shown in Table \ref{tab:catalog}. The full catalogues of three clusters NGC 288, NGC 1851, and, NGC 5466 will be made available online through Vizier catalogue service. The catalogues of rest of the clusters will be released to the community in the future as they are being analysed for science cases discussed in Section \ref{sec:sc_case}.

\subsection{FUV-optical CMDs}\label{sec:cmds_cluster}
Here, we describe the PM cleaned FUV-optical CMDs of 11 GCs, out of which 8 are reported for the first time from the UVIT observations. Whereas, three are already studied earlier \citep{Subramaniam2017, Sahu_ngc288, Sahu_ngc5466}. 
The F148W vs F148W$-$G CMDs of 11 GCs are shown in Figures \ref{fig:fuv_cmds_1}. Among our studied sample, we detect an extended HB comprising the largest number of HB and EHBs (165) in NGC~6205 with $T_{\rm eff}$ ranging from 22,000-33,000~K. We detect the second largest population of HB stars in NGC~5272 with $T_{\rm eff}$ between 6,500-28,500~K. This cluster also contains the largest number of FUV bright BSSs (35), whereas, NGC~6205 contains the least (10). We found the largest population of pHBs in NGC~6205 spanning 1-2 mag above the ZAHB in FUV-optical CMD. NGC~362 contains the largest population of WDs (24), with the brightest WDs with temperatures 90,000~K, whereas, the faintest ones have T$_{\rm eff} \sim$ 30,000~K as inferred from the WD cooling curves.

We also show the F148W vs F148W$-$F169M CMD of NGC~6205 as an example in Figure~\ref{fig:fuv_cmds_3}. In only FUV CMD, the EHB stars deviate from the usual BHB distribution and turn fainter obeying the ZAHB models. We notice one mag dip in the F148W magnitude of HB distribution beyond (F148W$-$F169M, T$_{\rm eff}) \sim$ (0.02, 14,400~K) from the ZAHB models. The models also indicate that EHBs become fainter in F148W mag at color, (F148W$-$F169M) $\sim -$0.05 corresponding to T$_{\rm eff}$ range of 30,000-34,000~K. This is roughly the region where the BHk candidates lie. We note that the ZAHB models of PGPUC and BaSTI are not in agreement for HB stars with F148W$-$F169M $>$ 0.25 corresponding to T$_{\rm eff} <$ 8,000~K. From the figure, its evident that the PGPUC models agree well with the observed HB distribution (Figure~\ref{fig:fuv_cmds_3}) as compared with the BaSTI models. However, the deviation of BaSTI models are within the 3$\sigma$ photometric errors at the fainter end of the HB distribution. For detailed description of individual cluster CMDs, we refer the readers to Appendix~\ref{sec_apendix:fuv_cmd}.\\

\begin{figure*}
\centering
\includegraphics[width=0.7\textwidth]{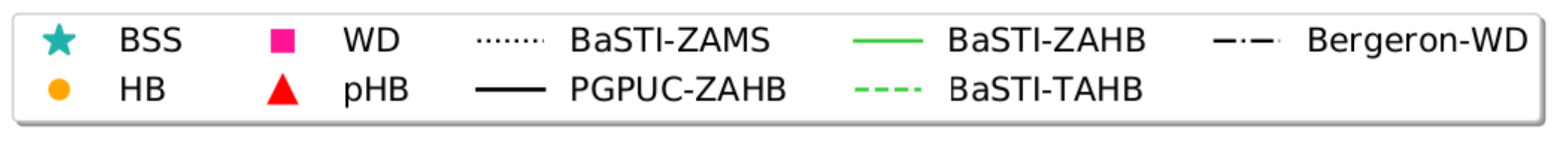}\\
\includegraphics[width=0.4\textwidth]{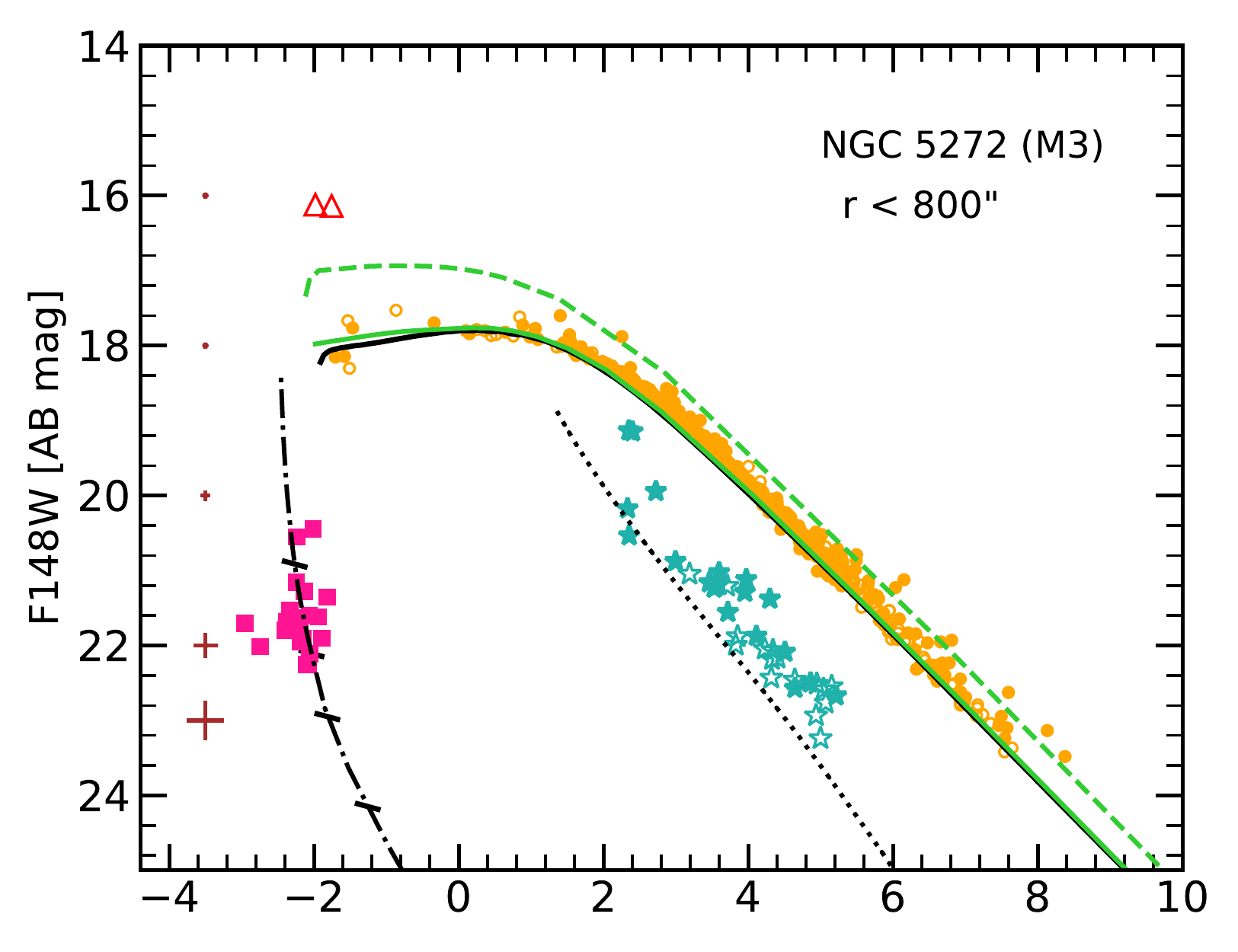}
\includegraphics[width=0.4\textwidth]{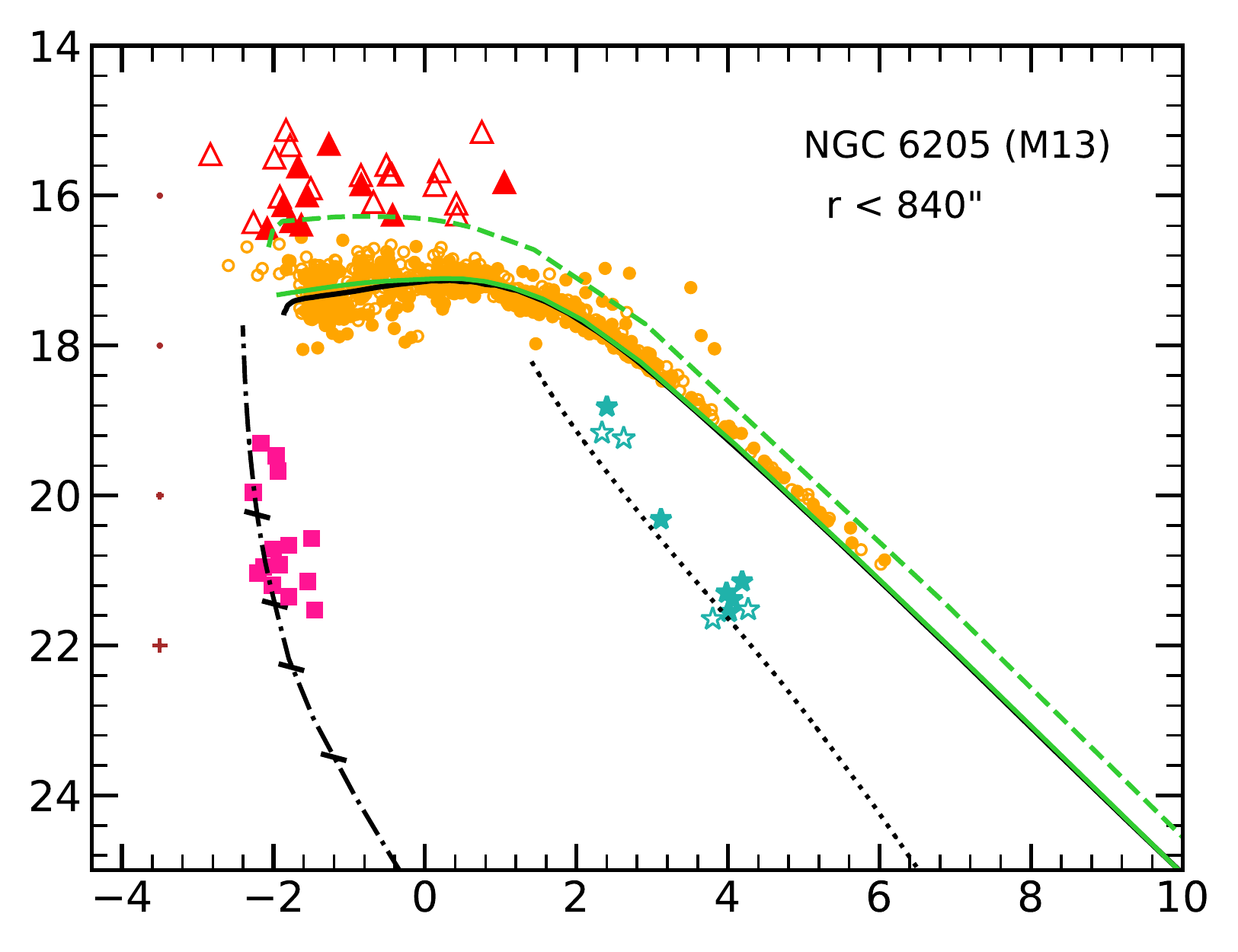}
\includegraphics[width=0.4\textwidth]{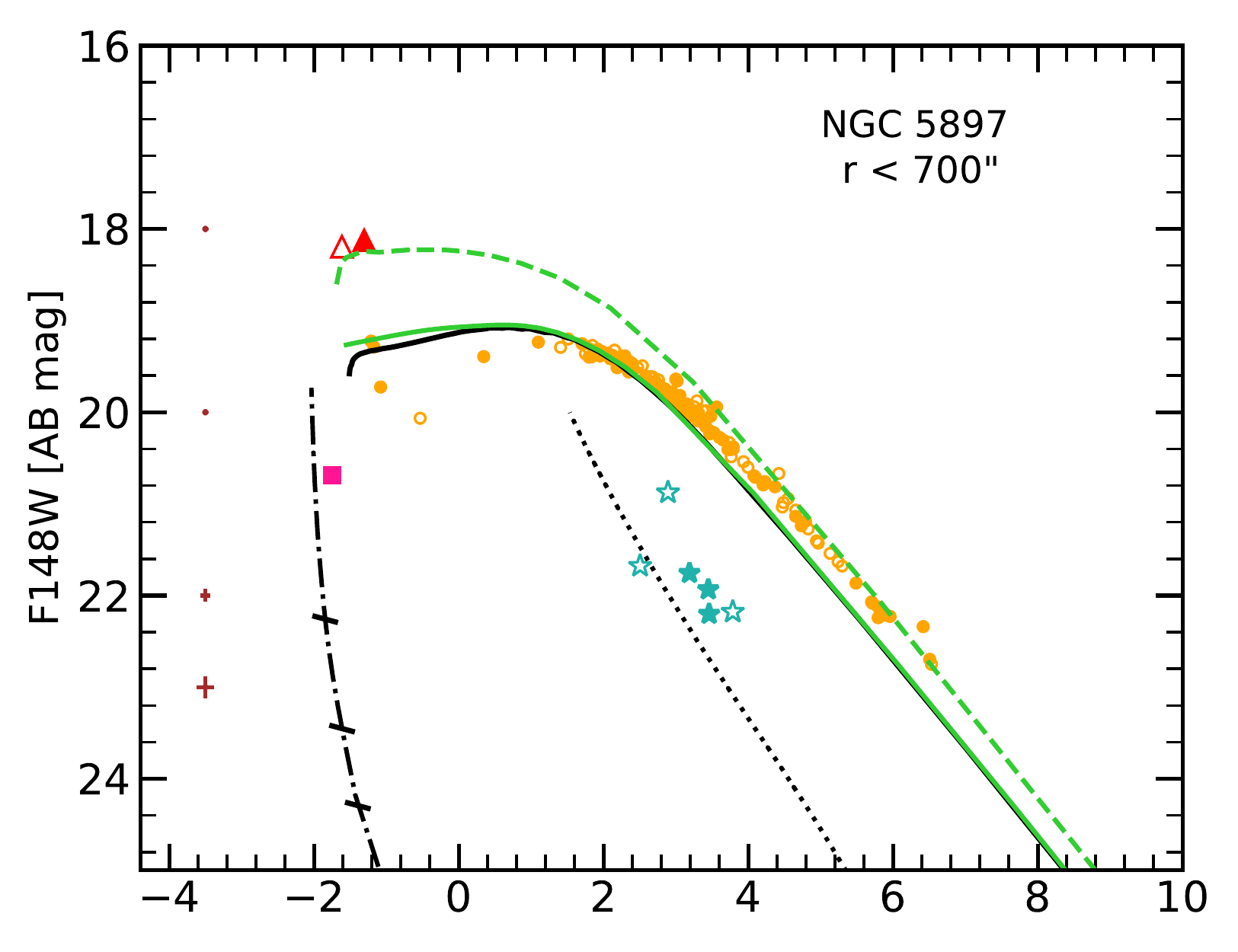}
\includegraphics[width=0.4\textwidth]{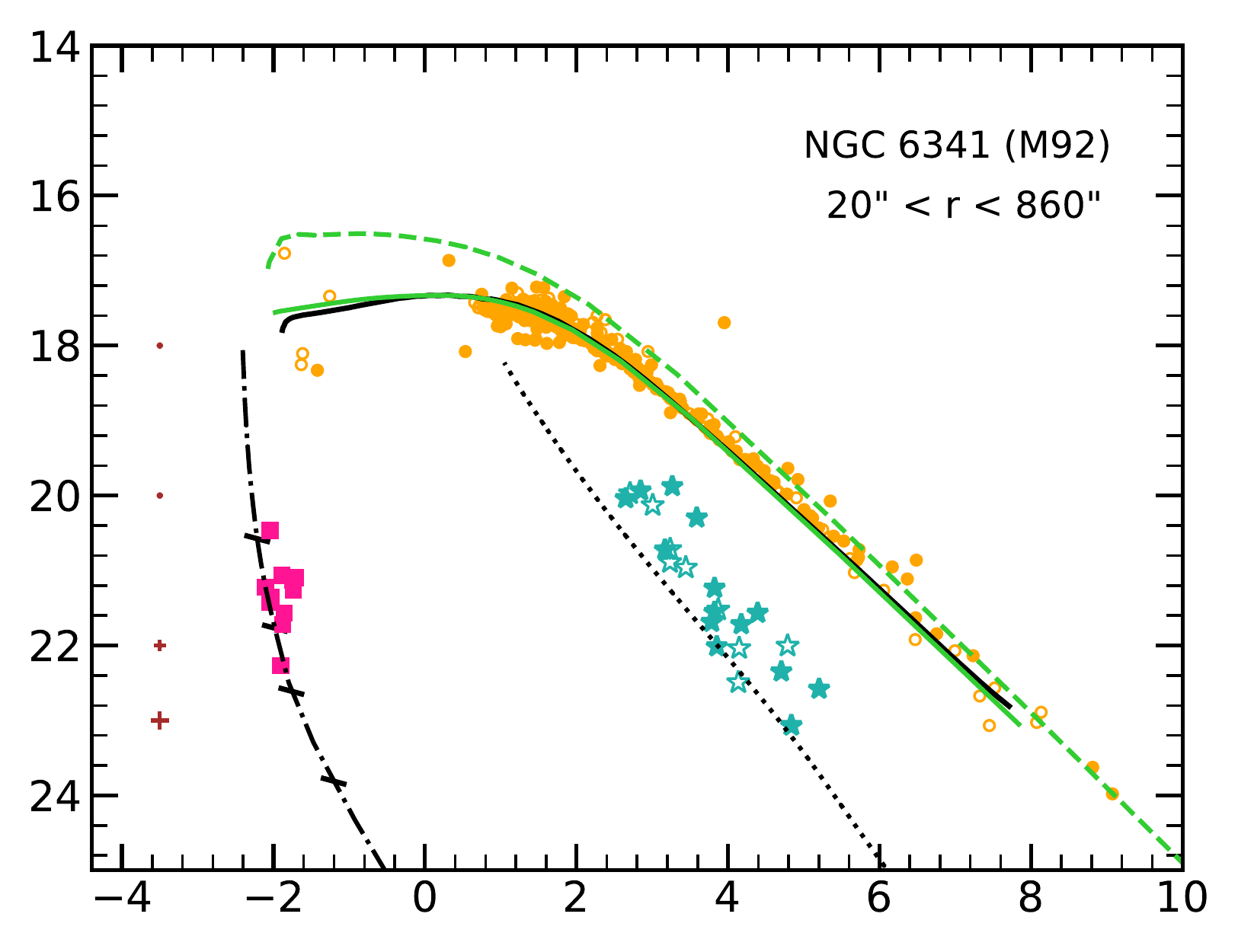}
\includegraphics[width=0.4\textwidth]{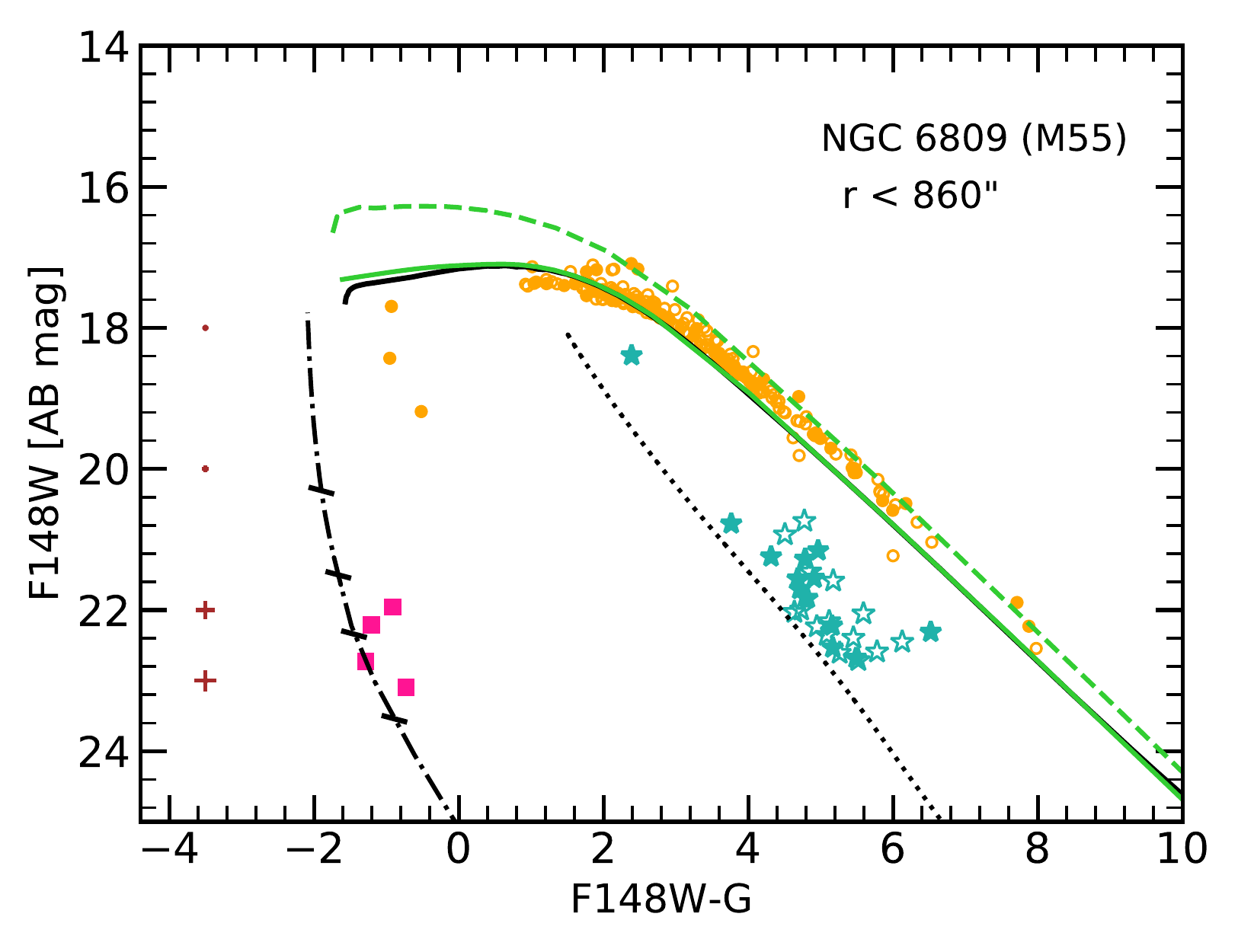}
\includegraphics[width=0.4\textwidth]{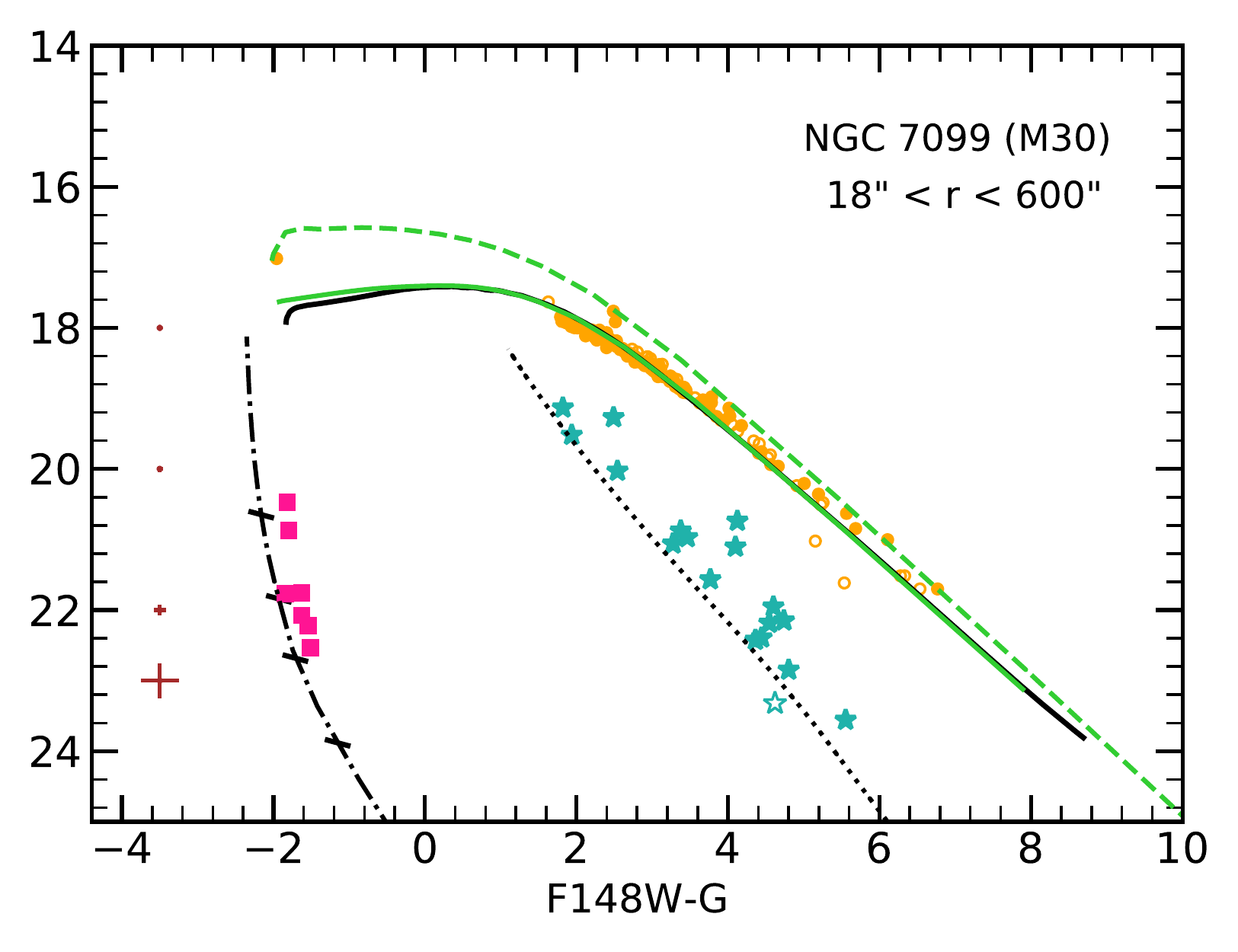}
\caption{F148W vs F148W$-$G CMDs of NGC~5272, NGC~6205, NGC~5897, NGC~6341, NGC~6809 and NGC~7099 (from top to bottom). The filled and open symbols denote the UVIT-\textit{HST} and the UVIT-\textit{Gaia} EDR3 cross-matched sources, respectively. The models are over-plotted by considering the reddening and distance modulus of each cluster from \citep{Harris1996}. The error bars (median) are shown in brown color on the left side of each plot. The abbreviations for the type of stars and stellar evolutionary models are defined in the text. }
\label{fig:fuv_cmds_1}
\end{figure*}
  
\renewcommand{\thefigure}{\arabic{figure} (Cont.)}
\addtocounter{figure}{0}

\begin{figure*}
\centering
\ContinuedFloat
\captionsetup{list=off}
\includegraphics[width=0.4\textwidth]{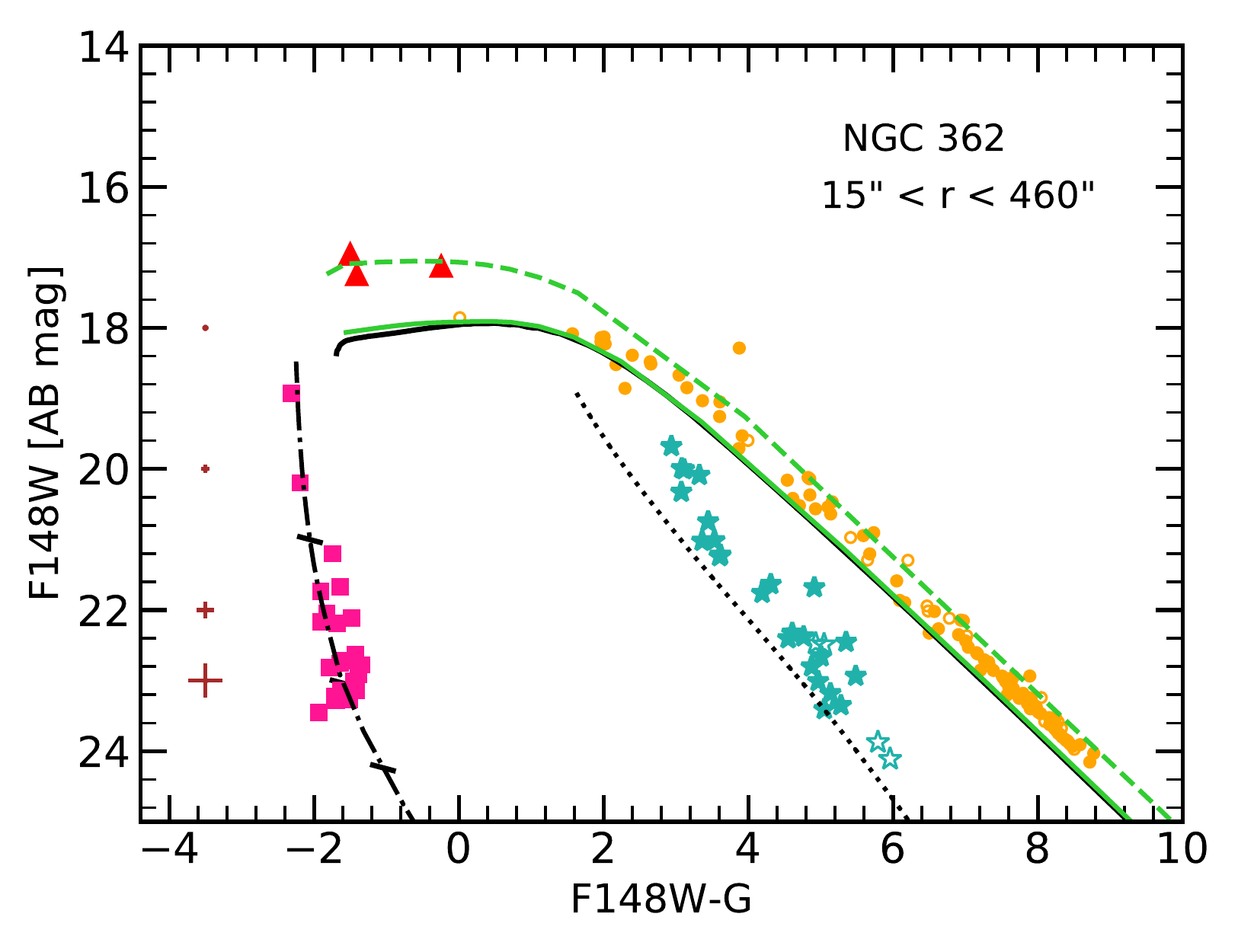}
\includegraphics[width=0.4\textwidth]{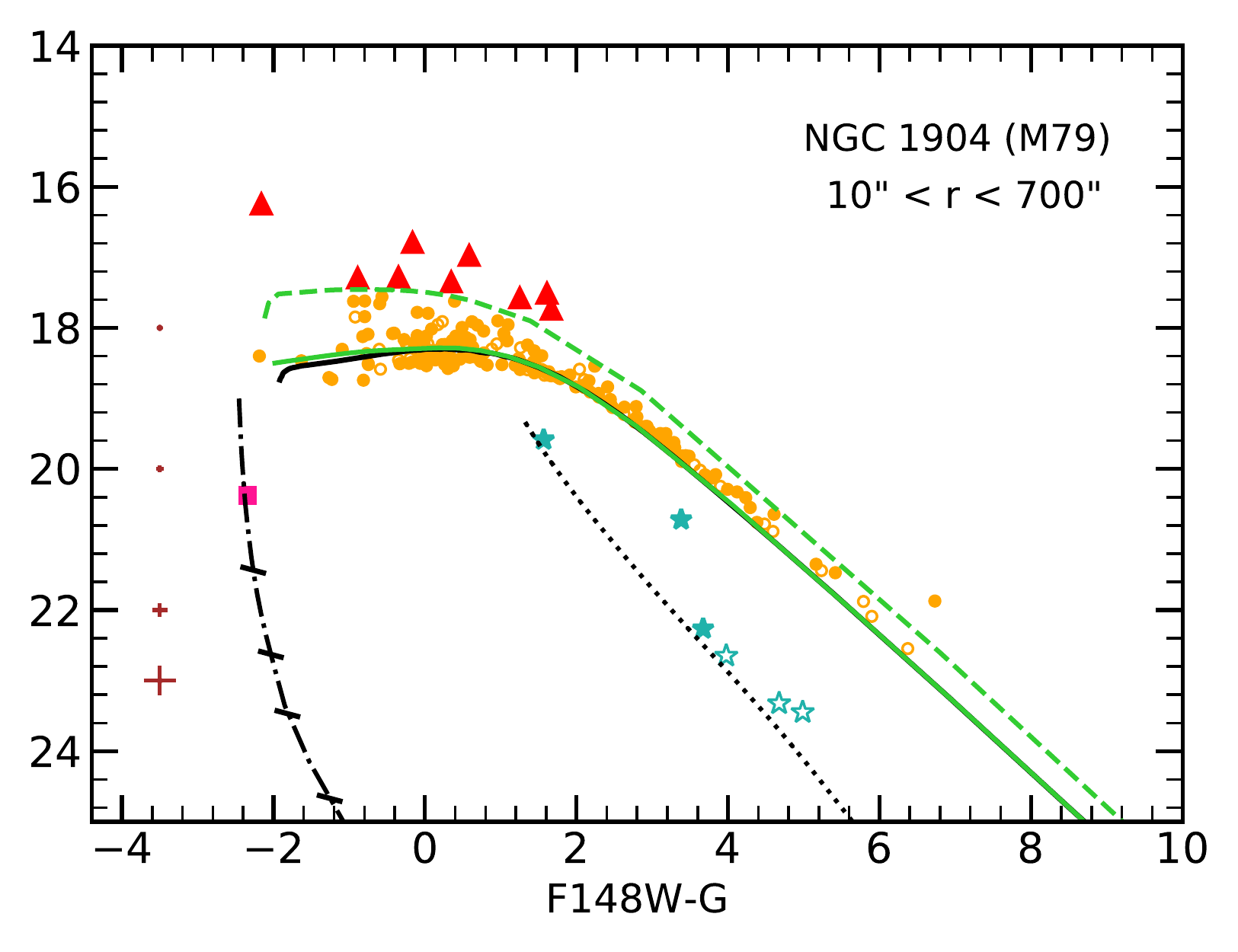}
\includegraphics[width=0.32\textwidth]{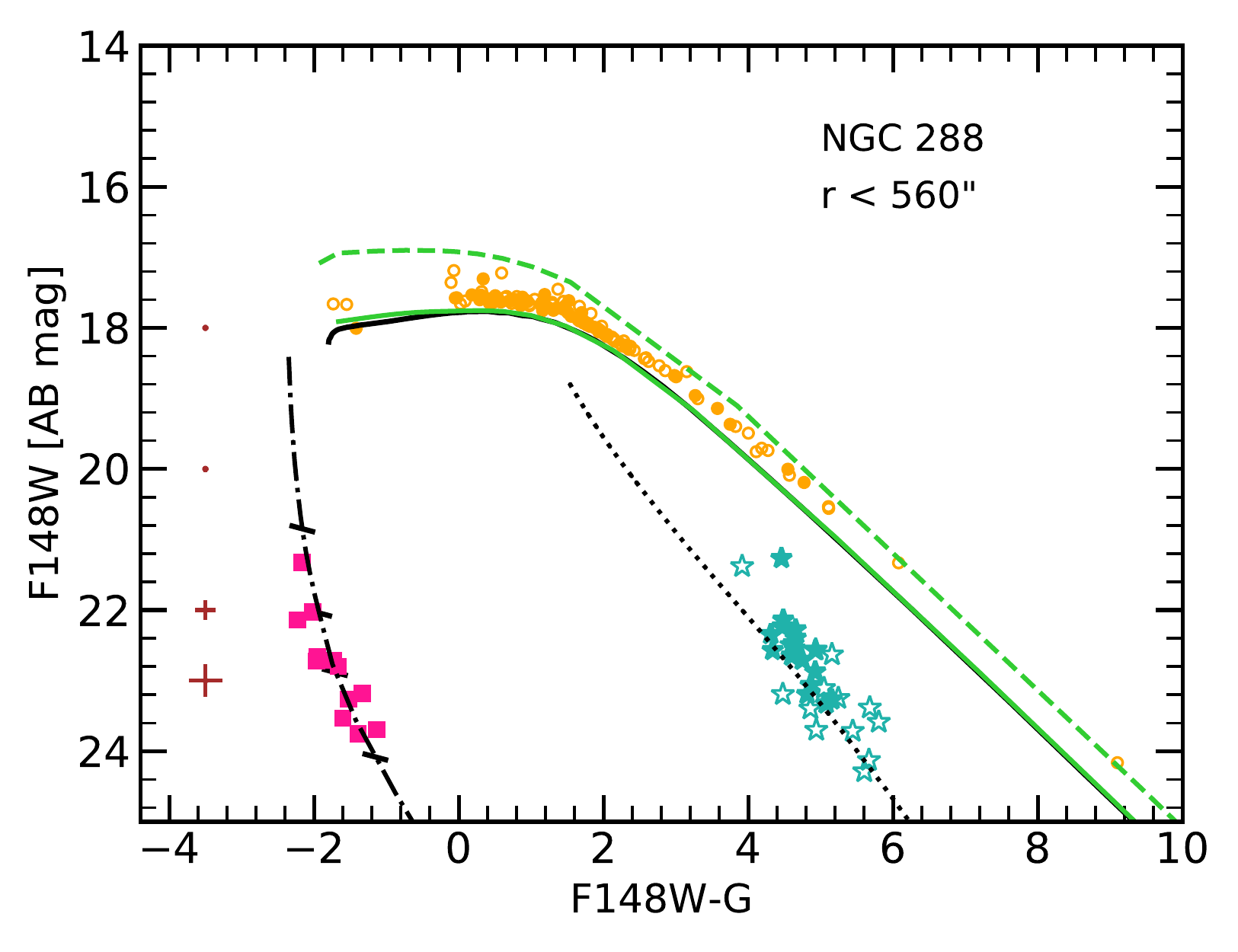}
\includegraphics[width=0.32\textwidth]{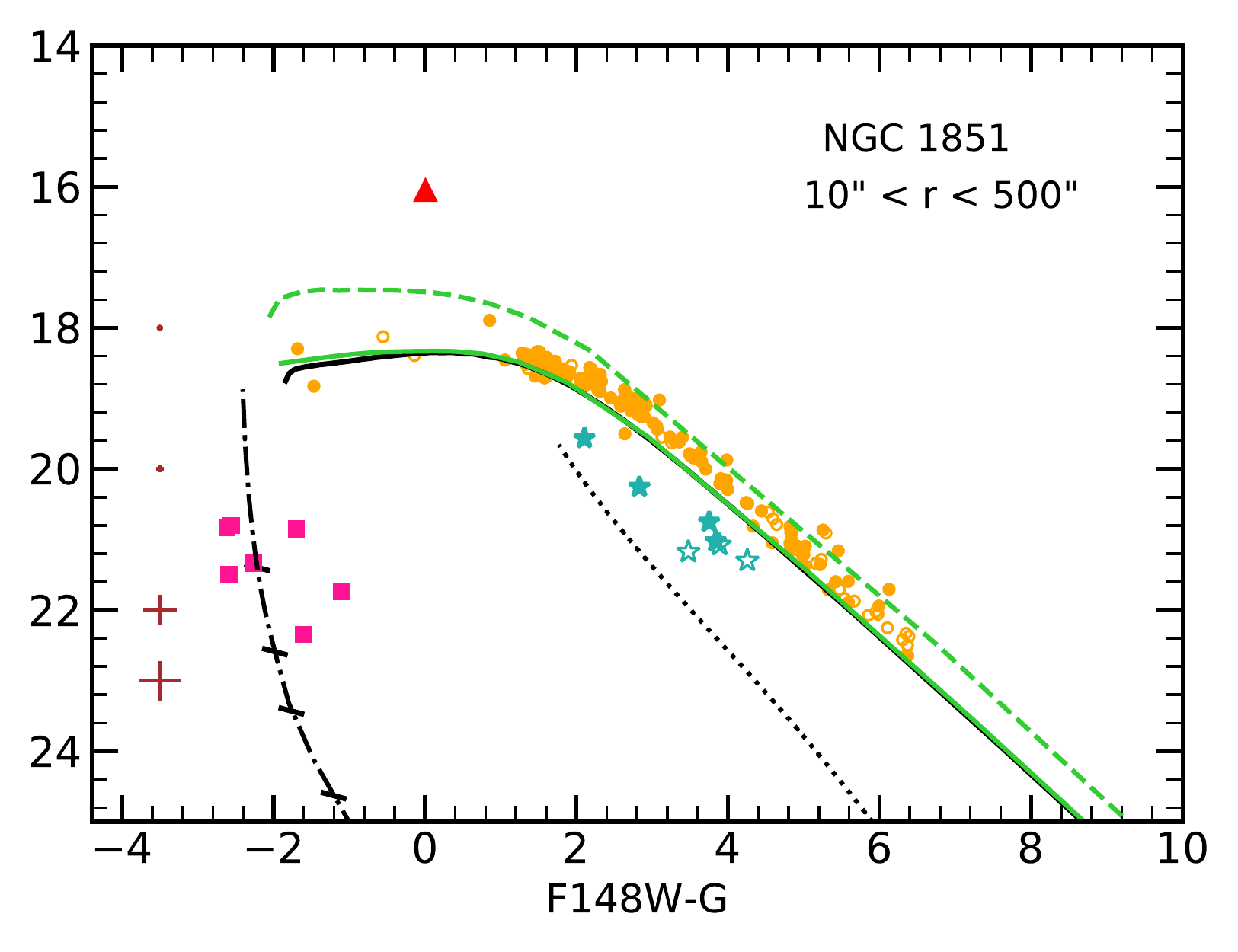}
\includegraphics[width=0.32\textwidth]{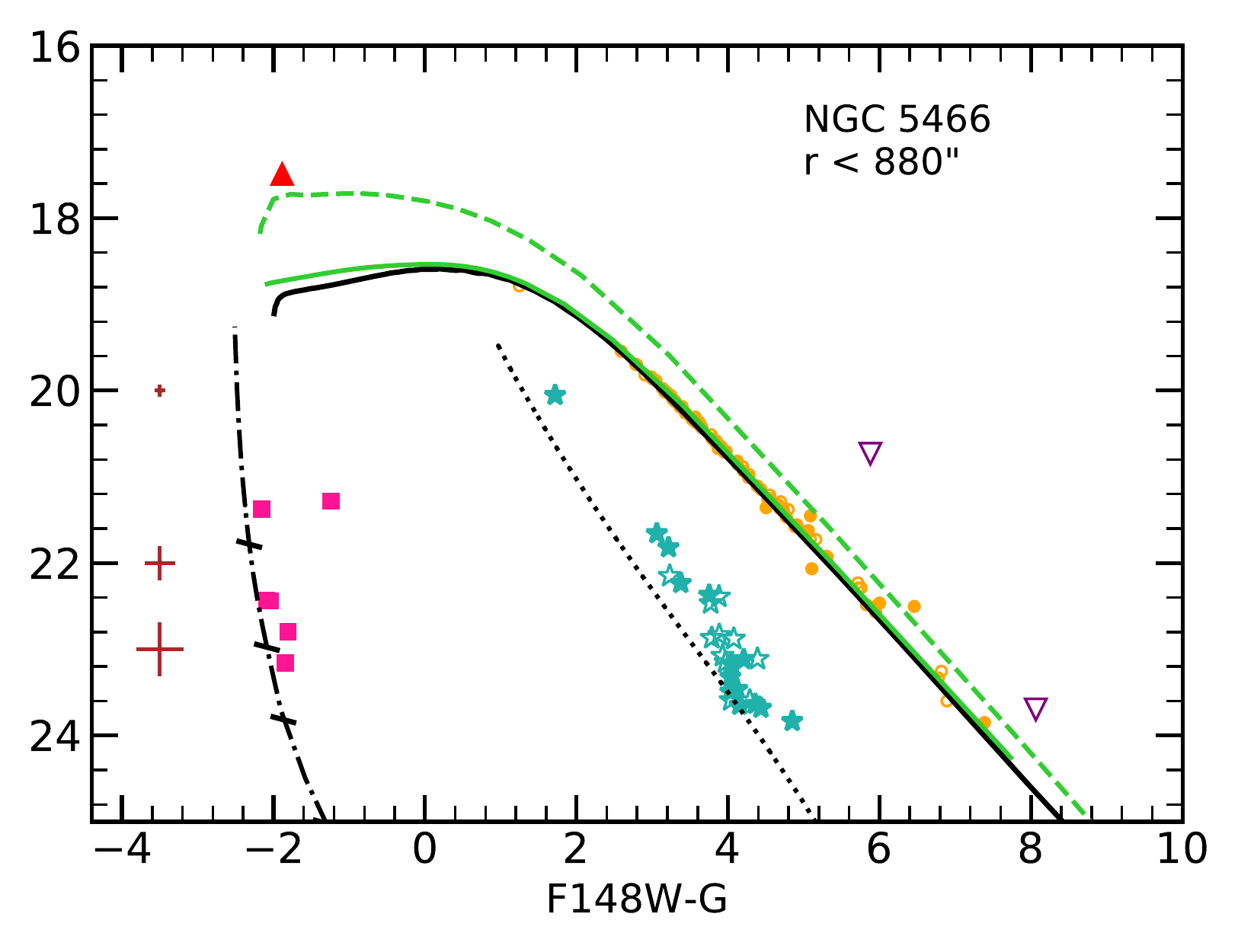}
\caption{F148W vs F148W$-$G CMDs of NGC~362, NGC~1904, NGC~288, NGC~362, NGC~1851 and NGC~5466.}
\end{figure*}

\renewcommand{\thefigure}{\arabic{figure}}

\begin{figure}
\centering
\includegraphics[width=0.9\columnwidth]{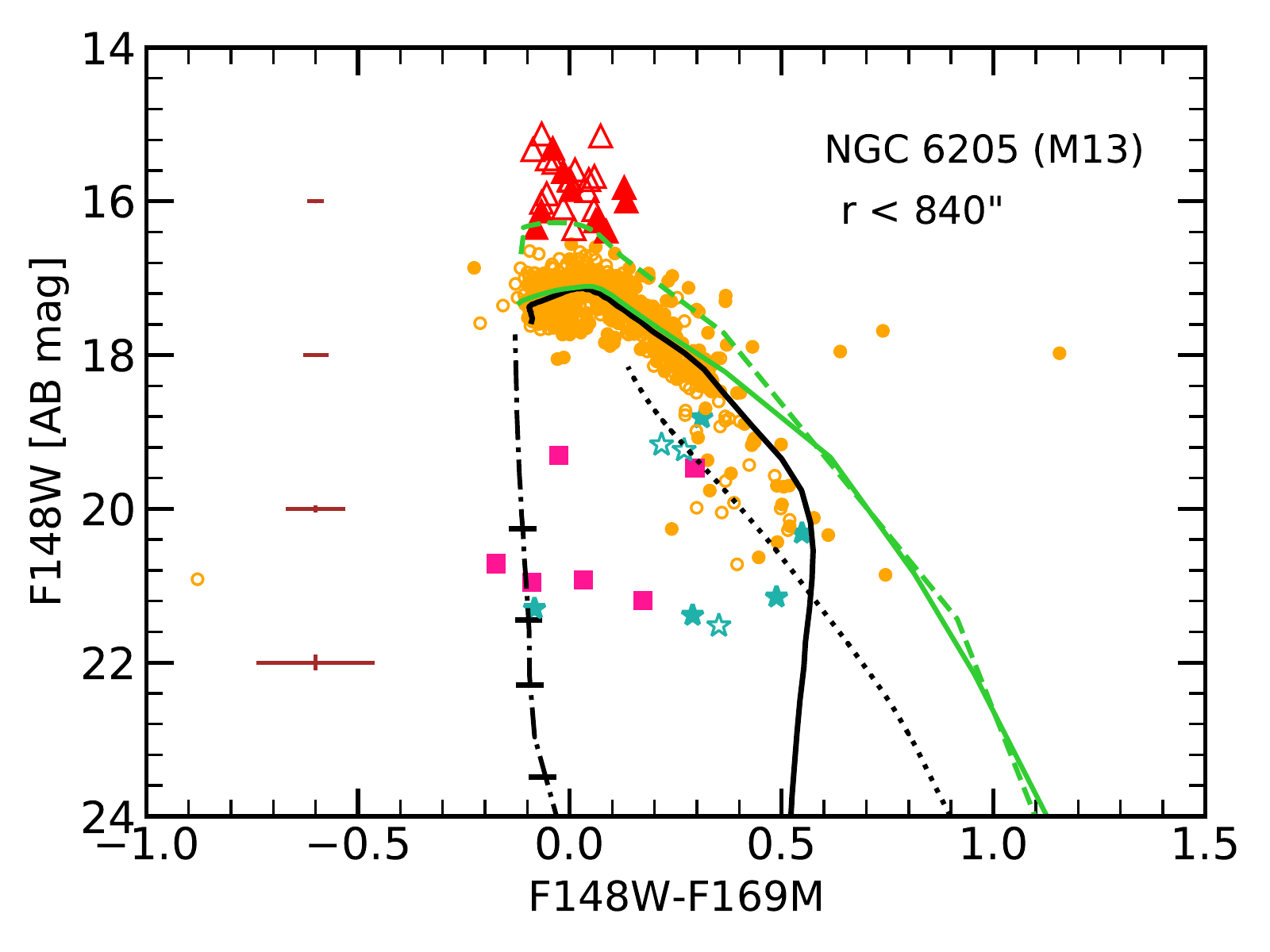}
\caption{F148W vs F148W$-$F169M CMDs of NGC~6205. The error bars are shown in brown color. Refer to Figure~\ref{fig:fuv_cmds_1} for description of symbols.}
\label{fig:fuv_cmds_3}
\end{figure}

\begin{table*}
\centering
\begin{threeparttable}
\addtolength{\tabcolsep}{-0.5pt}
\caption{Total number of sources detected by cross-matching the UVIT with the \textit{HST} and \textit{Gaia} EDR3 data along with their classification.}
\begin{tabular}{lcccc|ccc|cccc|ccc|cc}
\hline
\hline
Cluster & \multicolumn{4}{c|}{ UVIT$_{\mathrm{F148W}}$-\textit{\textit{HST}}}& \multicolumn{3}{c|}{ UVIT$_{\mathrm{F148W}}$-\textit{Gaia} EDR3} & \multicolumn{4}{c|}{Total} & \multicolumn{3}{c|}{Variables\tnote{*}}   &\multicolumn{2}{c}{ UVIT$_{\mathrm{F148W}}$-UVIT$_{\mathrm{F169M}}$}\\\hline
& WD & HB & BSS & pHB & HB & BSS & pHB & HB & BSS & pHB & EHB & RRL & SX Phe & E & \textit{HST} & \textit{Gaia} EDR3 \\\hline
NGC~288	&	12	&	42	&	17	&	-	&	90	&	14	&	-	&	132	&	31	&	-	&	3	&	1	&	5	&	-	&	57	&	97	\\
NGC~362	&	25	&	88	&	24	&	3	&	23	&	4	&	-	&	111	&	28	&	3	&	-	&	22	&	-	&	-	&	104	&	10	\\
NGC~1851	&	7	&	116	&	4	&	1	&	58	&	3	&	-	&	174	&	7	&	1	&	2	&	19	&	-	&	-	&	119	&	53	\\
NGC~1904	&	1	&	148	&	3	&	9	&	53	&	3	&	-	&	201	&	6	&	9	&	4	&	5	&	-	&	-	&	154	&	53	\\
NGC~5272	&	20	&	233	&	18	&	-	&	165	&	17	&	3	&	398	&	35	&	3*	&	5	&	147	&	5	&	-	&	209	&	138	\\
NGC~5466	&	5	&	31	&	14	&	1	&	56	&	11	&	-	&	87	&	25	&	1	&	-	&	18	&	2	&	2	&	45	&	63	\\
NGC~5897	&	1	&	71	&	3	&	1	&	108	&	6	&	1	&	179	&	9	&	2	&	3	&	9	&	-	&	-	&	-	&	-	\\
NGC~6205	&	14	&	368	&	6	&	10	&	384	&	4	&	17	&	752	&	10	&	27	&	165	&	7	&	2	&	-	&	385	&	400	\\
NGC~6341	&	11	&	219	&	15	&	-	&	136	&	9	&	-	&	355	&	24	&	-	&	5	&	15	&	2	&	-	&	-	&	-	\\
NGC~6809	&	4	&	61	&	13	&	-	&	190	&	15	&	-	&	251	&	28	&	-	&	2	&	10	&	16	&	1	&	77	&	202	\\
NGC~7099	&	7	&	104	&	17	&	-	&	72	&	1	&	-	&	176	&	18	&	-	&	1	&	4	&	2	&	1	&	119	&	69	\\
\hline
\label{tab:number_stars}
\end{tabular}
\begin{tablenotes}\footnotesize
    \item[*] Total number of variables detected by cross-matching the UVIT-\textit{HST} and the UVIT-\textit{Gaia}~EDR3 sources with \cite{Clement2001}.  RRL stands for RR Lyrae, SX Phe for SX Phoenicis, and E for eclipsing binaries.
\end{tablenotes}
\end{threeparttable}
\end{table*}

Detailed analysis of HB morphology and BSSs with UVIT was carried out in our previous studies for the three clusters, NGC~288 \citep{Sahu_ngc288}, NGC~1851 \citep{Subramaniam2017, Singh2020}, and NGC~5466 \citep{Sahu_ngc5466}. In this work, we have updated the FUV-optical catalogues with the HUGS data and the \textit{Gaia}~EDR3 PM analysis (lower panels of Figure~\ref{fig:fuv_cmds_1}). In addition, we have also included the WD detections from the UVIT-\textit{HST} common fields. The hottest WDs detected in these clusters have $T_{\rm eff}$ varying from 50 to 70~kK. These three clusters along with the other eight have been used for the rest of the analysis below.

\section{Stacked FUV-optical CMDs}\label{sec:stack_cmd}
Using \textit{GALEX} observations, \cite{Schiavon2012} generated the stacked FUV vs FUV$-$NUV CMDs of 23 GCs and found well-populated HB and UV bright stars. They showed that this stacked CMD is especially useful for identifying PAGBs that are otherwise sparse in number (due to fast evolutionary timescales), and do not form a well-defined sequence in individual UV CMDs. However, without PM information they were unable to separate the WD and BSS sequences. These sequences were contaminated by background objects. 

\begin{figure*}
\centering
\includegraphics[width=0.9\textwidth]{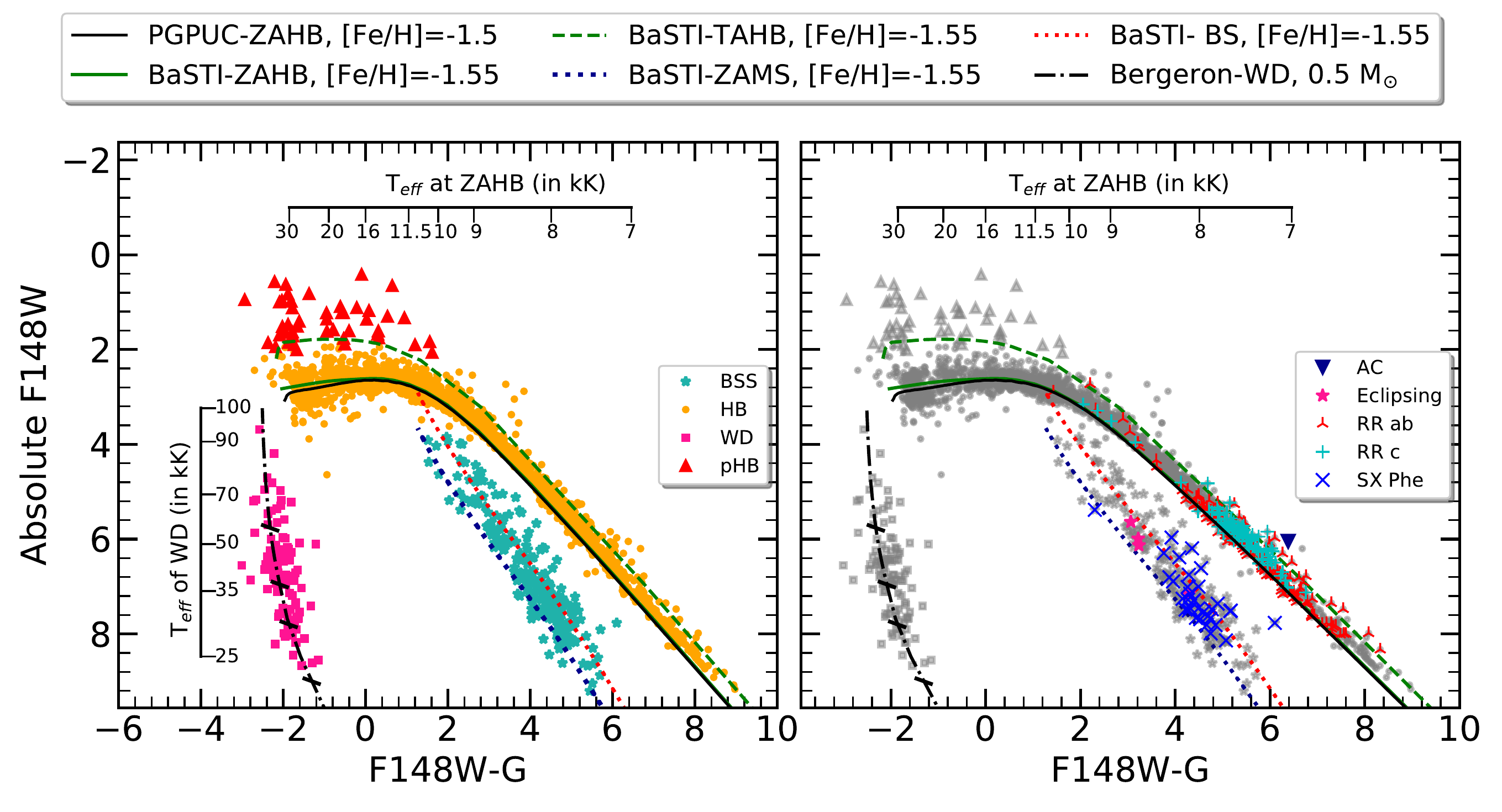}
\caption{Stacked absolute F148W vs F148W$-$ G CMD of 11 GCs after correcting for reddening and distance modulus. Both figures include the UVIT-\textit{HST} and the UVIT-\textit{Gaia} EDR3 common detections. For reference, ZAMS is shifted upward by 0.75, which denotes equal mass binary sequence (BS) (red dotted line). The effective temperatures (in 10$^{3}$K) at ZAHB and WDs from the models are also marked in the figures. The horizontal marks over WD models correspond to cooling ages of 1, 5, 10, and 25 Myrs respectively (from top to bottom). Right Panel: All the FUV detected variables: RRLs, SX Phes, Eclipsing, and Anomalous Cepheid (AC), where grey dots are the total number of detected sources from both the UVIT-\textit{HST} and the UVIT-\textit{Gaia} EDR3.}
\label{fig:stack_cmd_1}
\end{figure*}

\subsection{HB morphology}
The FUV bright population of 11 GCs comprises of 2,816 HB stars where 1,335 are the UVIT-\textit{Gaia} common detections. Among the HB population, 190 are EHB candidates and 257 are known RRLs. As shown in Figure~\ref{fig:stack_cmd_1}, the stacked HB distribution spans around 10~mag in color and 6~mag in the F148W magnitude and 10~mag in F148W$-$G color which is mostly populated by BHB and EHB stars. The PGPUC-ZAHB for [Fe/H] $= -1.5$, and the BaSTI-ZAHB and TAHB models for [Fe/H] $= -1.55$ dex are shown in Figure~\ref{fig:stack_cmd_1}. To check for peculiarities in the HB distributions, we plotted the effective temperature scale at ZAHB from PGPUC models. The HB distribution starts deviating from the usual diagonal sequence at 11,500~K which corresponds to the well-known Grundahl-jump (G-jump) where the atomic diffusion sets in \citep{Grundahl1999}.   

\begin{figure}
\centering
\includegraphics[width=\columnwidth]{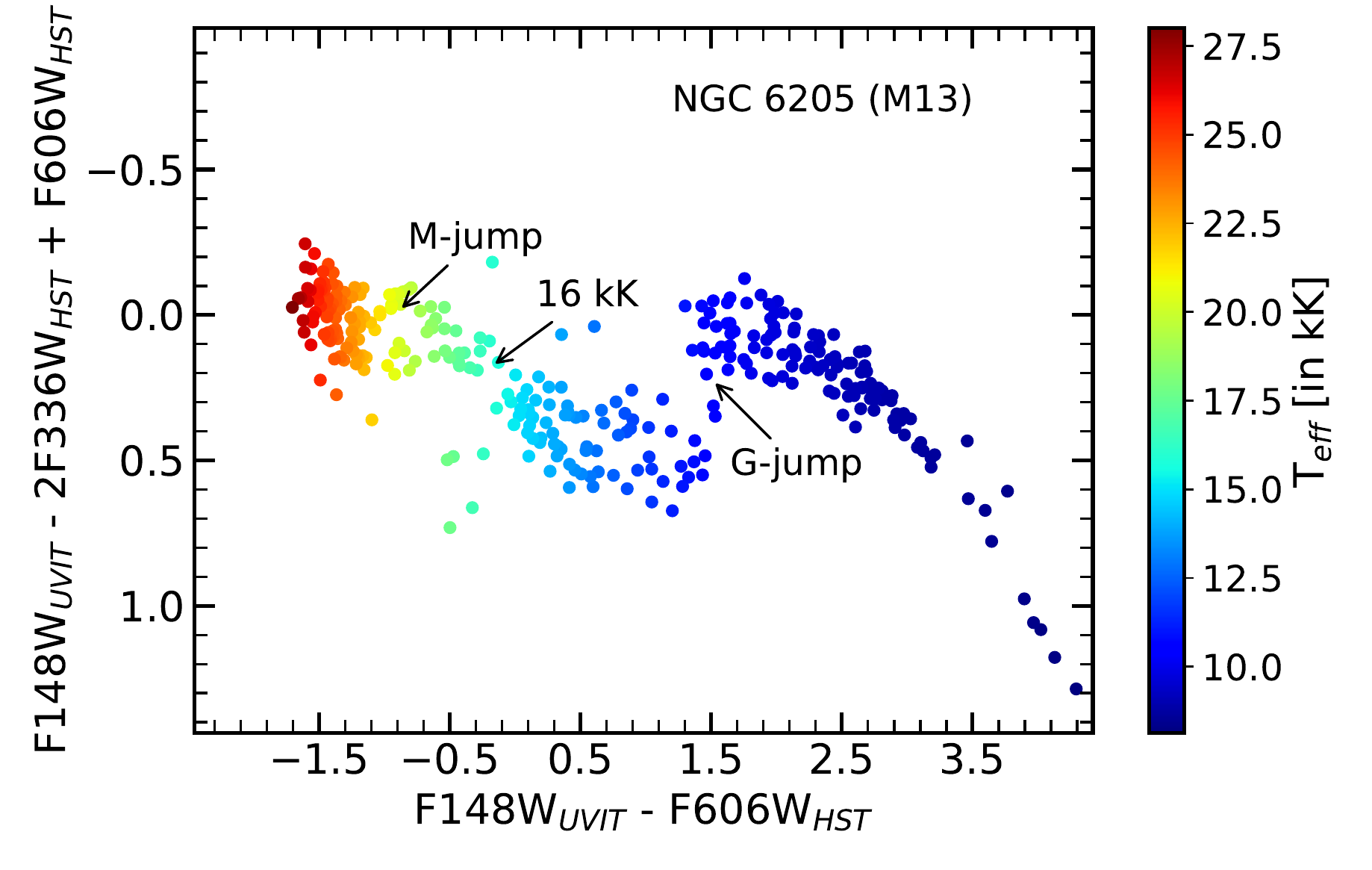}
\caption{Pseudo-color vs color plot i.e. (F148W$-$F606W) vs (F148W $-$ 2.F336W $+$ F606W) for the UVIT-\textit{HST} common HB sources of NGC~6205. The G-jump \citep{Grundahl1999} and M-jump \citep{Momany2004} corresponding to $T_{\rm eff}$ at $\sim$ 11,500~K and 21,000~K, respectively, are also marked in the figure. The color-bar shows the T$_{\rm eff}$ (in kK) of HB stars estimated from F148W$-$G color vs $T_{\rm eff}$ relation obtained from PGPUC models.}
\label{fig:pseudo_color}
\end{figure}

\begin{figure}
\centering
\includegraphics[width=0.9\columnwidth]{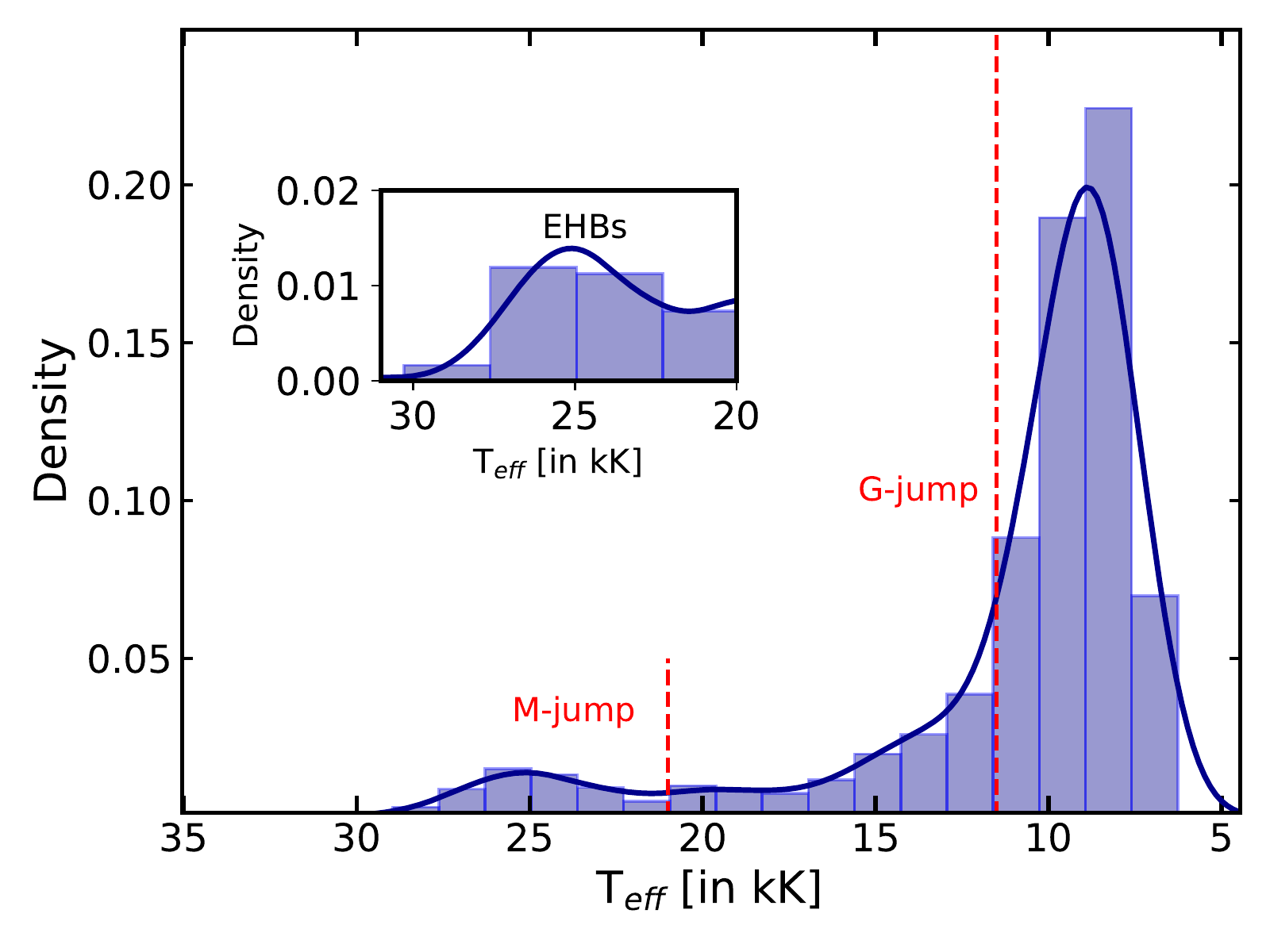}
\caption{Histogram of T$_{\rm eff}$ (in kK) for all the UVIT detected HB stars with G-jump and M-jump marked as red dashed lines. The EHB candidates are the stars that are hotter than the M-jump shown in the inset.}
\label{fig:hist_hb}
\end{figure}

\cite{Brown2016} used three filter combinations of the \textit{HST} (F275W, F336W, F438W) to create a pseudo color vs color plot and studied the discontinuities in HB distributions for 45 GCs. Similarly, we used the multi-wavelength observations by combining the UVIT/F148W with F336W and F606W filters of the \textit{HST} from the HUGS survey to study the HB gaps/peculiarities in FUV CMDs. The filters were chosen keeping in mind their sensitivity to surface gravity, $T_{\rm eff}$ and He depletion in the HB stars. The wide baseline F148W$_{\rm UVIT}$-F606W$_\textit{HST}$ is sensitive to $T_{\rm eff}$ and He variations, whereas, F336W is sensitive to the surface gravity of the HB stars \citep{Brown2016}. The pseudocolor-color stacked plot for the UVIT-\textit{HST} common detections of NGC~6205 with most extended HB is shown in Figure~\ref{fig:pseudo_color} where the y-axis is the difference of two colors, (F148W$-$F336W) and (F336W$-$F606W), and x-axis is F148W$-$F606W corrected for reddening. The effective temperatures of HB stars estimated from (F148W$-$G) vs $T_{\rm eff}$ relation, is shown in the color bar. We note that the stars in the pseudo-color plot bends towards fainter magnitude around the G-jump at 11,500~K \citep{Grundahl1999} and show gaps at around 16,000~K and M-jump (21,000~K) \citep{Momany2004}. The pseudocolor plot of the rest of the clusters is given in Figure~\ref{fig:pseudo_color_all}. We found clear gaps in the HB distribution at (F148W$-$G)$ \sim -1.2$, and, $-0.4$. To identify the gaps more prominently, we plotted the $T_{\rm eff}$ histogram of the UVIT detected HB stars for all the clusters in our sample (Figure~\ref{fig:hist_hb}). The HB distribution shows a peak at (F148W$-$G)$_{\rm redd} \sim$ 2.5 corresponding to $T_{\rm eff}$ $\sim$ 9,000~K. The histogram shows a dip at $T_{\rm eff}$ $\sim$ 11,500 and 21,000~K which corresponds to G-jump and M-jump respectively. The stars bluer than the M-jump are classified as candidate EHBs. A significant population of EHB stars (156) show a peak at (F148W$-$F606W)$\sim -$1.5 corresponding to T$_{\rm eff} \sim$ 25,000~K. 

We detected 46 pHB candidates in 11 GCs.
The pHB candidates are 1-2~mag brighter than the ZAHB in F148W and are bluer than F148W$-$G $\sim$ 2 mag in the stacked CMD (Figure~\ref{fig:stack_cmd_1}). They include mostly the AGBm and a few P(e)AGB candidates that will evolve towards AGB or WDs depending on the envelope mass after exhaustion of core He in the HB stars. \citep{Schiavon2012} provided a catalogue of UV bright stars using FUV vs FUV$-$NUV stacked CMDs of 44 GCs. By comparing the UVIT detections with their catalogue of pHB stars, we found only 9 such candidates as opposed to 19 stars for 8 GCs. The difference might be due to the membership cut-off which was not included in their studies. 

\subsection{BSSs}
We detected 221 FUV BSSs in 11 GCs. As shown in Figure~\ref{fig:stack_cmd_1}, the BSSs occupy a F148W magnitude range ($\sim$ 4-9) that spans around 5~mag in brightness. 

For reference, we show the ZAMS models and equal mass binary sequence in Figure~\ref{fig:stack_cmd_1}. Among the total number of FUV BSSs, we noted that around 70\% of them lie between the ZAMS and binary sequence. Whereas, rest of them populate the regions on the redder side of the binary sequence that are possibly evolving away from the MS towards giant phases. 
A comparison with the ZAMS models shows that the BSSs located at the brightest end of their distribution have apparent masses $\sim$ 1.5 M$_{\odot}$, which is twice the mass of MS turn-off mass (M$_{TO}$) for GCs. 

\subsection{WDs}
We detected 106 WDs in 11 GCs. As shown in Figure~\ref{fig:stack_cmd_1}, the WDs occupy similar absolute F148W magnitude range as BSSs ($\sim$ 4-9) spanning around 5~mag in brightness. The WDs are 7-8~mag bluer than the BSS sequence. Bergeron WD models of DA type and mass $0.5 M_{\odot}$ for different $T_{\rm eff}$ are shown in the vertical bar. We were able to detect WDs as faint as F148W$_{\rm abs}\sim$ 8.5 mag corresponding to T$_{\rm eff}$ and cooling age of around 25,000~K and 17 Myr respectively.

\subsection{Variables}
We detected 296 FUV bright variables in 11 GCs which comprise 257 RRLs, 34 SX Phes, 4 eclipsing binaries (EB), and 1 Anomalous Cepheid (AC), where the variables classification is taken from \cite{Clement2001}. Their distribution in F148W vs (F148W$-$G) stacked CMDs are shown in right panel of Figure~\ref{fig:stack_cmd_1}. In the figure, most of the FUV bright variables lie in magnitude range 5 $<$ F148W $<$ 8.5 in FUV CMDs. RRLs lying within this range cover more than 3 magnitude with effective temperatures varying from $\sim$ 6,200-8,300~K. They comprise of a mixed population of RRab and RRc, where RRc stars lie on the bluer end of the distribution similar to their locations in the optical CMDs. Since RRLs are sampled at random phases, a given star is plotted at a magnitude equal to the mean of the time series of observations for that star. Thus, we found few of them to lie at the brighter end of HB distribution. These RRLs are from clusters NGC~5272, NGC~362, and NGC~1904, which have the shortest exposure times among our sample. The SX Phe variables cover around 2~mag in FUV with most of them lying at the fainter end of the BSSs distribution in the stacked CMDs. 3 EBs are lying at F148W = 6.5 and 1 EB lying at F148W = 7.2 hiding below the SX Phe. They are located near the equal mass binary sequence as shown in the right panel of Figure~\ref{fig:stack_cmd_1}.

\section{Results} \label{sec:results}

\begin{table}
\centering
\begin{threeparttable}[t]
\caption{Parameters of clusters used for HB analysis. The fourth column denotes L extension parameter of HB and its corresponding uncertainty. See Section~\ref{sec:results} for more details.}
\begin{tabular}{lccc}
\hline
\hline
Cluster &  log$\rho_{c}$ \tnote{1} & $\delta Y_{max}$ \tnote{2} & $L_{F148W-G/F606W}$ \\
\hline
NGC~288	&	1.78	&	0.016	$\pm$	0.012	&	2.93	$\pm$	0.19	\\
NGC~362	&	4.74	&	0.026	$\pm$	0.008	&	2.66	$\pm$	0.16	\\
NGC~1851	&	5.09	&	0.025	$\pm$	0.006	&	2.29	$\pm$	0.1	\\
NGC~1904	&	4.08	&	-			&	3.92	$\pm$	0.07	\\
NGC~5272	&	3.57	&	0.041	$\pm$	0.009	&	3.85	$\pm$	0.05	\\
NGC~5466	&	0.84	&	0.007	$\pm$	0.024	&	1.37	$\pm$	0.05	\\
NGC~5897	&	1.53	&	-			&	3.01	$\pm$	0.05	\\
NGC~6205	&	3.55	&	0.052	$\pm$	0.004	&	4.47	$\pm$	0.03	\\
NGC~6341	&	4.3	&	0.039	$\pm$	0.006	&	2.89	$\pm$	0.05	\\
NGC~6809	&	2.22	&	0.026	$\pm$	0.015	&	2.36	$\pm$	0.06	\\
NGC~7099	&	5.01	&	0.022	$\pm$	0.01	&	1.25	$\pm$	0.04	\\
\hline
\label{tab:correlation}
\end{tabular}
\begin{tablenotes}\footnotesize
    \item[1] logarithm of central luminosity density (in units of L$_{\odot}$/pc$^{3}$) from \cite{Harris1996} (2010 edition).
    \item[2] Maximum internal He variation \citep{Milone2018}.
    
\end{tablenotes}
\end{threeparttable}
\end{table}

\subsection{FUV Color extension of HB}\label{sec:l_hb}
\cite{Brown2016} studied the HB morphologies of 44 GCs and inferred that He enhanced second generation (2G) stars populate the bluer end of the HB region. 
In a recent study, using \textit{HST} photometry, \cite{Milone2018} derived the maximum He variation ($\delta Y_{max}$) between 2G and 1G stars in a sample of 57 GCs. They found that $\delta Y_{max}$ positively correlates with the color extension of HB derived from F275W$-$F814W. Recently, \cite{Tailo2020} derived the mass-loss parameters of 57 GCs and found a positive correlation of the mass-loss difference ($\delta \mu_{e}$) between 2G and 1G stars with $\delta Y_{max}$. They suggested that both mass-loss and He enhancement are the main second parameters influencing the HB morphology.

\begin{figure*}
\centering
\includegraphics[width=\textwidth]{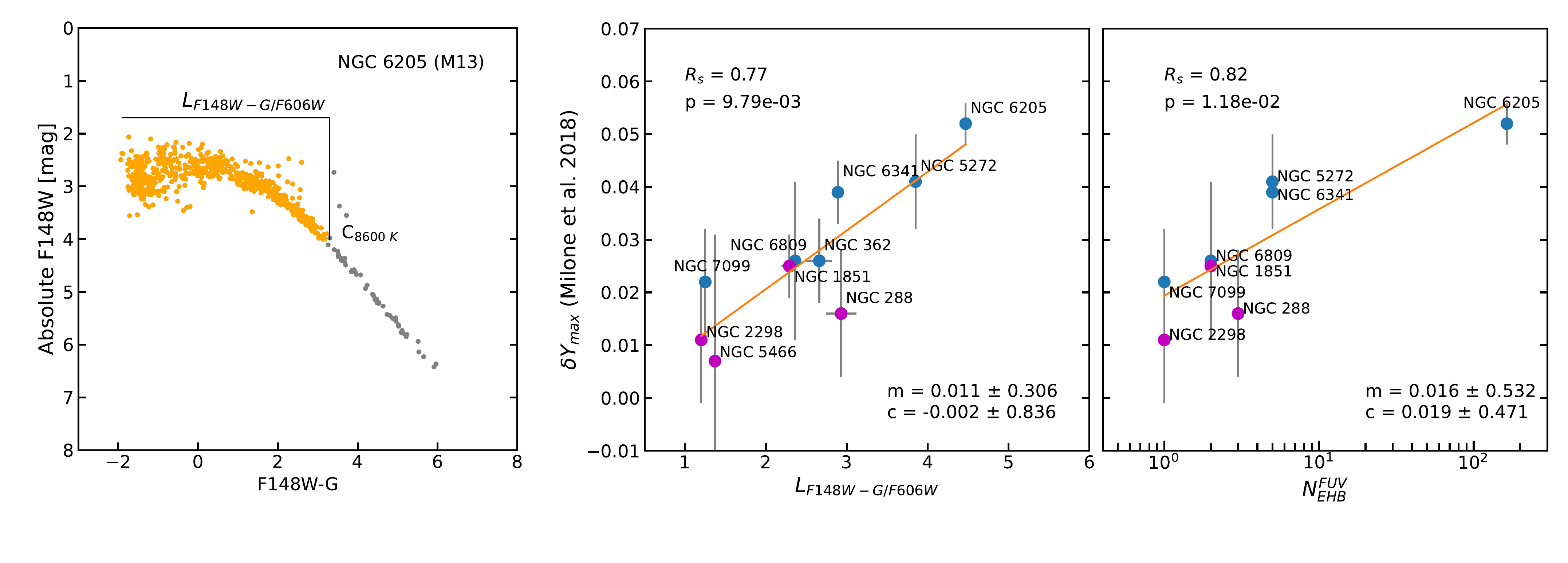}
\caption{Derivation of L extension parameter shown for cluster NGC~6205 as an example in the leftmost panel. HB color extension $L_{F148W-G/F606W}$ and number of EHB stars detected in FUV ($N^{FUV}_{EHB}$) as a function of maximum He variation within each GC ($\delta Y_{max}$) \citep{Milone2018} in the middle and right panels, respectively. The blue dots are the clusters studied for the first time in this work and green dots are clusters from our previous works. The Spearman's rank correlation coefficients and p-values are provided in the Figure.}
\label{fig:hb_corr}
\end{figure*}

We explored the effect of He enrichment in the HB morphology of 11 GCs by studying their correlation with He variations in GCs. \cite{Chung2018} using Yonsei Evolutionary Population Synthesis (YEPS) He enhanced models in different passbands demonstrated that FUV-V color is most sensitive to the variations of Y$_{ini}$ in HB stars.  Thus, for our analysis, we studied the relation using the FUV$-$G color extension of HB. The $\delta Y_{max}$ values were adopted from \cite{Milone2018} for 9 clusters in our sample, that are varying from 0.007-0.052 (Table~\ref{tab:correlation}). These values were not available for NGC~1904 and NGC~5897. We chose the F148W$-$G/F606W plane which provides a wider color baseline than NUV and has the maximum sensitivity to the effective temperature variations in HB stars. The HB color extension parameter is denoted as $L_{F148W-G/F606W}$ and their corresponding uncertainties are determined by following the method of \cite{Milone2014}. Similar to their definition of $L_{F275W-F814W}$, $L_{F148W-G/F606W}$ is defined as the difference between 96th and 4th percentile of the color F148W$-$G/F606W distribution of HB stars, by excluding the stars redder than the RRL strip. We calculated the errors by performing 1000 iterations on the sample of HB stars in FUV using bootstrap statistics, where the 68.27th percentile of the bootstrap measurements was considered as the corresponding uncertainty for each cluster. For calculating the L extension parameter, we chose HB stars with T$_{\rm eff} \geq 8,600$ K as a reference point for comparing different clusters in our sample as shown in the leftmost panel of Figure \ref{fig:hb_corr}, which corresponds to reddening corrected color F148W$-$G $\sim$ 3.4. The motivation to choose this reference point is that the effects of different metallicities are not present for stars hotter than 8,600~K as found from the PGPUC and BaSTI-IAC models. This reference point was also chosen by \citep{Brown2016} for studying the HB morphologies of different clusters. The values of the L parameter and their errors are listed in Table~\ref{tab:correlation}.

We checked the correlation of $L_{F148W-G/F606W}$ with $\delta Y_{max}$ in a sample of 9 clusters with FUV observations. These clusters are mostly metal-poor ($\rm [Fe/H] < -1.5$ dex) and populated with only BHB stars (except NGC 362 and NGC 1851). 
The cluster NGC~2298 from our recent work \citep{2021ApJ...923..162R} is also included in the analysis whose $\rm L_{F148W-G} \sim 1.2 \pm 0.07$ and $\Delta Y_{max}\sim 0.011 \pm 0.012$ \citep{Milone2018}.
As shown in the left panel of Figure~\ref{fig:hb_corr}, we find a strong correlation of the L parameter with $\delta Y_{max}$ where the Spearman's rank correlation coefficient (R$_{s}$) is close to 0.79 with p-value=$6.5\times10^{-3}$. From the figure, we note that the value of the L extension parameter increases with increasing $\delta Y_{max}$ within each GC with the clusters with higher values having more extended HB. The results are in fair agreement with \cite{Tailo2020} suggesting that the He enhancement plays an important role in shaping the bluer end of the HB morphology in GCs. The detailed HB morphologies of these clusters will be studied in the future to check the existence of multiple stellar populations (MSPs, \cite{Bastian2018} and the references therein).
 
\subsection{FUV detected EHB stars}
In the massive clusters ($> 10^5 M_{\odot}$) which host a large number of EHB and BHk stars, such as $\omega$ cen and NGC~2808, spectroscopic studies \citep{Moehler2011, Brown2012} have shown that a significant population of EHBs stars are He-rich which supports the late hot-flash scenario. On a different perspective, \cite{Antona2002} and \cite{Lee2005} suggested He self-enrichment scenario where EHB stars are the result of the normal evolution of He-enhanced sub-populations formed from the ejecta of massive AGB stars. This was supported by the multiple splits in the MS found in the GC CMDs from the \textit{HST} photometry of clusters hosting EHB stars \citep{Piotto2015}. These studies were mostly biased towards clusters with uniformly populated HB extending to EHB stars. Using the FUV-optical CMDs, we were able to detect sparsely populated EHB stars in clusters mostly hosting only BHB stars (Table~\ref{tab:number_stars}). We studied their relation with the He variation to explore the He self-enrichment scenario for the EHB formation.  
 
We considered 7 clusters in our sample hosting EHB stars to check their correlation with $\delta Y_{max}$ \citep{Milone2018}. The number of FUV detected EHBs ($N^{FUV}_{EHB}$) as a function of $\delta Y_{max}$ are shown in the right panel of Figure~\ref{fig:hb_corr}. We find a significant positive correlation of $N^{FUV}_{EHB}$ with $\delta Y_{max}$ with $R_{s} \sim$ 0.73 and 96\% confidence interval denoting that the number of EHB stars increases with increasing $\delta Y_{max}$. This suggests that the evolution of He enhanced 2G MS stars would have led to the formation of EHB/BHk stars in GCs. However, we need a larger sample of GCs with EHB stars to confirm this result. Our future studies will aim at deriving the atmospheric parameters of the EHB stars using SEDs and comparing them with both the hot-flashers and He-enriched models.

\section{Science Cases from GlobULeS catalogue}\label{sec:sc_case}
Below we highlight the potential science cases from the GlobULeS project that we intend to pursue in the future:

\subsection{He enrichment in HB stars}
\begin{figure}
\centering
\includegraphics[width=\columnwidth]{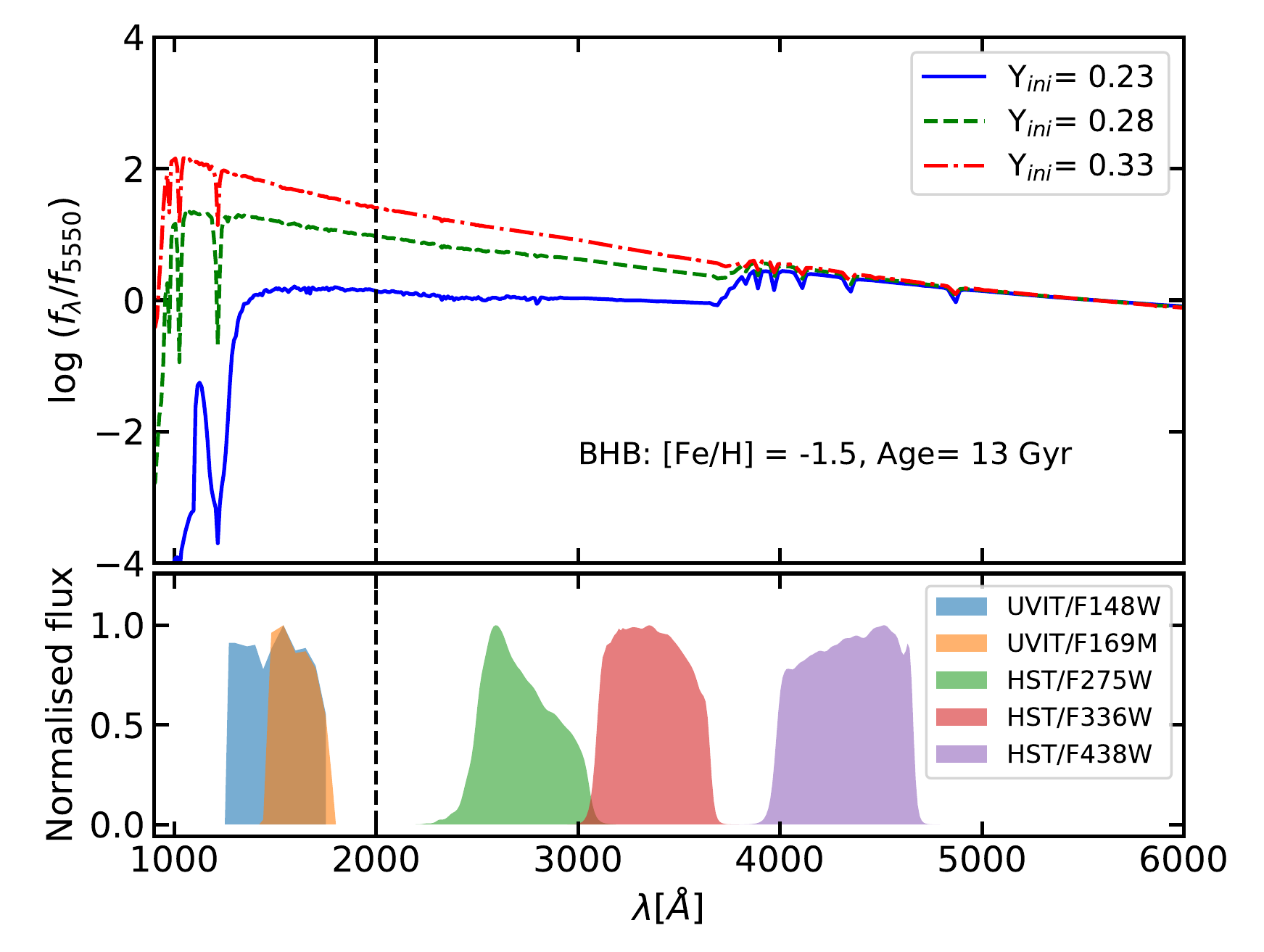}
\caption{Top panel: Synthetic SEDs of BHB stars from YEPS models \citep{Chung2018} for $\rm [Fe/H] = -1.5$ dex and age 13 Gyr corresponding to three different Y$_{ini}$ abundances 0.23, 0.28 and 0.33. The Y-axis is the logarithm of model flux normalised to the flux at 5550 \AA. Bottom panel:  Effective area curves (normalised) for the two UVIT filters (F148W, F169M) and three \textit{HST} filters (F275W, F336W and F438W) as a function of wavelength.}
\label{fig:hb_he_model}
\end{figure}

GCs host complex HB morphologies that cannot be explained by only one parameter, metallicity, that mostly contributes to the fraction of BHB and RHB stars. Additional parameters are required to explain the clusters with different HB morphologies but with no variation in metallicity or age. This is known as the ``second parameter problem'' (refer to review papers by \cite{Catelan2009, Gratton2010, Dotter2013} and references therein for more details). One such important second parameter is He abundance \citep{Sandage1967}. Figure~\ref{fig:hb_he_model} shows SEDs of HB stellar populations as a function of wavelength for a metallicity of $-$1.5 dex, age = 13 Gyr and different He abundances \citep{Chung2018}. The UVIT filter effective areas for two FUV filters (F148W and F169M) along with the \textit{HST} filters are also compared in the figure. We note that {\textit the flux change among the He-normal and He-rich populations is maximum in FUV wavelengths as compared with the NUV and optical. Being most sensitive to He abundance variations,}  \cite{Dalessandro2010, Dalessandro2013} suggested that FUV-optical CMDs are optimal diagrams for studying their effect in BHB stars with T$_{\rm eff} >$ 10,000~K. Combining FUV and optical observations from the \textit{HST}, they investigated the role of different He abundances in the HB distribution and computed the maximum He values in various GCs (NGC~2808, NGC~6205, NGC~1904). However, these studies covered only central regions of GCs. The GlobULeS catalogue will assist in identifying the sub-populations with synthetic HB models and in comparing their spatial distributions over the cluster region extending to a 10$'$ radius from the cluster centre. The strong correlation of HB color extension with He variations, derived from the UVIT/FUV-optical color for EHB stars, indicates the significance of our studies in the context of MSPs in GCs.

\subsection{Evolutionary status of pHB stars}
The most luminous and hot population among the UV bright stars are pHB stars. \cite{Schiavon2012} classified post-core-He burning stars of 44 GCs into AGBm, P(e)AGB  using pHB evolutionary models of different masses. However, membership analysis was not performed. Comparing it with the GlobULeS sample, we note that NGC~6205 was not included in their study. 

\cite{Moehler2019} calculated the expected number of pAGB stars using stellar evolutionary models in 17 GCs and found it to be in good agreement with the observed number. However, they suggested that these comparisons are affected by low number statistics. Thus, a large sample of GCs with pHB stars is required to check their expected number from models. Our survey will help in improving the statistics. The advantage of using the UVIT to study pHBs is presented for GC NGC~2808 by \cite{Deepthi2021}. From GlobUleS, we have identified pHB candidates in 7 GCs along with their membership using the \textit{HST} and \textit{Gaia}~EDR3. Among the GlobULeS sample, 4 clusters (NGC~1851, NGC~5272, NGC~6205, and NGC~7099) are in common with \cite{Moehler2019}, where we detect the previously reported 3 PAGB stars in the clusters that are heavily saturated in the UVIT/F148W band. The FUV filter data points will be crucial for deriving the parameters of pHB stars (such as luminosity, T$_{\rm eff}$) and inferring their evolutionary status.

\subsection{FUV bright BSSs}
Using FUV observations, we have detected very bright BSSs (absolute F148W $<$ 5.5~mag) that probably have masses twice the mass of MSTO. The parameters and formation mechanism of such bright BSSs are not well-understood. \cite{Sills1999} compared the observed BSS distribution of NGC~5272 with theoretical CMDs generated from simulations for different assumptions and inferred that bright BSSs can be reproduced by increasing their He content resulting from mixing and rotation. Similarly, \cite{Ferraro2003} spotted very bright BSSs in NUV-optical CMDs of NGC~5272 and NGC~6341. Comparing their results with theoretical collisional models, they suggested that the presence of these BSSs could indicate a continuous BSS formation or they might belong to a different distribution involving triple collisions. \cite{Nathan2007} investigated the relationship of BSSs with their cluster environments and suggested that the clusters with the highest collisional rates could host the most massive BSSs due to the increased rate of three-body encounters. From the \textit{HST}/FUV observations, \cite{Knigge2008} identified a super-massive BSS (M$\sim 1.9 M_{\odot}$) in 47 Tuc using SED fits, pointing out that formation of such BSSs might require the involvement of three progenitors.

From our study of BSS distribution in FUV-optical CMDs of 11 GCs, we note that the FUV bright BSSs are mostly found in high-density ($\rho_{c} \geq 10^{3.6} L_{\odot}/pc^{3}$) ( Table~\ref{tab:correlation}) and dynamically more relaxed clusters, with N$_{relax} \geq$ 2. The examples are NGC~362, NGC~6341, and NGC~7099 that are also core-collapsed. The stellar encounter rates ($\Gamma$) of such clusters are high with log $\Gamma>$ 2 \citep{Bahramian2013}. On the other hand, the presence of very bright BSSs is negligible in low-density clusters (NGC~288, NGC~5466). These clusters have a high specific frequency of BSSs ($N_{BSS}/N_{HB}>0.2$) compared to other clusters in our sample.

It suggests that the clusters with high stellar encounter rates are hosting the FUV brightest BSSs. This bright BSSs are preferentially found in the central regions of the clusters suggesting that the multiple collisions in the dense core regions might be responsible for their production.  
However, to study a specific correlation of the BSSs with cluster parameters, a complete sample of BSSs down to MSTO is required. This has been achieved using the NUV-optical CMDs from the HST observations in the recent works, but these studies were restricted to only the half-light radius of GCs. Future studies are planned in combining the HST data with the Swift/UVOT observations to cover the outer regions as well.

The GlobULeS catalogue of BSSs will also be crucial to derive the properties of FUV bright BSSs and identify those showing ultraviolet excesses using SEDs. \cite{Knigge2008} and \cite{Sahu_ngc5466} analysed this FUV excess and suggested that it is due to the presence of WD companions. The BSS-WD systems suggested by their studies are the only two such systems known to date in GCs. More discoveries of such hot companions and detailed spectroscopic diagnosis using high-spectroscopic data from large facilities (e.g. VLT/ESO); as well as theoretical modeling will have important implications for our understanding of the binary evolution of BSS stars and the astrophysical processes involved (mass transfer, the role of metallicity, etc.). 

\subsection{Study of WDs}
The WD sequence for the inner cluster region (the UVIT-\textit{HST}) is prominent in the stacked FUV-optical CMD (Left panel of Figure~\ref{fig:stack_cmd_1}. 
We checked whether the number of WDs detected with the UVIT/F148W in FUV-optical CMDs matched the theoretical expectations. The expected number of WDs ($N_{WD}$) that are brighter than the UVIT/F148W detection limit of each GC (see Table~\ref{tab:gc_uvit_survey_phot}) was calculated following the relation $N_{WD}/N_{HB} = \tau_{WD} /\tau_{HB}$ (Eq.3 of \cite{Knigge2002}), where $N_{HB}$ and $\tau_{HB}$ are the number of HB stars and their corresponding lifetime, whereas $N_{WD}$ is the number of WDs brighter than a given magnitude and $\tau_{WD}$ is the corresponding cooling time down to that magnitude. We used HB stars as reference considering $\tau_{HB}=10^{8}$ years as the HB lifetime \citep{Iben1991}. 
We selected the number of HB stars ($N_{HB}$) lying in \textit{HST} FOV from F275W vs F275W$-$F336W \textit{HST} CMDs to avoid incompleteness.

The cooling ages of WDs ($\tau_{WD}$) corresponding to the UVIT magnitude detection limit were considered from Bergeron WD models of DA spectral type and mass 0.5 M$_{\odot}$. According to the models, the WD temperatures of 11 GCs lie in the range 23,000-45,000~K with their cooling ages varying from $\sim$ 4 to 25 Myrs. Comparing with the predictions, we found that 50-70\% of the WDs are detected with the UVIT/F148W in three GCs, $>90\%$ in three of them, while the rest of them have 30-40\% detections excluding NGC~1904. For the regions not covered with the \textit{HST}, deep UBVRI observations of GCs \citep{Stetson2019} will be utilised to cross-match and identify the hot WDs. Since these regions are not affected by crowding, we expect the number of WDs brighter than the detection limit to be in line with the predicted number. The WD catalogue from GlobULeS will be useful to derive the atmospheric parameters and check for binarity using SEDs and models \citep{Knigge2008}.

\section{Summary and conclusion} \label{sec:conc}
We present the first results for eight GCs from the UV survey of GCs (GlobULeS) carried out using UVIT onboard AstroSat. We also present some initial analysis of 11 GCs after including data of 3 more GCs previously observed using UVIT.
The FUV-optical CMDs of proper motion members were constructed by combining \textit{AstroSat}/UVIT observations with \textit{HST} HUGS survey and \textit{Gaia} EDR3 data. For the regions not covered by \textit{HST}, the membership probabilities were derived using Gaia EDR3. The FUV-optical CMDs mainly consist of HBs, pHBs,  BSSs, and WDs. We used several stellar evolutionary models such as PGPUC, BaSTI-IAC, and Bergeron models for HBs, BSSs, and WDs respectively, and found it to match well with the observed CMDs. Overall, we have 1747 (1481 HBs, 134 BSSs, 107 WDs, 25 pHBs) common sources with UVIT-HST cross-match and 1443 (1335 HBs, 87 BSSs, 21 pHBs) from the UVIT-Gaia EDR3 cross-match (for the regions not observed by HST). A total of 190 EHB candidates in 11 GCs were identified photometrically based on the effective temperatures obtained from color (F148W$-$G) vs T$_{\rm eff}$ relation of HB models, with the largest population being in NGC 6205. Thus, our study highlights the importance of combining multi-wavelength observations covering FUV (from UVIT) to optical for the exploration of UV bright stars. 

To highlight the important features of UV CMDs, we created a stacked FUV-optical CMD of members from 11 GCs. The stacked CMD shows a dominant population of HBs spanning T$_{\rm eff}$ range 6,000-32,000~K followed by the BSSs being as bright as BHBs in a few clusters. The WDs span a temperature range of 25,000-90,000~K corresponding to the cooling age range of 1-25 Myr, as per the Bergeron DA WD models. We also detect 296 known variables in FUV (RRL, SX Phes) distributed mostly in the fainter end of the BHB and BSS distribution, respectively. To identify the discontinuities present in the HB distribution of the UVIT-HST common sources, we used three filter combinations of UVIT/F148W and HST (F336W and F606W) for creating pseudo-color diagrams. The diagrams show detectable gaps at T$_{\rm eff} \sim$ 11,700 and 21,000~K corresponding to the well-known G-jump and M-jump, respectively. We also found that the color (F148W-G) extension of HB distribution of 10 GCs for stars hotter than 8,600~K is strongly correlated to the maximum internal He variation within each GC \citep{Milone2018}, thus showcasing the larger sensitivity of FUV-optical color to the He enrichment when compared to NUV, as well as to studies that use only optical bands. 

We discuss the important science cases that will be pursued in the future using the Globules survey products, with a particular focus on the EHB stars, pHB stars, FUV bright BSSs, and WDs. The FUV catalogues from the GlobULeS survey will provide broader opportunities to reveal the nature of poorly studied hot stellar populations in GCs which will aid in constraining the stellar evolutionary models in UV. The initial analysis of the identified UV peculiar and exotic stars from Globules catalogues will serve as important targets to the upcoming spectroscopic facilities in different wavebands including the UV missions that are being planned.

\section{Data availability}
The UVIT photometry data underlying this work will be shared at reasonable request to the corresponding author. The raw UVIT data files of the clusters can be downloaded from the Indian Space Research Organisation (ISRO) Science Data Archive for AstroSat Mission \footnote{\url{https://astrobrowse.issdc.gov.in/astro_archive/archive/Home.jsp.}}.

\section{Acknowledgements}
We thank the referee for valuable comments and suggestions which helped in improving the manuscript.
We thank Pierre Bergeron for providing the WD cooling models in UVIT filters. We would like to thank Francesco Ferraro and Barbara Lanzoni for useful discussions and suggestions. SS \& AS acknowledges the support from SERB POWER fellowship grant SPF/2020/000009. NWCL gratefully acknowledges the generous support of a Fondecyt Iniciaci\'on grant 11180005, as well as support from Millenium Nucleus NCN19-058 (TITANs) and funding via the BASAL Centro de Excelencia en Astrofisica y Tecnologias Afines (CATA) grant PFB-06/2007.  NWCL also thanks support from ANID BASAL project ACE210002 and ANID BASAL projects ACE210002 and FB210003. THP gratefully acknowledges support in form of FONDECYT Regular project (No.~1201016) and the ANID BASAL project (FB210003). C.C. acknowledges the support provided by the National Research Foundation of Korea (2022R1A2C3002992).

This publication uses UVIT data from the Astrosat mission of the ISRO, archived at the Indian Space Science Data Centre (ISSDC). The UVIT project is a result of collaboration between IIA, Bengaluru, IUCAA, Pune, TIFR, Mumbai, several centres of ISRO, and CSA. This publication uses UVIT data processed by the payload operations center by the IIA. This work has made use of data from the European Space Agency (ESA) space mission Gaia (\url{https://www.cosmos.esa.int/gaia}), being processed by the Gaia Data Processing and Analysis Consortium (DPAC). Funding for the DPAC is provided by national institutions, in particular the institutions participating in the Gaia MultiLateral Agreement (MLA). The Gaia archive website is (\url{https://archives.esac.esa.int/gaia}). This research made use of the Aladin sky atlas developed at CDS, Strasbourg Observatory, France \citep{Bonn2000}. This research also made use of Topcat \citep{Taylor2005}, Matplotlib \citep{Hunter2007}, NumPy \citep{van2011}, Scipy \citep{Oliphant2007, Millman2011}, Astropy \citep{astropy2013, astropy2018}, and Pandas \citep{mckinney2010}.

\bibliographystyle{mnras}
\bibliography{refs}

\newpage{}
\appendix

\section{}
\begin{figure}
\centering
\includegraphics[width=\columnwidth]{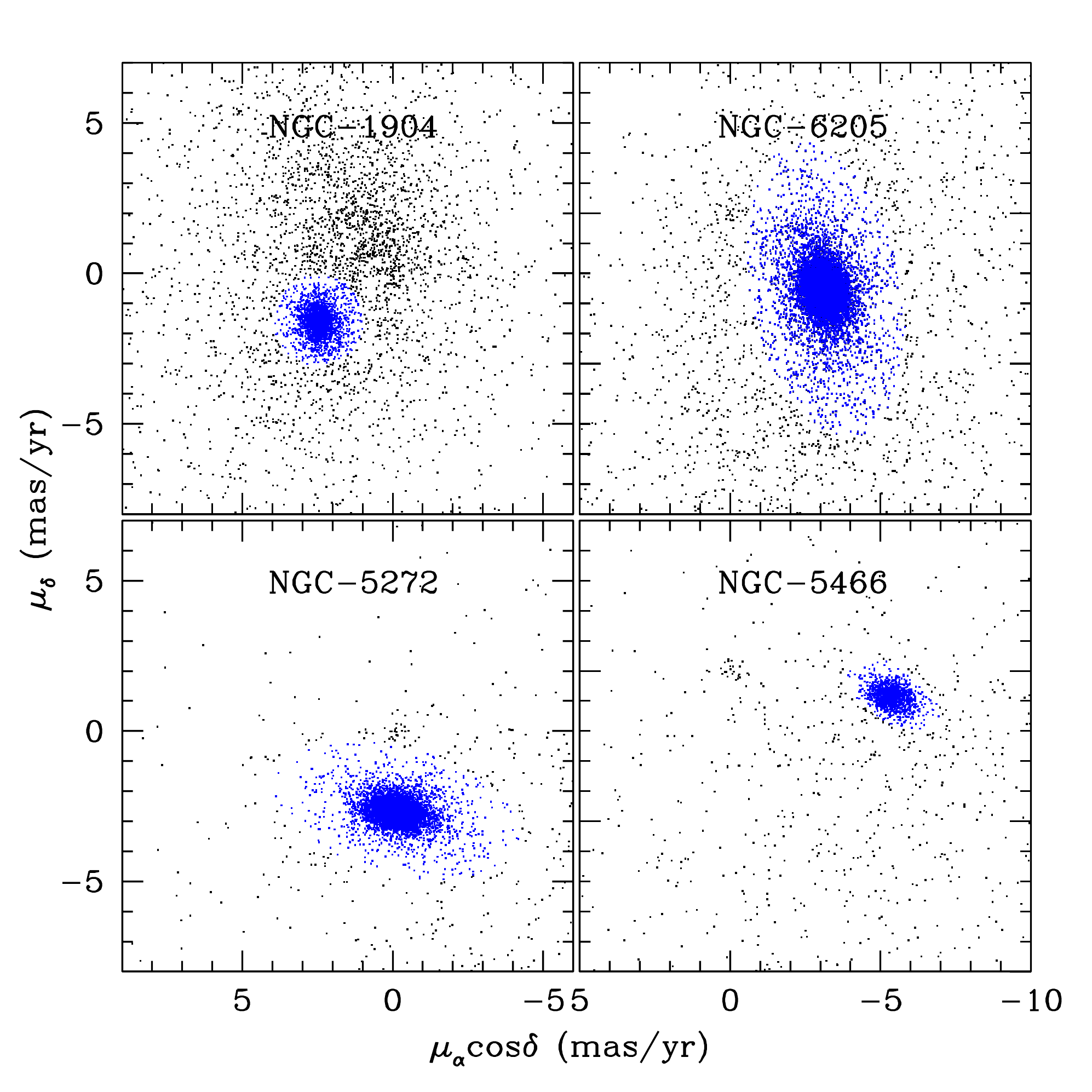}
\includegraphics[width=\columnwidth]{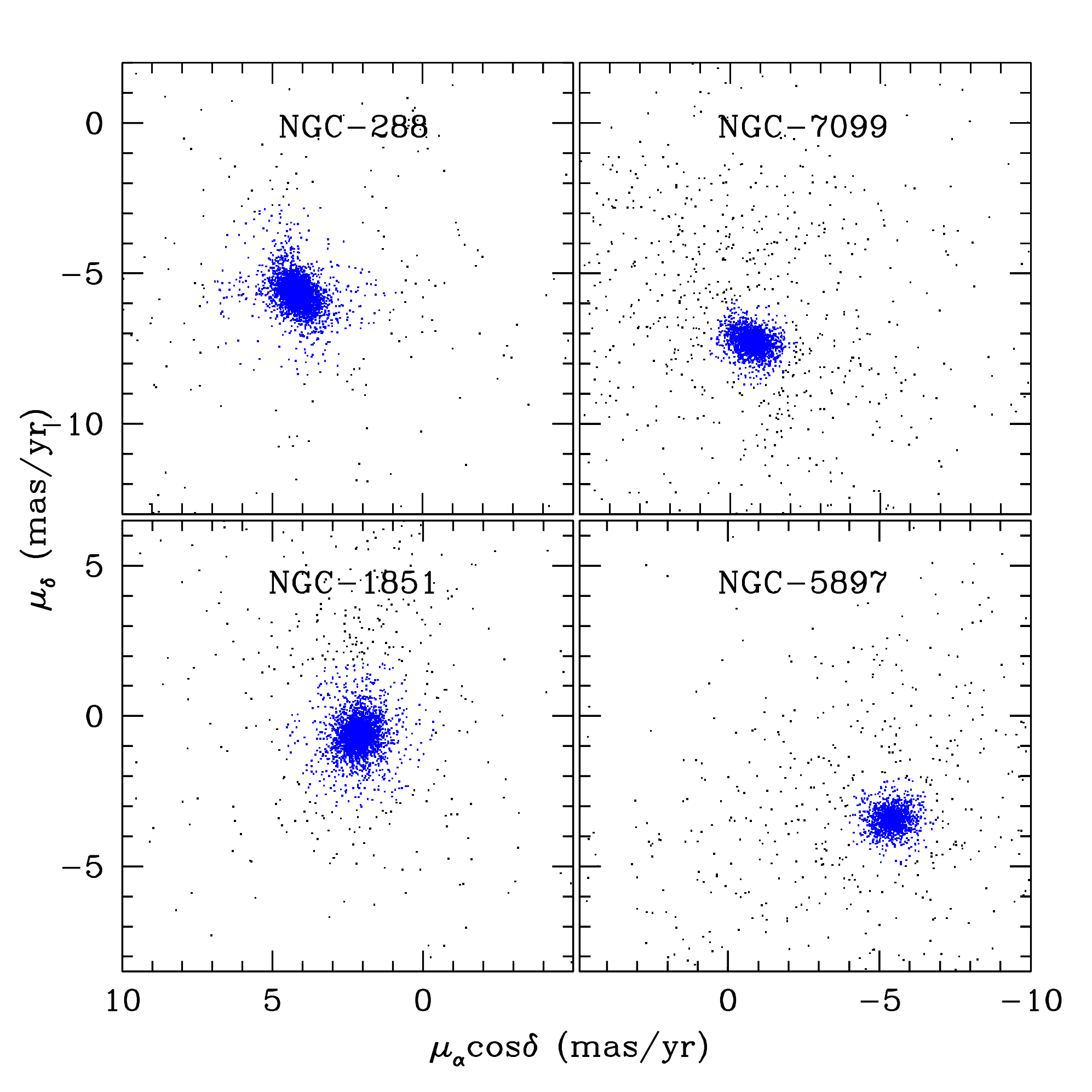}
\includegraphics[width=\columnwidth]{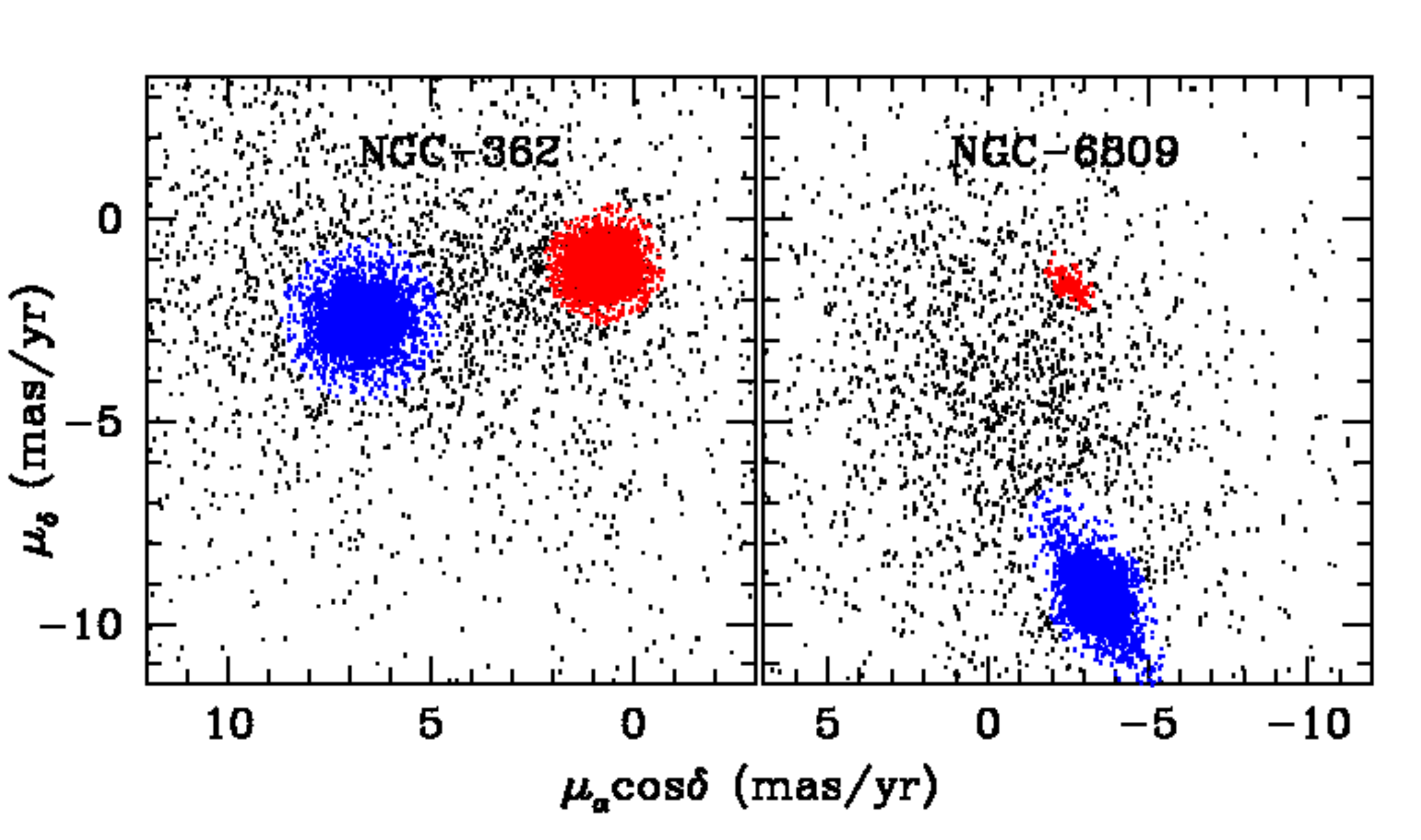}
\caption{VPDs of 10 clusters using \textit{Gaia} EDR3. The blue and black dots are same as described in Figure~\ref{fig:vpd_ngc6341}. The red dots in the VPDs of NGC~362 and NGC~6809 are the SMC field and Sagittarius background sources respectively. }
\label{fig:vpds_gaiadr3}
\end{figure}

\begin{figure}
\centering
\includegraphics[width=0.8\columnwidth]{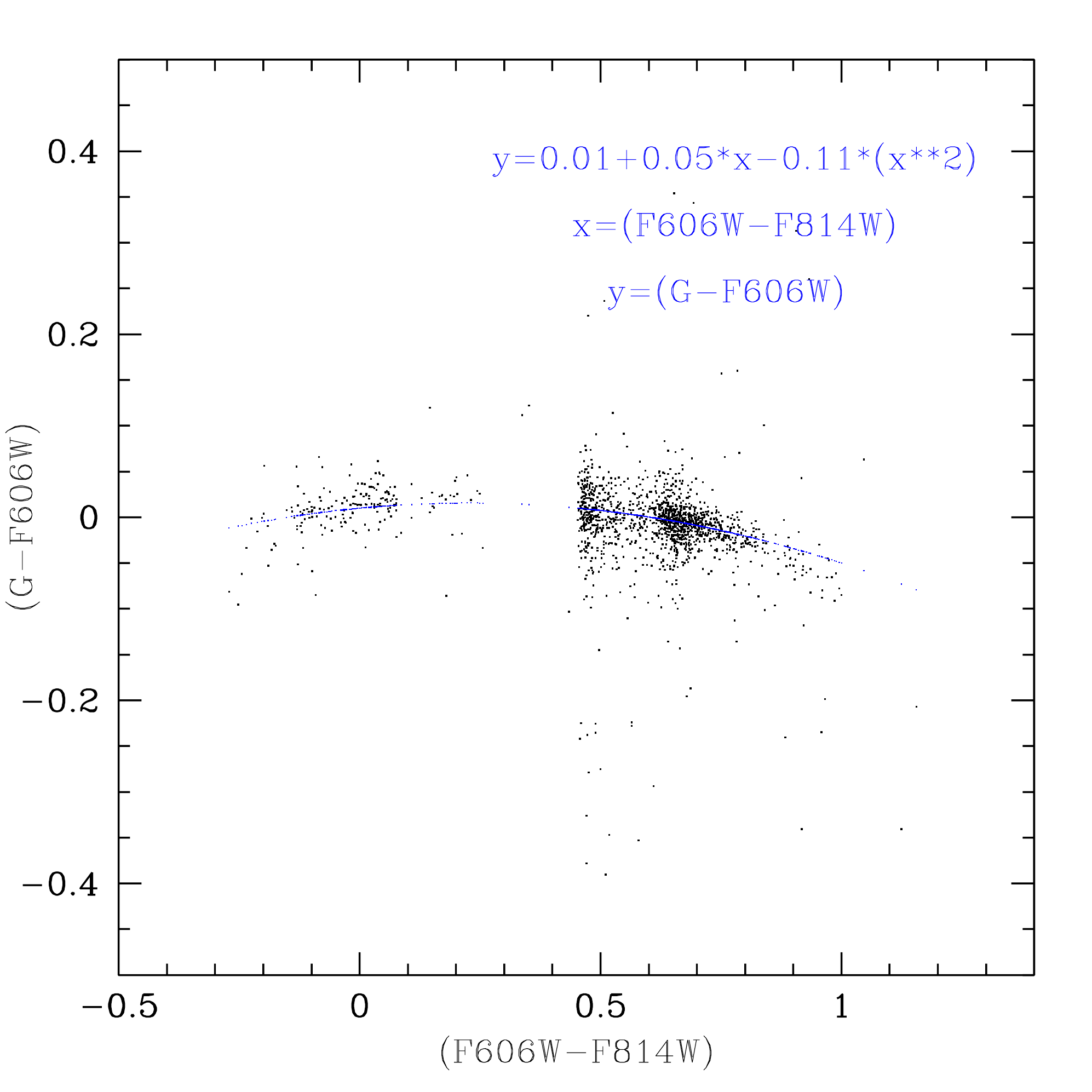}
\caption{Transformation relation from plot of the HST/F606W filter to Gaia EDR3 G band.}
\label{fig:hst_g_relt}
\end{figure}

\begin{figure}
\centering
\includegraphics[width=0.8\columnwidth]{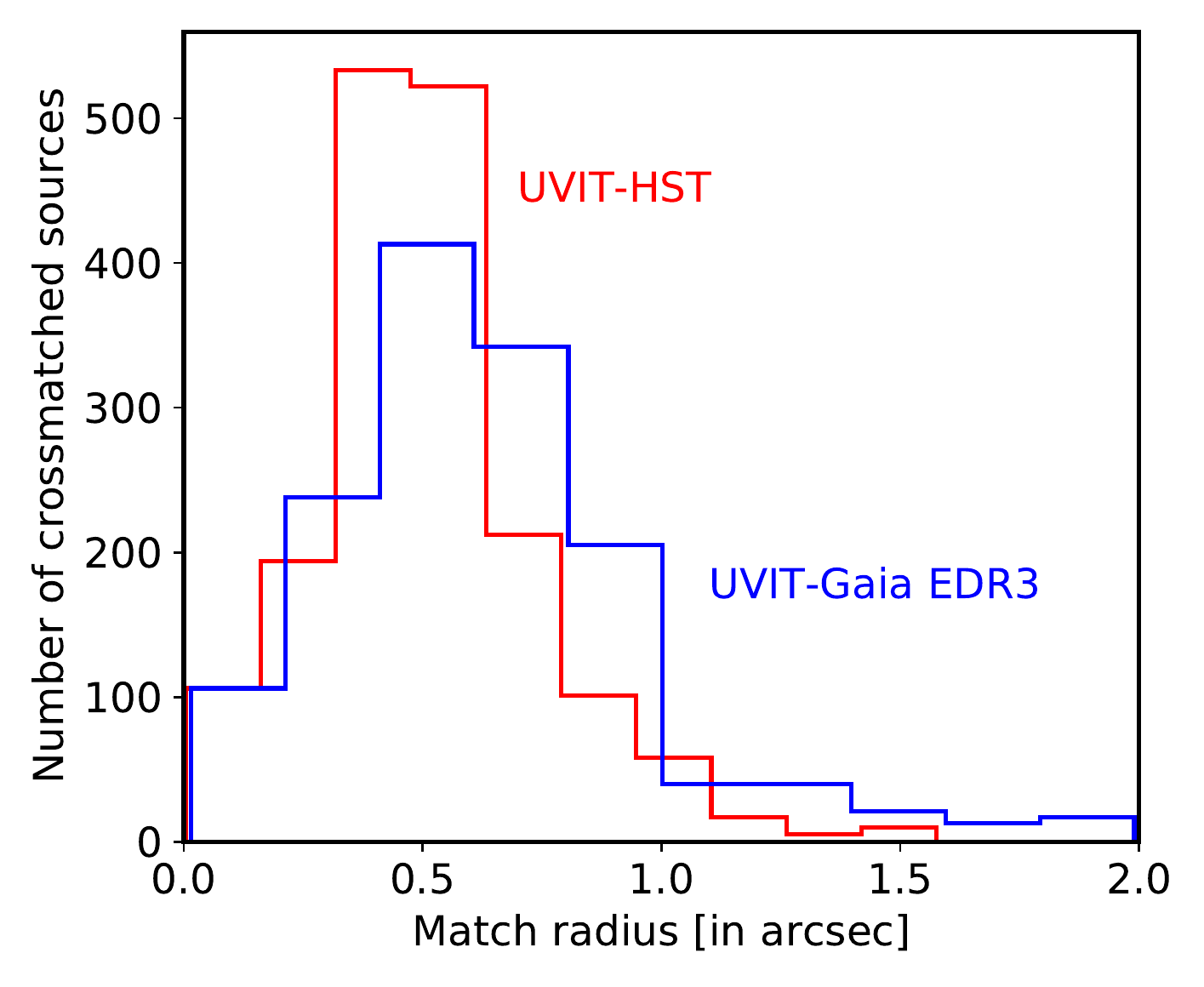}
\caption{Number of cross-matched UVIT sources with the \textit{HST} and \textit{Gaia} EDR3 for 11 clusters with match radius $<$2''.}
\label{fig:sep}
\end{figure}

\begin{figure}
\centering
\includegraphics[width=\columnwidth]{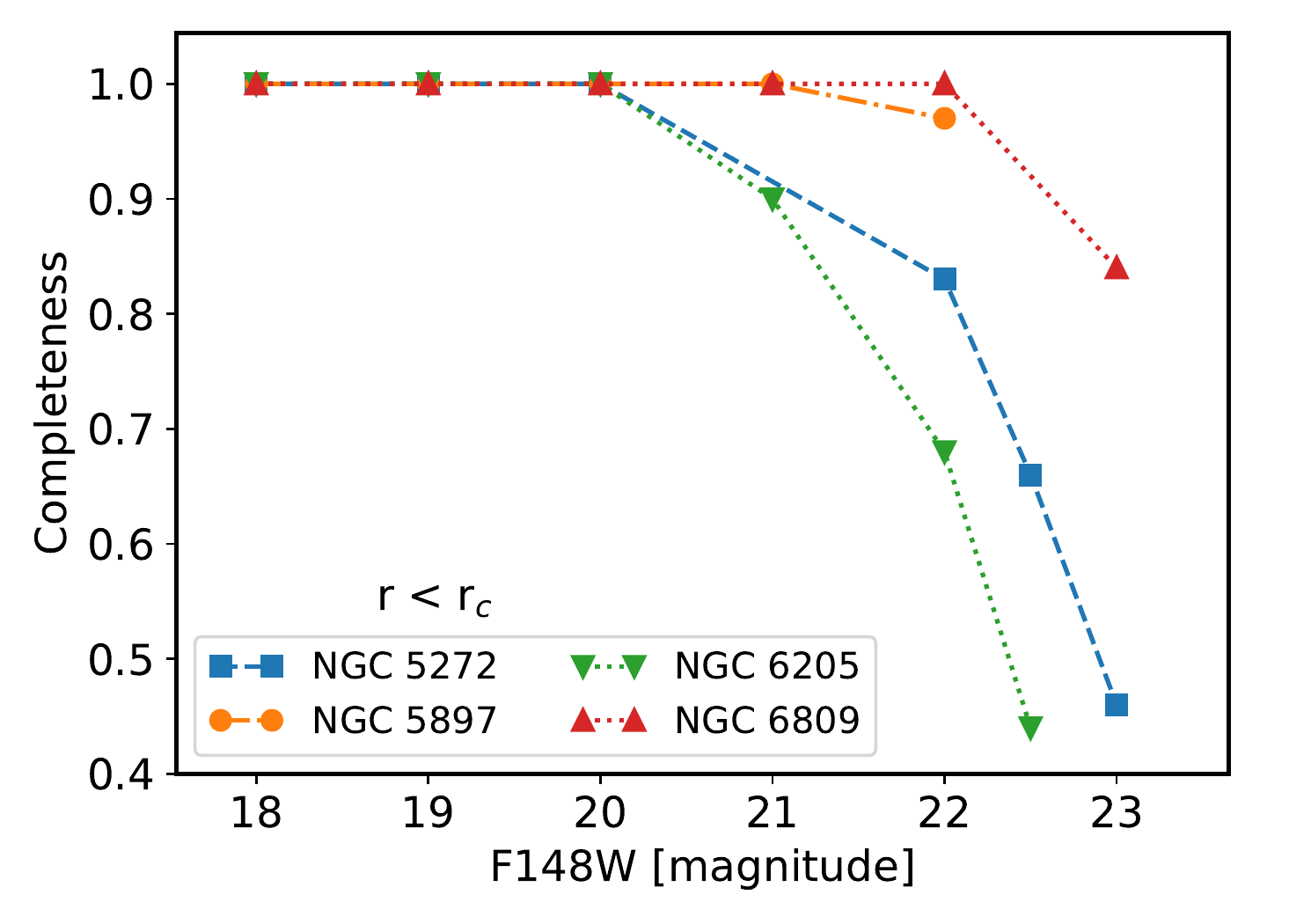}
\includegraphics[width=\columnwidth]{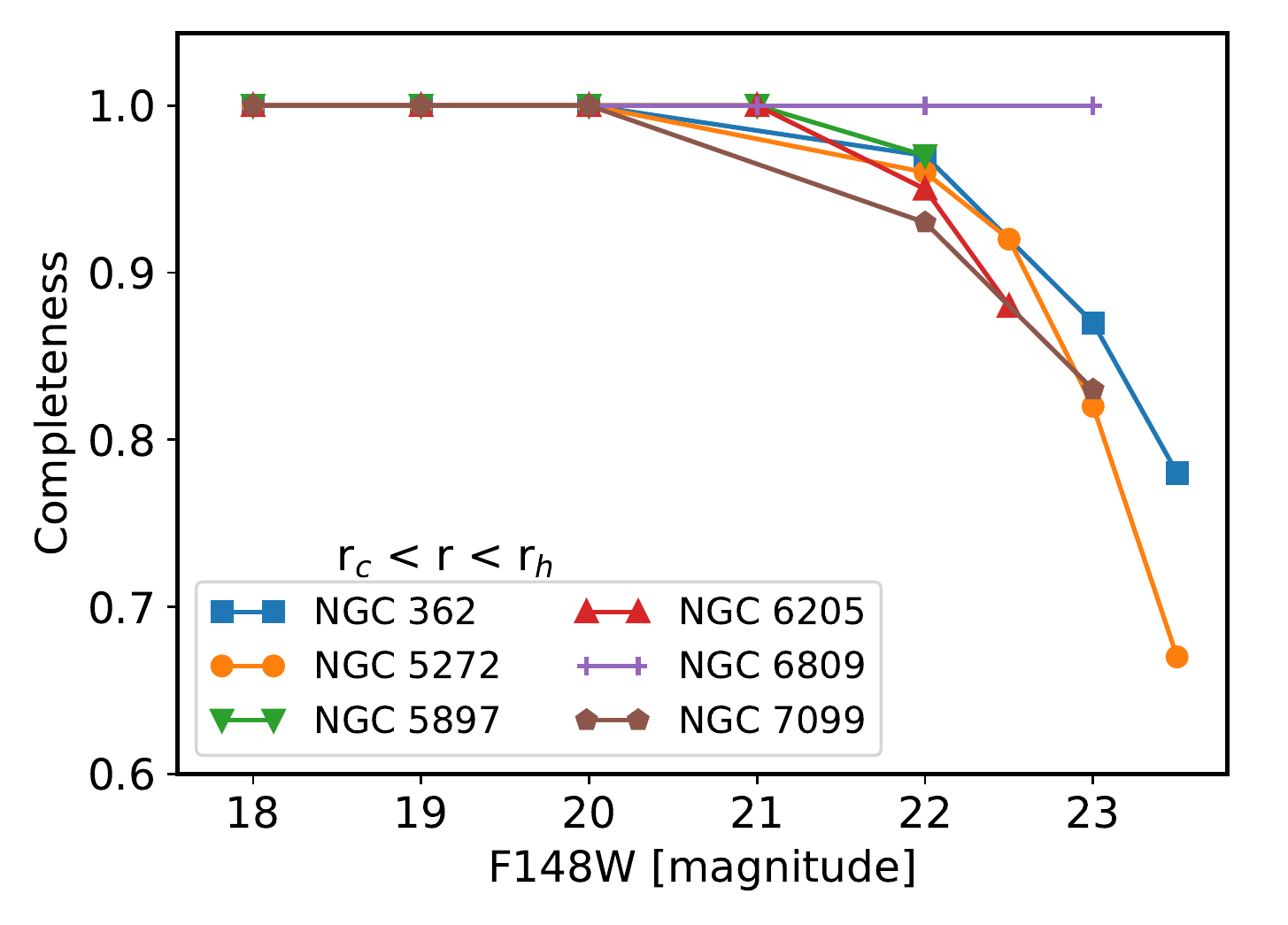}
\caption{Completeness plot as a function of magnitude for the sources located inside the $r_{c}$ and within $r_{c}$ to $r_{h}$ of the clusters for comparison are shown in upper and lower panels respectively. The cluster cores of NGC 7099 and NGC 362 are not resolved by UVIT.}
\label{fig:ast_all}
\end{figure}

This section shows the VPDs of 10 clusters (Fig~\ref{fig:vpds_gaiadr3}), a plot of transformation between the \textit{HST} and \textit{Gaia} passbands (Fig~\ref{fig:hst_g_relt}), and a plot of common stars between the UVIT, the \textit{HST} and \textit{Gaia}~EDR3 for 11 clusters (Fig~\ref{fig:sep}). The completeness plot of clusters as a function of magnitude within core radius and half-light radius of clusters are shown in Fig~\ref{fig:ast_all}.

\section{}\label{sec_apendix:fuv_cmd}
\begin{figure*}
\centering
\includegraphics[width=\linewidth]{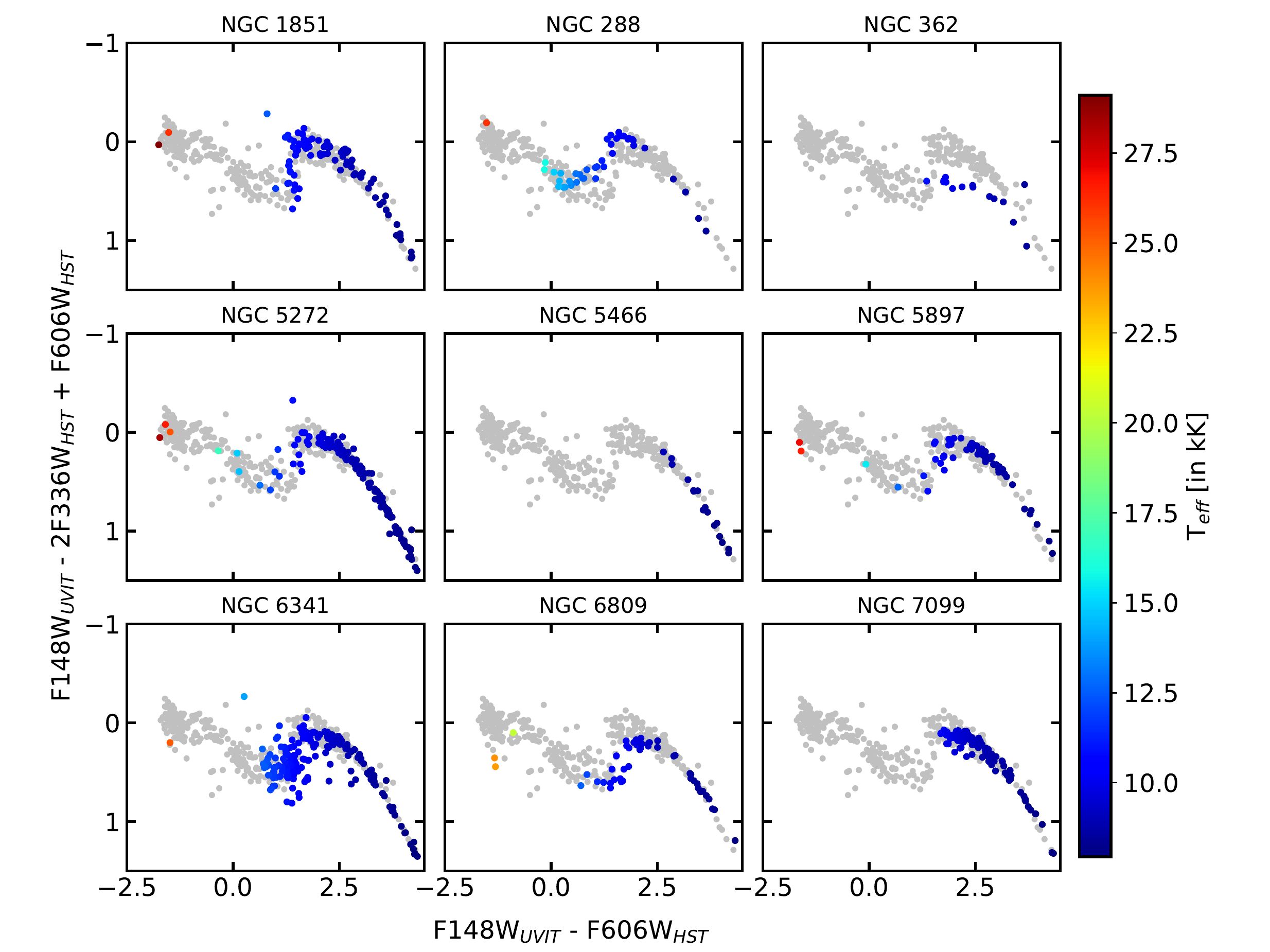}
\caption{Pseudo color-color plot (F148W$-$2F336W$+$F606W) vs (F148W$-$F606W) of 9 clusters for the UVIT-\textit{HST} common detected HB sources corrected for reddening. The grey dots are detections from NGC~6205 cluster shown for reference. The color-bar shows the $T_{\rm eff}$ (in kK) of HB stars estimated from F148W$-$G color vs $T_{\rm eff}$ relation obtained from the PGPUC models.}
\label{fig:pseudo_color_all}
\end{figure*}

This sections describes in detail the FUV-optical (Figure \ref{fig:fuv_cmds_1}), FUV-FUV CMDs (Figures \ref{fig:fuv_cmds_3} and \ref{fig:fuv_cmds_4}) and shows the pseudo-color plots (Figure \ref{fig:pseudo_color_all}) of eight GCs.

\subsection{NGC~5272 (M3)}
There are 227 BHB stars with $T_{\rm eff}> 8,000$ K uniformly populating the ZAHB track and following the ZAHB models. This cluster is known to harbor a large population of RRLs (241, \cite{Clement2001}), where we detect 149 (103 RR0, 44 RR1) of them in the F148W filter. We also identified 5 EHBs with temperatures ranging from $25,000 < T_{\rm eff} < 29,000$ K, among which 3 are located below the ZAHB model and thus are BHk candidates. We detected three pHB sources, where the brightest FUV source among them is a well-known PAGB star VZ 1128. This source has been studied extensively in UV \citep{Chayer2015}. We have not plotted it in the UV CMDs as it is heavily saturated in our FUV exposures. 

The BSSs span more than 4~mag in F148W with some of them being as bright as the BHB stars. Two of the brightest BSSs are located on the redder side of the ZAMS model extension. In F148W vs F148W$-$F169M CMDs, we find the BHBs nicely follow the models till F148W = 20 mag, whereas, at F148W $\sim$ 21 mag, around 85 of the HB stars deviate from the ZAHB model \ref{fig:fuv_cmds_4}. These stars show a scatter of around 6~mag in F148W$-$F169M that are larger than the photometric errors and are classified as RRLs \citep{Clement2001}. The WD sequence in FUV-optical CMDs follows the DA cooling models with the hottest ones having T$_{\rm eff} \sim$ 65,000~K.\\

\subsection{NGC~6205 (M13)}
Among the detected EHBs, 14 lying outside the UVIT-\textit{HST} overlap region are non-members according to PM from \textit{Gaia} EDR3. The temperatures of the EHBs range from 22,000 $< T_{\rm eff}<$ 33,000~K with the peak lying at color F148W$-$G$\sim -1.5$ corresponding to T$_{\rm eff} \sim$ 25,000~K. There are 555 BHB stars with temperatures lying between 22,000 $> T_{\rm eff} >$ 8,000~K. A comparison with HB models shows that a bunch of stars fall below the ZAHB model line, thus suggesting that there are a significant number of BHk stars among the EHB population.  

The WDs detected in this cluster are as hot as 75,000~K. This is one of the clusters in our sample having fewer FUV bright BSSs (only 10) that spans around 2.5 magnitude in F148W as compared to 4 magnitude in the second parameter pair cluster M3. In F148W vs F148W$-$F169M CMD, the bright BSSs follow the BSS model sequence with one of them showing a significant shift towards the WD model sequence.

\subsection{NGC~5897}
This is the most heavily reddened cluster in our sample with low central concentration (c=0.86 \cite{Harris1996}). This cluster's HB is mostly populated by BHB stars (158) in FUV-optical CMDs with colors ranging from $-0.13 < F148W-G < 4.4$ corresponding to $16,000 > T_{\rm eff} > 8,000$~K. In the UVIT-\textit{Gaia} detections, all stars have P$_{\mu} > $ 70\% except 8 (5 HB, 3 BSS) which do not have PM measurements available. We note that the BHB stars lying at the redder side of $F148W-G \sim 2$ deviate from the ZAHB models. This arises from an assumption of a constant extinction coefficient in FUV over the entire HB distribution, which is more prominent in a higher reddened cluster.  We report the detection of 3 EHB candidates (the UVIT-\textit{HST}) T$_{\rm eff}>$ 26,000~K and only one WD candidate with a temperature of 85,000~K. We also found 2 pHB candidates lying just above the end of the HB track. We detect 9 BSSs spanning around one mag in F148W. 

\subsection{NGC~6341 (M92)}
This is the most metal-poor cluster ($\rm [Fe/H] = -2.31$ dex) in our sample. With UVIT, the inner 20$''$ from the cluster centre is unresolved in FUV which mostly comprises the core ($r \sim 1.3 r_{c}$). After checking the sources that lie in this unresolved region in F275W vs F275W$-$F336W CMD (HUGS), we note that we are missing around 20 HB source detections in this field in the UVIT/F148W. Beyond $r > 20''$, there are 301 BHB stars with colors ranging from 0.2 $< (F148W-G) <$ 4.2 corresponding to effective temperatures 14,000 $> T_{\rm eff} >$ 8,000~K. The BHB stars clump roughly around the G-jump \citep{Grundahl1999}) corresponding to color (F148W$-$G) $\sim$ 1.4. We have also found 5 EHBs having T$_{\rm eff} >$ 23,000~K, where 4 are from the UVIT-\textit{Gaia} common field. We do not detect any pHB stars. The BSSs span more than 3 magnitude in F148W with the brightest one lying 2 magnitude below the ZAHB. Most of the BSSs are located on the redder side of the ZAMS. We also detect 3 SX Phe variables. The WDs detected are as hot as 60,000~K.\\

\subsection{NGC~6809 (M55)} 
This is the second most heavily reddened cluster in our sample (E(B-V) = 0.08). The FUV CMD of this cluster is mainly populated by 220 BHB stars (173 from the UVIT-\textit{Gaia} detections) lying in the color range 0.5 $< (F148W-G) <$ 4.4 corresponding to effective temperatures 13,000 $>$ T$_{\rm eff} >$ 8,000~K. We also detect 2 EHBs having temperatures $>$ 23,000~K. These stars lie below the ZAHB and thus can be BHk candidates. Similarly, there is also one HB star with temperature  $\sim$ 21,000~K. This star is fainter by 2 magnitude in F148W from the ZAHB and can be a candidate BHk. Due to higher reddening similar to NGC~5897, the HB distribution of the cluster deviate from the ZAHB models for colors redder than $F148W-G \sim 2$. In F148W vs F148W$-$F169M CMD, these three stars are fainter than the ZAHB. The hottest detected WD in this cluster has T$_{\rm eff} \sim$ 30,000~K as inferred from the WD cooling curve.

In F148W vs F148W$-$G CMD, the BSSs span around 2 magnitude with the brightest BSS being 3.5 magnitude fainter than the bluest end of the ZAHB. This cluster hosts the largest number of BSS variables (17) in FUV in our studied sample. Among the detected BSSs (27), one is an eclipsing binary, and 16 are SX Phes. In the case of F148W vs F148W$-$F169M CMD, the BSSs are distributed around the FSPS-BSS model sequence with a spread of 0.8 magnitude in F148W$-$F169M color. Three of the BSSs lying in the UVIT-\textit{Gaia} region are showing a shift towards the WD cooling curve. It indicates that this BSSs might have a UV excess which could be due to a hot companion associated with them \citep{Sahu_ngc5466}.

\begin{figure*}
\centering
\includegraphics[width=0.7\textwidth]{Figures/legend_n.pdf}\\
\includegraphics[width=0.4\textwidth]{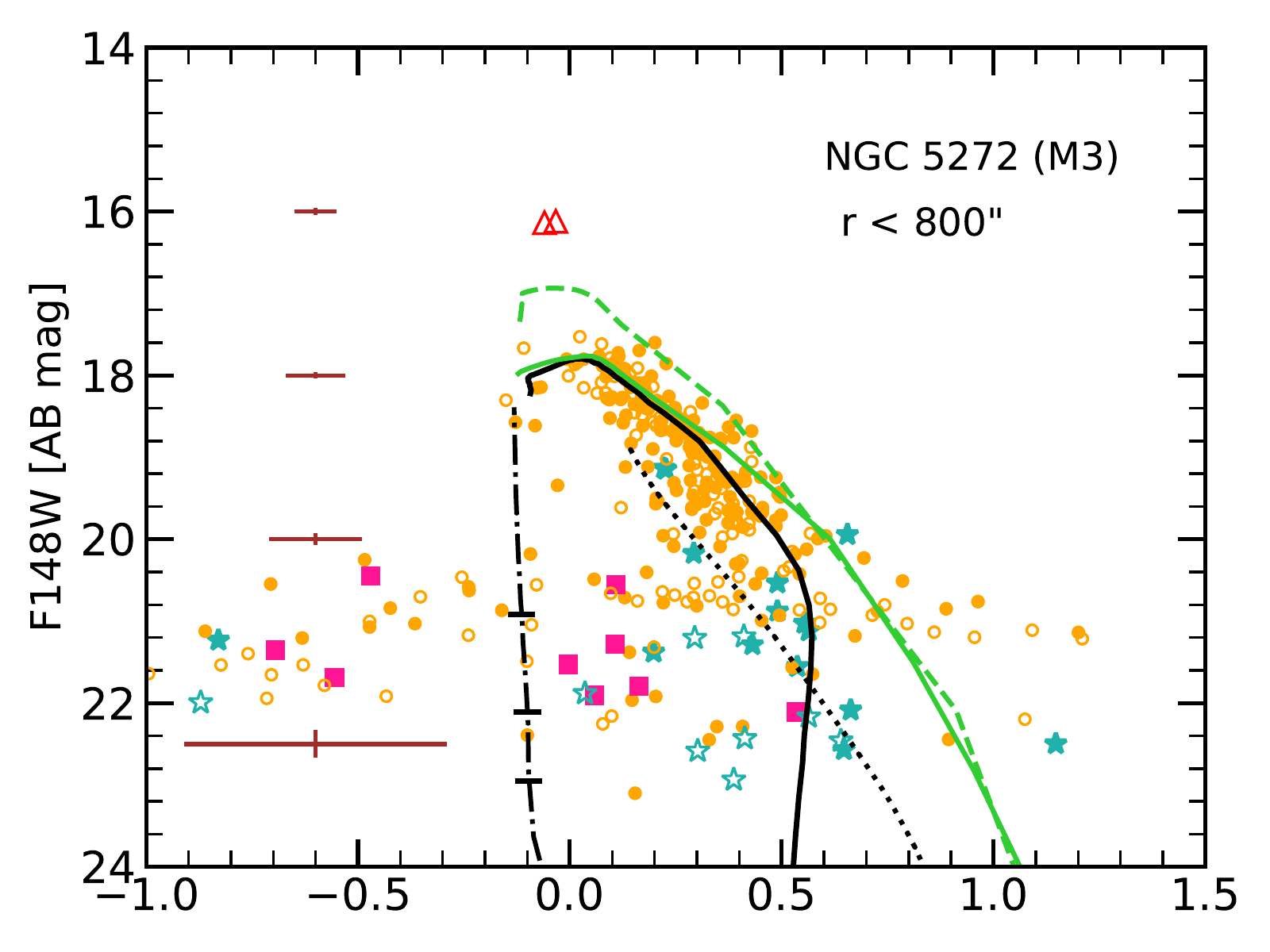}
\includegraphics[width=0.4\textwidth]{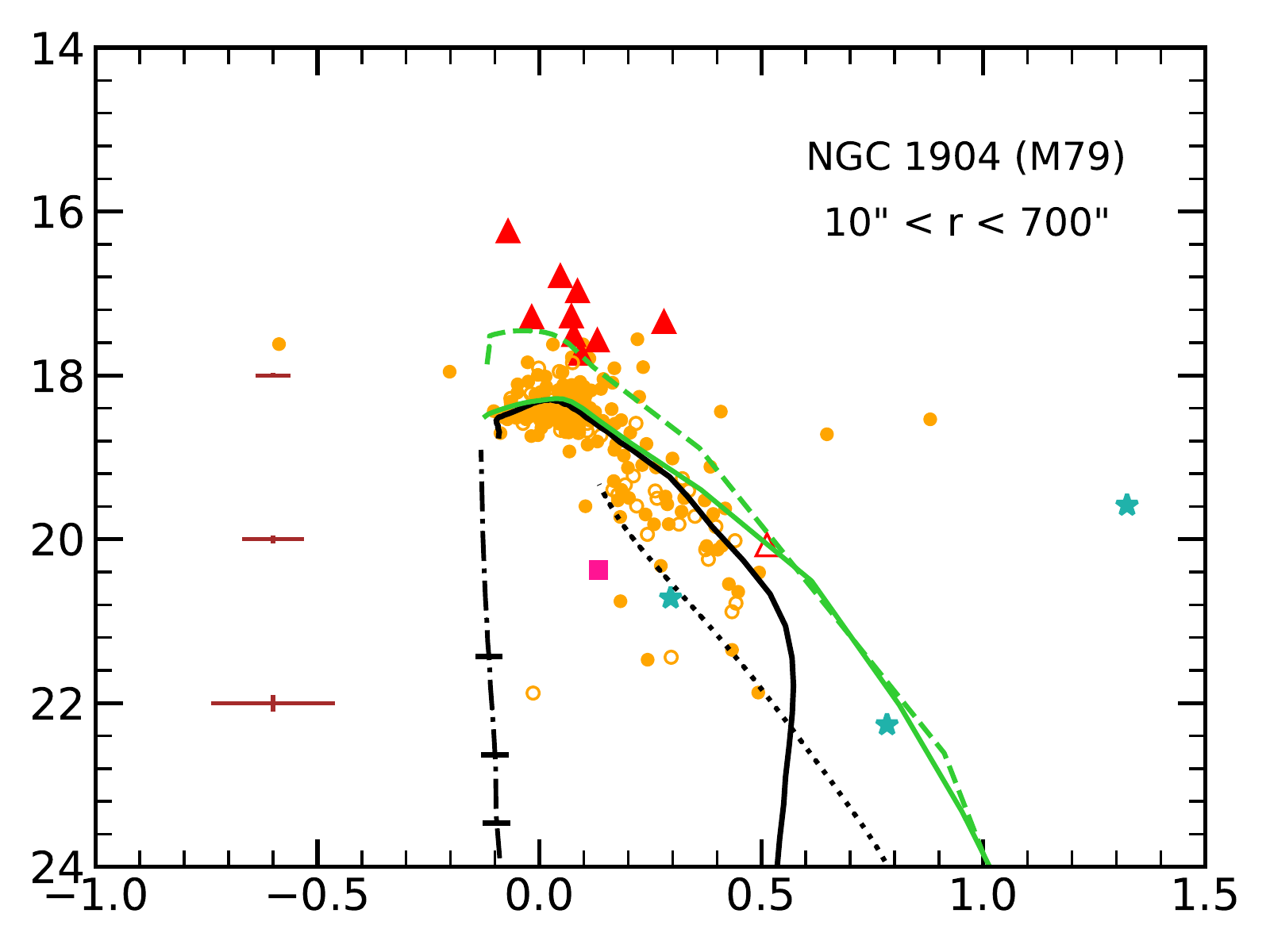}\\
\includegraphics[width=0.4\textwidth]{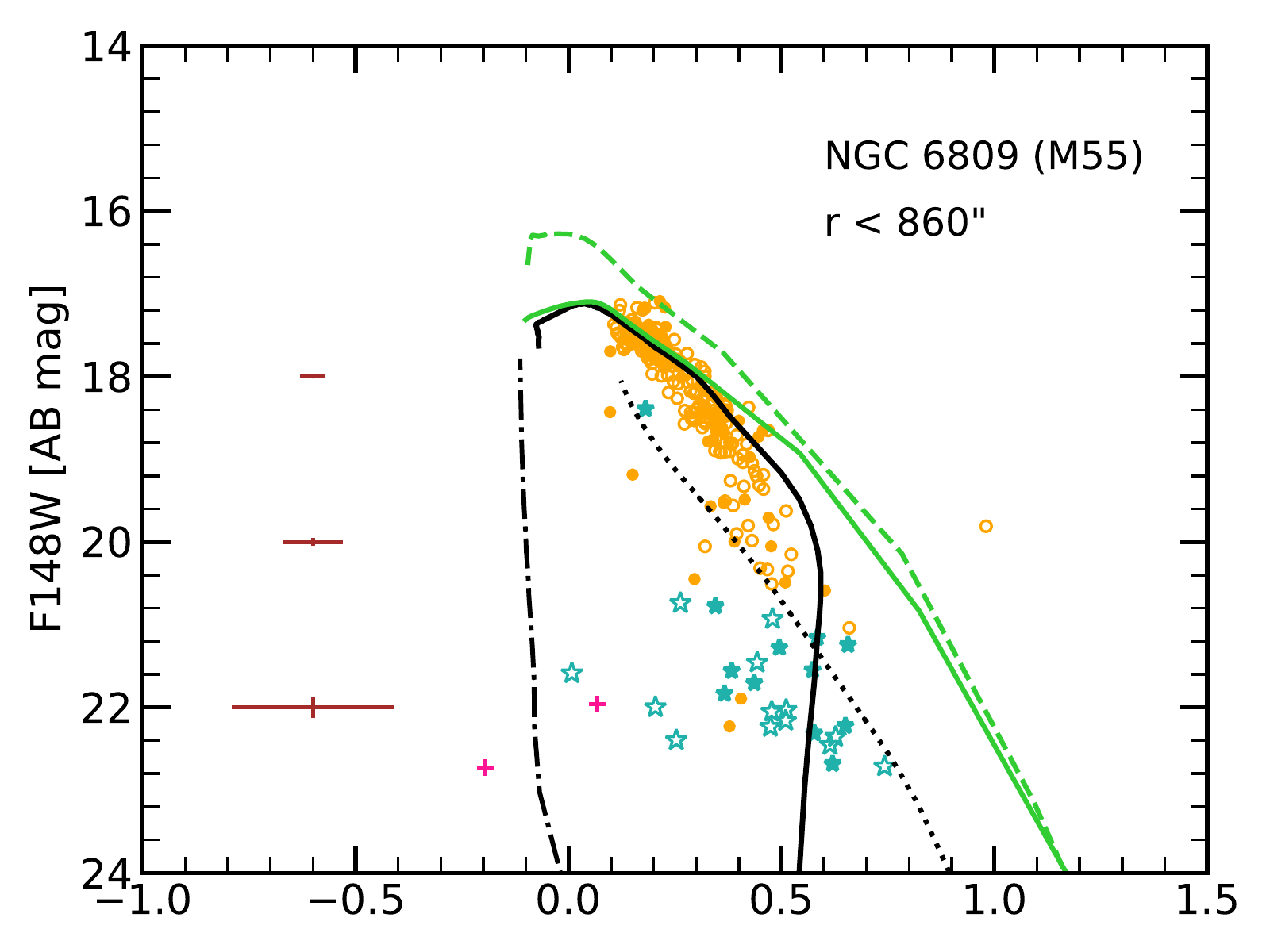}
\includegraphics[width=0.4\textwidth]{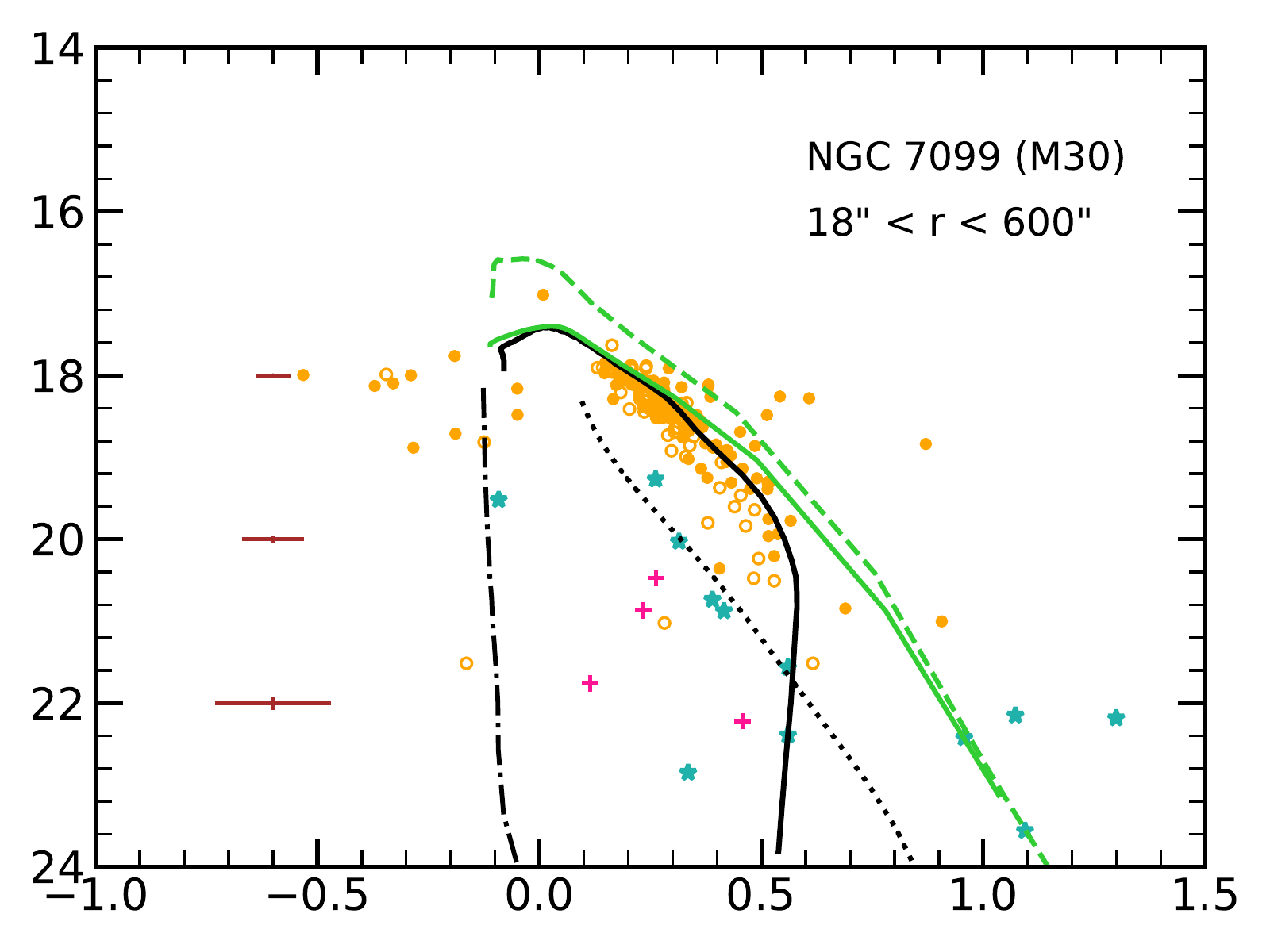}\\
\includegraphics[width=0.4\textwidth]{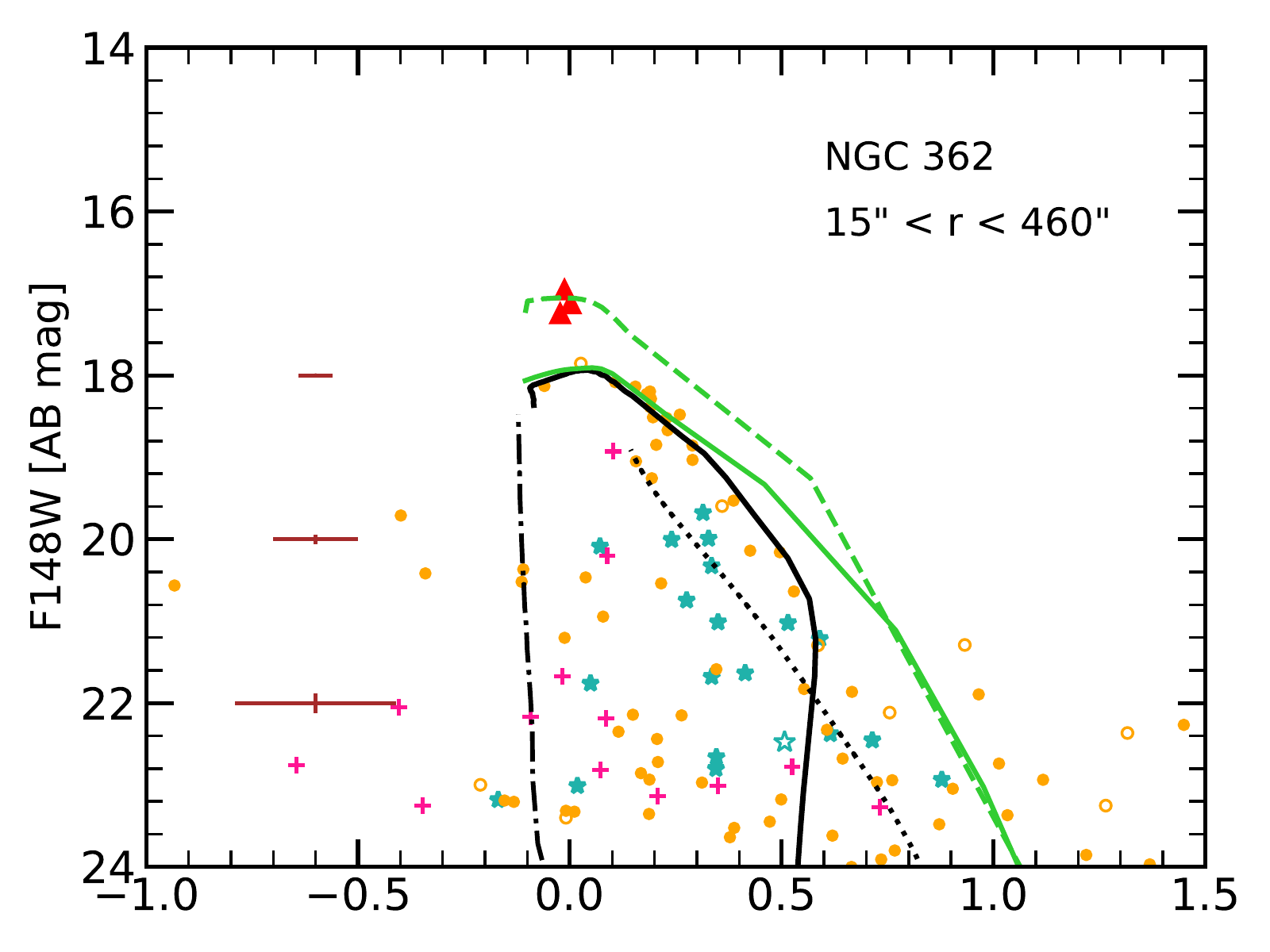}
\includegraphics[width=0.4\textwidth]{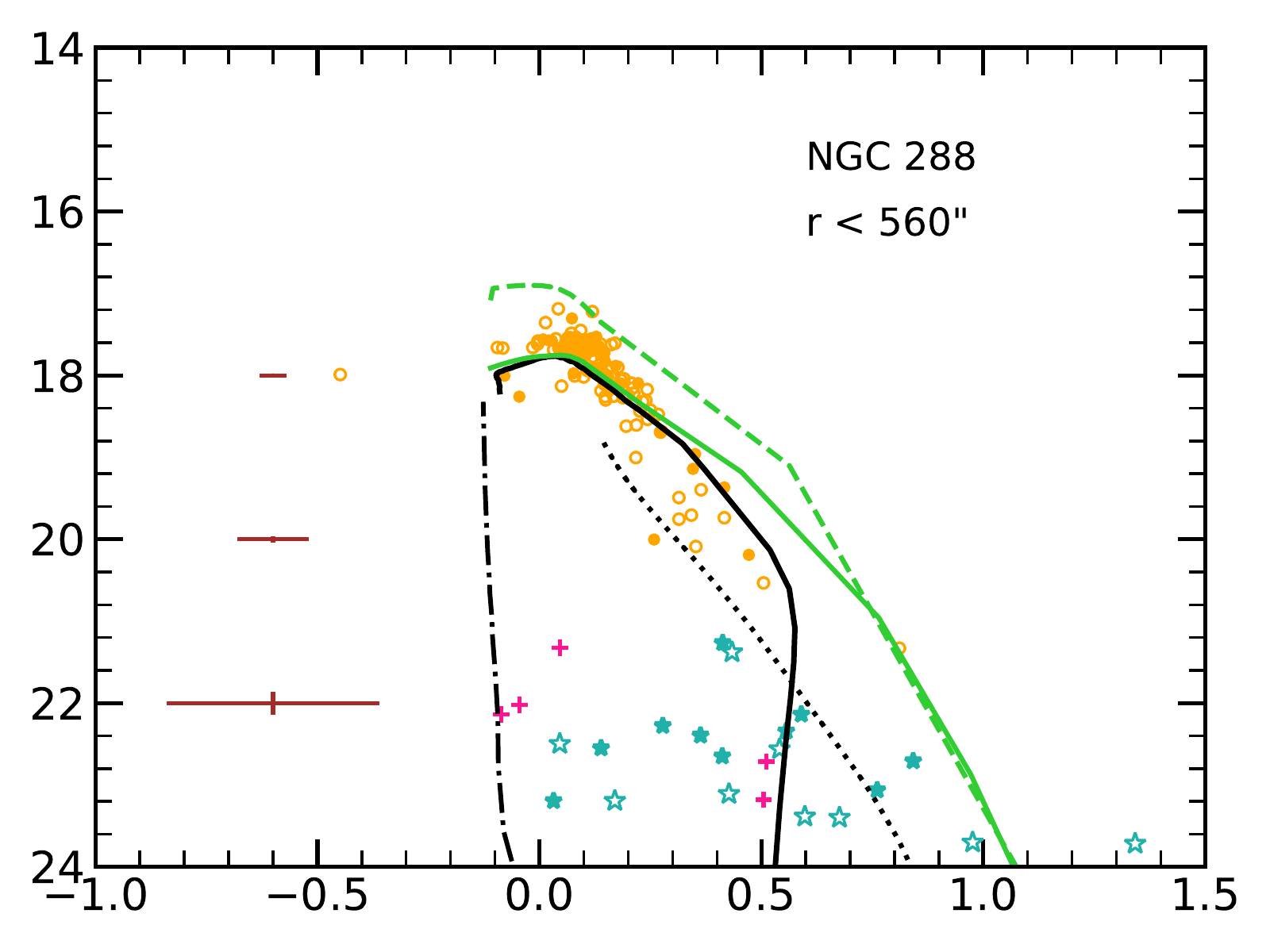}\\
\includegraphics[width=0.4\textwidth]{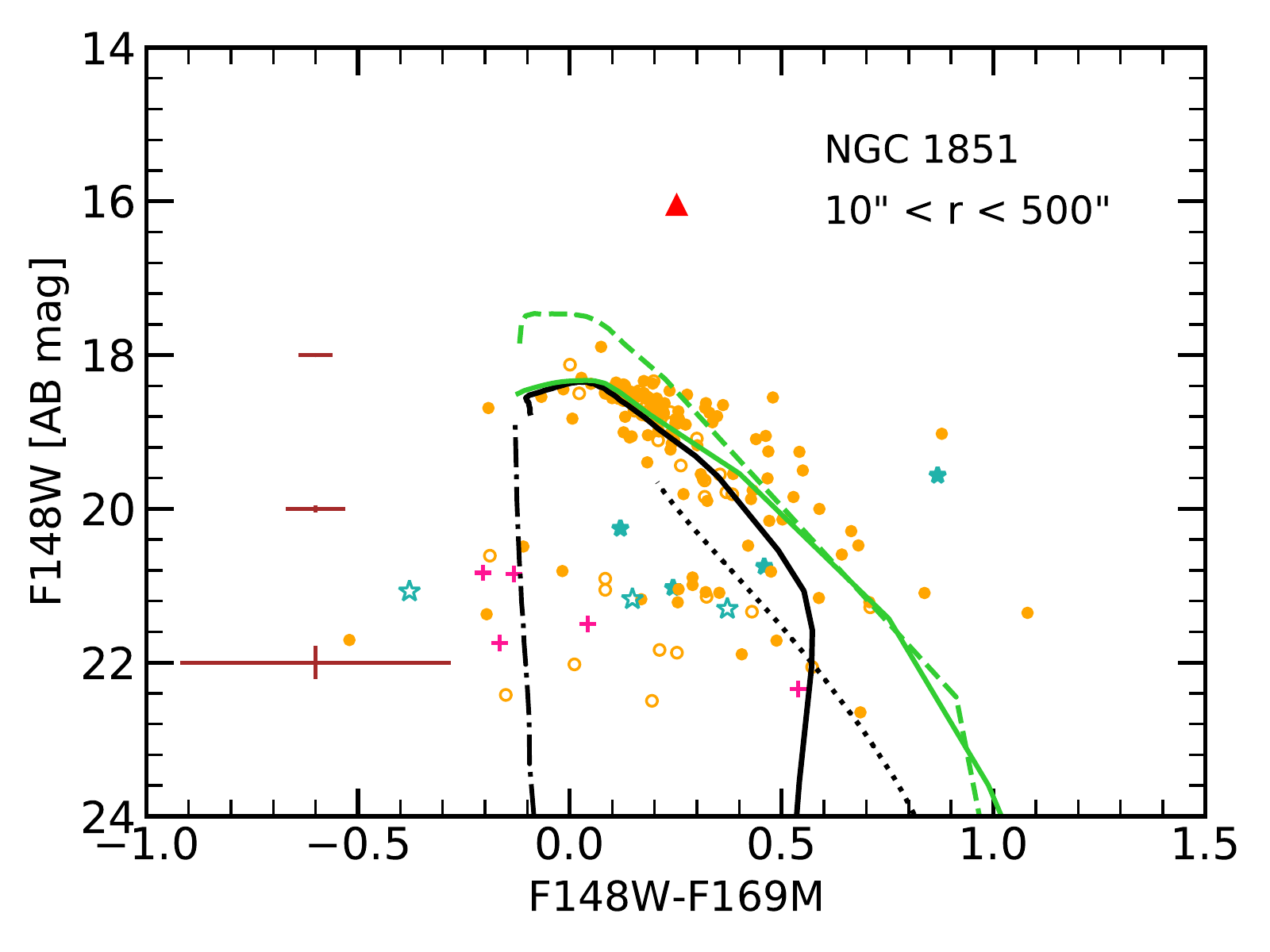}
\includegraphics[width=0.4\textwidth]{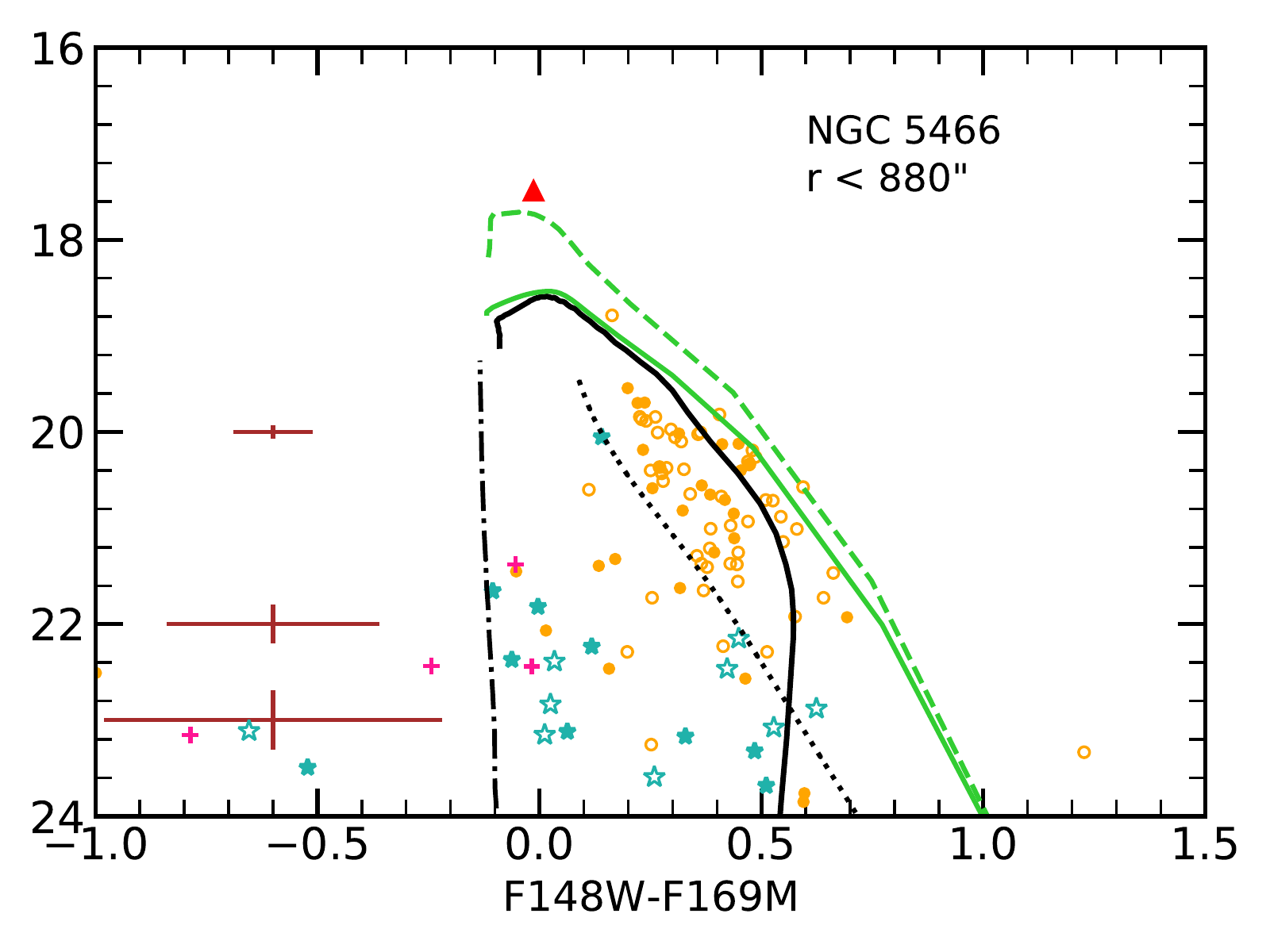}

\caption{F148W vs F148W$-$F169M CMDs of NGC~5272, NGC~1904, NGC~6809, NGC~7099, NGC~362, NGC~288, NGC~1851 and NGC~5466. The error bars are shown in brown color.}
\label{fig:fuv_cmds_4}
\end{figure*}

\subsection{NGC~7099 (M30)}
This metal-poor cluster has the highest central concentration (c = 2.5) in our studied sample. As this cluster has very tight core ($r_{c} = 0\farcm6$), the UVIT is unable to resolve the inner 18$''$ radius. Tallying this unresolved field with the HUGS catalogue, we found that there are around 15 HB and BSS sources located in this region that are not resolved in FUV. We detected 157 BHB stars in FUV with effective temperatures less than the G-jump. One HB is lying near the TAHB with $T_{\rm eff} >$ 34,000~K. It is possibly an EHB/AGBM candidate. We do not detect any pHB stars which is in agreement with \citep{Moehler2019}. In F148W vs F148W$-$F169M CMD, the BHBs form a very tight color sequence following the ZAHB models.

As this is a core-collapse cluster, almost all the FUV detected BSSs (16) are located in the UVIT-\textit{HST} region. The brightest BSSs lie around 1.5 magnitude below the ZAHB line with color $F148W-G = 1.94$. On comparing the distribution of BSSs in FUV-optical CMD with the two BSS sequences reported by \cite{Ferraro2009}, we note that the blue BSS sequence lie close to the ZAMS extension, whereas, few detected red BSSs lie 0.8 magnitude shifted from the ZAMS line. In F148W vs F148W$-$F169M CMD, we found that most of them lie around the BSS model sequence except one, which shifts towards the WD cooling curve. Among the detected BSSs, two are SX Phe variables and one is an eclipsing binary. The WDs detected in this cluster have effective temperatures ranging from 30,000 to 60,000~K.\\

\subsection{NGC~362}
The detected stars in this cluster lie in a region $15'' < r < 460''$. This cluster HB is mostly populated by RHB populations that show a very tight sequence in FUV-optical CMDs with few BHB stars. A comparison of the observed HB distribution with the PGPUC and BASTI models shows that most of them lie in ZAHB. We detected 27 BHBs (of which 8 are RRLs) spanning color range from $-0.25 < F148W-G < 4.6$ corresponding to effective temperatures 16,000 $>$ T$_{\rm eff} > 8,000$~K, in addition to 116 RHB stars. \cite{Dorman1997} studied this cluster with UIT and detected 36-43 blue HB stars using FUV-V CMDs where 4 are Supra-HB stars. The difference in BHB star detections from our study might be due to the number of stars we are missing in the central regions ($r < 15''$) which are not resolved by UVIT and due to membership cut-off. We detected 7 pHBs but found only 3 of them to be PM members.Whereas, four UVIT-\textit{Gaia} common detections are non-members according to \textit{Gaia} EDR3. 

We detected 28 FUV bright BSSs (24 with the UVIT-\textit{HST}) spanning more than 4 magnitude and found that majority of BSSs lie near the extension of ZAMS with a few brighter ones having redder F148W$-$G color. In F148W vs F148W$-$F169M CMDs, we detected BHB and a few RHB stars which nicely follow the PGPUC ZAHB track. HBs and BSSs show significant scatter at faint F148W magnitudes (F148W$>$21) which are arising due to large errors as shown in Figure~\ref{fig:fuv_cmds_4} for this cluster.\\

\subsection{NGC~1904 (M79)}
The UVIT field covers up to a radius of 3$'$ in the eastern side from the cluster centre. As this is a core-collapse cluster, the UVIT/FUV is unable to resolve its core (central 15$''$ region).  
This cluster mostly contains BHB (185) stars with colours varying from  $-0.87 < (F148W-G) < 4.4$ corresponding to temperatures, 19,800 $> T_{\rm eff} >$ 8,000~K as shown in Figure~\ref{fig:fuv_cmds_1}. Some BHB stars hotter are than the G-jump clump at around (F148W-G) $\sim$ 0.3~mag, corresponding to a mean $T_{\rm eff} \sim$ 14,400~K. This is roughly the temperature at which we start observing the BHB stars turning fainter in F148W vs (F148W$-$F169M) CMD (Figure~\ref{fig:fuv_cmds_3}), similar to NGC~ 6205. We detected 4 EHB stars satisfying our criteria of effective temperatures $>$23,000~K. We also detected nine pHB stars, all lying in the UVIT-\textit{HST} common field. The BSSs detected are very few (6), with one lying~2.5 mag above the MSTO i.e. near the brighter end of the ZAMS model sequence. \\

\section{}
This section shows the spatial distribution of HB and BSS stars in eight GCs detected in F148W filter of UVIT.
\begin{figure*}
\centering
\includegraphics[width=\linewidth]{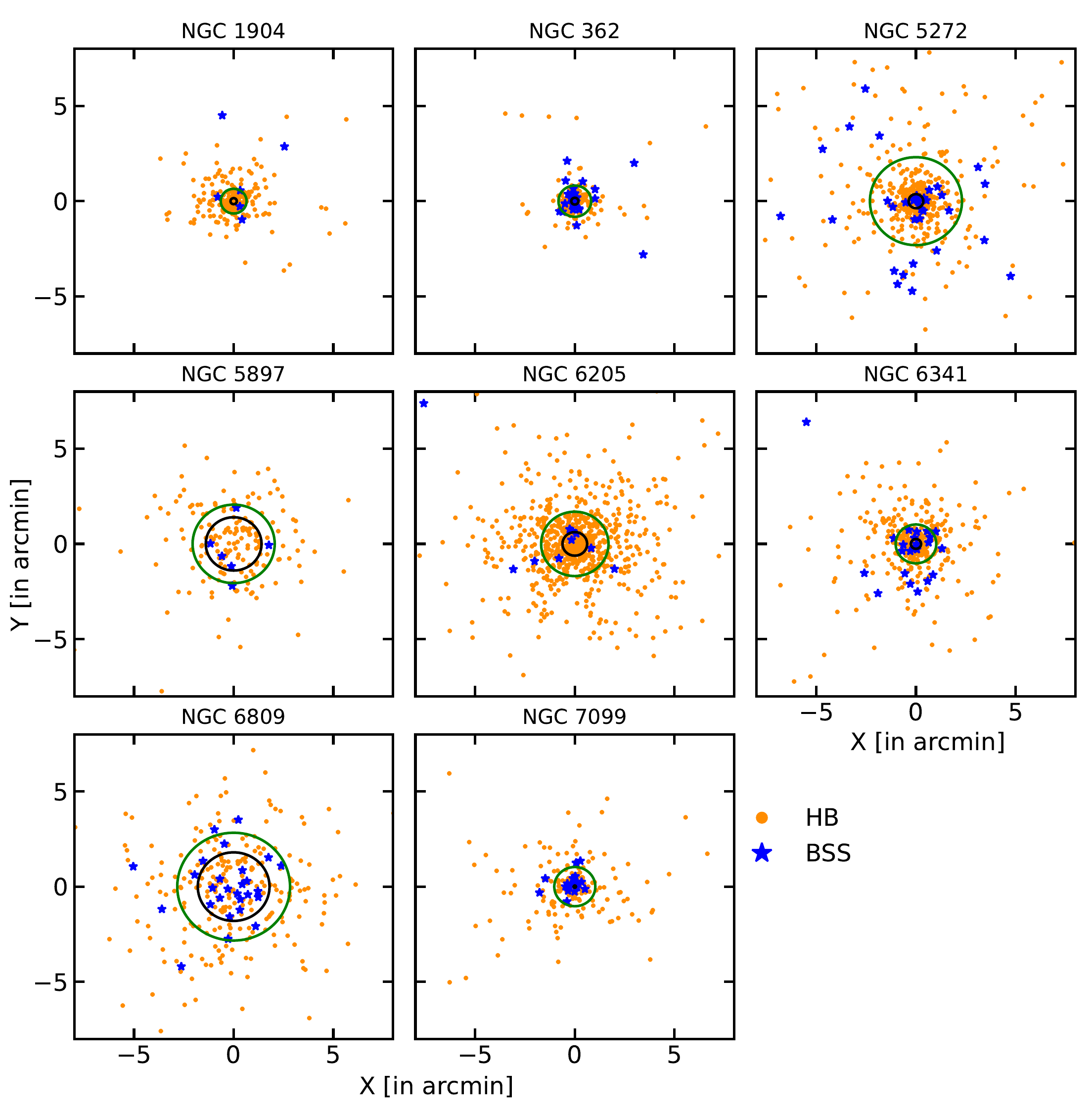}
\caption{Spatial distribution of HB and BSSs of 8 GCs detected in the UVIT/F148W marked in orange dots and blue stars respectively. For comparison, the core radius and half-light radius of the GCs \citep{Harris1996}, (2010 edition) are shown in black and green solid lines respectively.}
\label{fig:dist_plot}
\end{figure*}

\end{document}